\documentclass{article}

\usepackage{arxiv}

\usepackage[utf8]{inputenc} % allow utf-8 input
\usepackage[T1]{fontenc}    % use 8-bit T1 fonts
\usepackage{hyperref}       % hyperlinks
\usepackage{url}            % simple URL typesetting
\usepackage{booktabs}       % professional-quality tables
\usepackage{amsfonts}       % blackboard math symbols
\usepackage{nicefrac}       % compact symbols for 1/2, etc.
\usepackage{microtype}      % microtypography
\usepackage{lipsum}		% Can be removed after putting your text content
\usepackage{graphicx}
\usepackage{natbib}
\usepackage{doi}

\usepackage{comment}
\usepackage{amsmath}
\usepackage{amssymb}
\usepackage{xcolor}
\usepackage{soul}
\usepackage{multirow}
\usepackage[linesnumbered,ruled,vlined]{algorithm2e}
\usepackage{algorithmic}
\usepackage{subcaption}
\usepackage{dsfont}
\usepackage[normalem]{ulem}

\usepackage{xr}
\newcommand{\R}{\mathbb{R}}

\newcommand{\clr}{\mathrm{clr}}

\DeclareMathOperator*{\argmin}{arg\,min}

\newtheorem{corollary}{Corollary}[section]
\newtheorem{proposition}{Proposition}[section]
\newtheorem{definition}{Definition}[section]
\newtheorem{remark}{Remark}[section]
\newtheorem{lemma}{Lemma}[section]

\title{Robust functional PCA for relative data}

\author{ \href{https://orcid.org/0009-0009-0942-1622}{\includegraphics[scale=0.06]{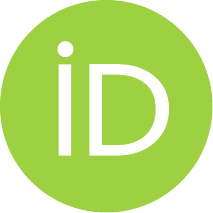}\hspace{1mm}Jeremy Oguamalam} \\
	Institute of Statistics and Mathematical Methods in Economics\\
	TU Wien\\
	Vienna, Austria\\
\And
	\href{https://orcid.org/0000-0002-8014-4682}{\includegraphics[scale=0.06]{orcid.pdf}\hspace{1mm}Peter Filzmoser} \\
	Institute of Statistics and Mathematical Methods in Economics\\
	TU Wien\\
	Vienna, Austria \\
\And
	\href{https://orcid.org/0000-0002-1847-6598}{\includegraphics[scale=0.06]{orcid.pdf}\hspace{1mm}Karel Hron} \\
	Department of Mathematics and Statistics\\
	Palacký University Olomouc\\
	Olomouc, Czechia\\
\And
	\href{https://orcid.org/0000-0003-0682-6412}{\includegraphics[scale=0.06]{orcid.pdf}\hspace{1mm}Alessandra Menafoglio} \\
	Department of Mathematics\\
	Politecnico di Milano\\
	Milan, Italy \\
 \And
	\href{https://orcid.org/0000-0003-0329-0595}{\includegraphics[scale=0.06]{orcid.pdf}\hspace{1mm}Una Radojičić} \\
	Institute of Statistics and Mathematical Methods in Economics\\
	TU Wien\\
	Vienna, Austria \\      
}

\hypersetup{
pdftitle={A template for the arxiv style},
pdfsubject={q-bio.NC, q-bio.QM},
pdfauthor={David S.~Hippocampus, Elias D.~Striatum},
pdfkeywords={First keyword, Second keyword, More},
}

\begin{document}
\maketitle

\begin{abstract}
	This paper introduces a robust approach to functional principal component analysis (FPCA) for relative data, particularly density functions. While recent papers have studied density data within the Bayes space framework, there has been limited focus on developing robust methods to effectively handle anomalous observations and large noise. To address this, we extend the Mahalanobis distance concept to Bayes spaces, proposing its regularized version that accounts for the constraints inherent in density data. Based on this extension, we introduce a new method, robust density principal component analysis (RDPCA), for more accurate estimation of functional principal components in the presence of outliers. The method's performance is validated through simulations and real-world applications, showing its ability to improve covariance estimation and principal component analysis compared to traditional methods.
\end{abstract}

% keywords can be removed

\section{Introduction}
Functional data analysis (FDA) has developed rapidly over the past decade (e.g., \cite{silverman2005}), yet many applied settings involve relative functional data—functions where the key information lies not in their absolute magnitude, but the dynamics or distributional pattern of the measured quantity carries relevant information. Examples include age-specific fertility or mortality rates, spectral curves in materials science, and normalized sensor signals, where data are typically rescaled and interpreted in terms of relative variation across a domain.\\
These data often share structural features with probability density functions (PDFs)—non-negativity, unit integral, and a relative (compositional) structure. While PDFs are a motivating case, the broader goal is to analyze functions where the shape or internal distribution carries the signal of interest, and scale is irrelevant or removed. For the simplicity of the notation, we refer to such functional data as \textit{densities}, emphasizing that the probability density functions are just one example of the functional data we consider in this context.\\    
To analyze such data, we use compositional data analysis (CoDA) extended to infinite dimensions, modeling functions as infinite-dimensional compositions in a Bayes space, a Hilbert space respecting their relative nature~\citep{egozcue2006, vanboogaart2010}. Within this framework, simplicial functional PCA (SFPCA)~\citep{hron2016} offers a dimension reduction tool tailored to relative functional data. However, SFPCA depends on the sample covariance operator and is sensitive to outliers, limiting its robustness in practice. To address this, robust alternatives using trimmed covariance estimators and center-outward orderings have been proposed. Depth measures such as spatial depth~\citep{Menafoglio2021} and others~\citep{fraiman2001, febrero2008, pintado2009, arribas-gil2014, Chakraborty2014} offer one route, while Mahalanobis-type distances—widely used in multivariate and functional settings~\citep{Stamm2013, galeano2015, ghiglietti2017statistical, berrendero2020}—provide another, such as the minimum regularized covariance trace (MRCT) estimator~\citep{oguamalam2024}. In this paper, we focus on the latter group of methods, extending the concept of regularized Mahalanobis distance for univariate functional data (later called RDMD) to the Bayes space. We then use this notion to construct a robust covariance estimator and a corresponding robust PCA method tailored for densities.

The rest of the paper is organized as follows: Section \ref{sec:preliminaries} discusses the concept of Bayes spaces and SFPCA and showcases potential problems that the outliers in the sample can cause. In Section \ref{sec:standardization} we propose a way of standardizing (whitening) relative functional data, thus extending and generalizing the notions of truncated- and $\alpha$-Mahalanobis distance from \cite{galeano2015} and \cite{berrendero2020}, respectively, to Bayes Hilbert spaces. In Section \ref{sec:RDPCA}, the novel, robust density PCA (RDPCA) method is introduced. Its performance is evaluated in a simulation study and real-data examples in Sections \ref{sec:RDPCA_simus} and \ref{sec:RDPCA_examples}, respectively. Proofs of technical results are given in Supplement Section \ref{supp:proofs}.

\section{Preliminaries} \label{sec:preliminaries}

In traditional (multivariate) compositional data analysis, the emphasis is on the information found in the relative relationships between variables. This approach led to the development of CoDA based on the Aitchison geometry \citep{pawlowsky15}. CoDA focuses on positive multivariate observations that are scale invariant, i.e. that can be arbitrarily rescaled equivalence classes of proportional vectors thus form the sample space of compositional data. Along with perturbation and powering operations these concepts form a linear space that operates on the sample space of compositions.

\subsection{Bayes space} \label{subsec:Bayesspace}

If we expand the horizon to infinite-dimensional compositions, we consider positive functions that are typically represented with unit integral constraint. Intuitively, such compositions can be thought of as density functions. Therefore, statistical concepts designed for unconstrained functional data need to be adapted, as they do not account for this feature, leading to potentially meaningless results when applied to densities \citep{delicado2011, hron2016, menafoglio2018}. Consequently, to work with functional compositions, it is necessary to establish the fundamental concepts of the underlying space structure.

The foundational results were initially presented by \cite{egozcue2006} and were subsequently extended by \cite{vanboogaart2010}, \cite{vanboogaart2014}, and \cite{talska2020}. The framework for probability density functions is based on a measurable space $(\Omega, \mathcal{A})$, typically set to $(\R, \mathfrak{B}(\R))$, with $\mathfrak{B}(\R)$ as the Borel $\sigma$-algebra on $\R$. Within this space, we consider the set of measures $\mu$ that are absolutely continuous with respect to a reference measure $\lambda$, here assumed to be the Lebesgue measure with compact support $I = [a,b] \subset \R$ and with length $\eta := b - a$. While the authors of \cite{vanboogaart2014} developed the theory of Bayes space for more general measures $\lambda$, in practice, commonly the Lebesgue measure is chosen. Furthermore, usually the measures $\mu$ are identified with their respective Radon-Nikodym derivative, density $f$. For these densities, a notion of equivalence in the Bayes sense for multiples is introduced, i.e $ f =_B g$ if $f = c \cdot g$ for some $c \in \R.$ W.l.o.g. within these $(=_B)$-equivalence classes, we take the element satisfying the unit integral constraint as the representative. 

To equip the sample space of these density functions with a structure of a Hilbert space on a given bounded domain,  \cite{egozcue2006} followed the ideas grounding the Aitchison geometry, and extended its operations $(\oplus, \odot)$ (a.k.a. perturbation and powering), and the inner product $\langle \cdot,\cdot\rangle$ to the infinite-dimensional setting by defining 
\begin{align*}
    (f \oplus g) (t) =_B f(t)g(t), \quad
    (\alpha \odot f)(t) =_B (f(t))^\alpha, \quad
    \langle f,g \rangle_{B^2} = \frac{1}{2\eta} \int_I \int_I  \text{log} \frac{f(t)}{f(s)}  \text{log} \frac{g(t)}{g(s)} \mathrm{d}t\, \mathrm{d}s,
\end{align*}
for densities $f, g$ and $\alpha \in \R$. In this context, negative perturbation $(\ominus)$ is defined as $f \ominus g = f \oplus (-1 \odot g) =_B f/g$. Endowing the subspace $\mathcal{B}^2(I)$ of positive densities on $I$ with a square-integrable logarithm \citep{vanboogaart2014} with this structure yields a separable Hilbert space. In practice, the centered logratio (clr) transformation 
\begin{align}\label{eq:fclr}
    \text{clr} (f) = \text{log} (f) - \frac{1}{\eta} \int_I (\text{log}(f))(t) \mathrm{d}t
\end{align}
is used to directly work in a space equipped with a Euclidean structure. Observe that all clr transformed densities satisfy the zero integral constraint.

Denoting $L_0^2(I)\subset L^2(I)$ the set of all square integrable functions on $I$ that additionally integrate to zero, it can be shown that the clr transformation is an isomorphism between $\mathcal{B}^2(I)$ and $L_0^2(I)$, with the corresponding inverse transformation $\clr^{-1}(f)=_B\exp{(f)}$, for $f\in L_0^2(I)$. The clr transformation \eqref{eq:fclr} enables thus the computation of perturbation, powering, and the inner product of densities through their transformed counterparts in the $L^2$ space. Let $f, g \in \mathcal{B}^2(I)$ and $\alpha,\beta \in \R$, then
\begin{align*}
\langle f, g \rangle_{B^2} =~ \langle \text{clr}(f), \text{clr}(g)\rangle_{L^2} ~~\text{and}~~
    \text{clr}((\alpha \odot f) \oplus (\beta \odot g)) =~ \alpha \cdot \text{clr}(f) + \beta \cdot \text{clr}(g),
\end{align*}
making clr transformation an isometry.

\subsection{Moments of random densities}

A Bayes space is a Hilbert space. Therefore, we use the notion of the mean and the covariance in a general Hilbert space structure to specify and inspect the properties of the Bayes mean density $\mu_{B^2,X}$ and Bayes covariance $C_{B^2,X}$ of 
$X\in\mathcal{B}^2(I)$; see e.g. \cite{kokoszka2017} for more detail.
For a random density $X\in \mathcal{B}^2(I)$, a \textit{Bayes mean density}  $\mu_{B^2,X}$ is an object in $\mathcal{B}^2(I)$ satisfying
\begin{equation}\label{eq:bayes_mean_1}
\langle \mu_{B^2,X}, f\rangle_{B^2}=\mathbb{E}(\langle X,f\rangle_{B^2}),
\end{equation}
for every $f\in \mathcal{B}^2(I)$. Alternatively, the Frechet mean approach used in \cite{petersen2022} can be used to obtain the same definition.  
%\paragraph{Bayes covariance operator}
Using the equivalent identity of covariance operators, we define the \textit{Bayes covariance operator} $C_{B^2,X}:\mathcal{B}^2(I)\to\mathcal{B}^2(I)$ of a random density $X\in\mathcal{B}^2(I)$ as a $\mathcal{B}^2(I)$ operator satisfying
$$
\langle C_{B^2,X}f,g\rangle_{B^2}=\mathbb{E}\left(\langle X\ominus\mu_{B^2,X},f\rangle_{B^2}\langle X\ominus\mu_{B^2,X},g\rangle_{B^2} \right),
$$
for every $f,g\in\mathcal{B}^2(I)$. We write $C_{B^2,X}=\mathbb{E}((X\ominus\mu_{B^2,X})\otimes_{B^2}(X\ominus\mu_{B^2,X})),$ 
where for $f,g\in \mathcal{B}^2(I)$ the outer product $f\otimes_{B^2} g:\mathcal{B}^2(I)\to \mathcal{B}^2(I)$ is the linear operator satisfying  $(f\otimes_{B^2} g)(h)=\langle g,h\rangle_{B^2} f$, for $h\in \mathcal{B}^2(I)$. We require the density variable $X$ to possess finite fourth moment, i.e., $\mathbb{E}[\|X\|^4_{B^2}]<\infty$. Under this assumption, the covariance operator is a symmetric positive-definite Hilbert-Schmidt operator and admits the eigen-decomposition
\[C_{B^2,X} f = \sum_{j=1}^{+\infty}\lambda_j\langle f, \xi_j\rangle \xi_j,\]
where $(\lambda_j, \xi_j)$ are the eigenpairs of $C_{B^2,X}$, conventionally ordered as $\lambda_1\geq\lambda_2\geq...\geq0$. Note that all the eigenvalues are non-negative and, under the conditions above, $C_{B^2,X}$ is trace-class, i.e., $\sum_{j=1}^{+\infty} \lambda_j <\infty$ (see, e.g., \cite{HorvathKokoszka2012}).
Lemma \ref{lemma:bayes_mean_cov} gives the connection between the Bayes mean density $\mu_{B^2,X}$ and covariance  $C_{B^2,X}$  and the “traditional" $L^2$-mean and the covariance of $\clr(X)$. The proof of Lemma \ref{lemma:bayes_mean_cov} can be found in Section \ref{supp:proofs} of the supplement. 

\begin{lemma}\label{lemma:bayes_mean_cov}
  Let $X\in\mathcal{B}^2(I)$ and $\clr(X)\in L^2(I)$ be its clr transformation. Then the  following identities hold:
  \begin{itemize}
      \item[i)]  $\clr(\mu_{B^2,X})=_{a.e}\mathbb{E}(\clr(X))$.
      \item[ii)] $   \clr(C_{B^2,X} f)=_{a.e.}C_{\clr(X)}\clr(f)$, for every $f\in\mathcal{B}^2(I)$.
      \item[iii)] Let $(\lambda_i,\zeta_i)$, $i\geq 1$ be the eigenpairs of $C_{B^2,X}$. Then, $(\lambda_i,\clr(\zeta_i)/\int_I\clr(\zeta_i(t))\mathrm{d}t)$, $i\geq 1$ are the eigenpairs of $C_{\clr(X)}$.
  \end{itemize}
 
\end{lemma}

\subsection{Simplicial functional PCA} \label{subsec:SFPCA}

As in the multivariate principal component analysis, functional PCA aims to search for the main modes of variability within a dataset, see e.g. \cite{silverman2005}. More recently, this concept was extended to the Bayes spaces setting in \cite{hron2016}. The authors coined the term \textit{simplicial functional PCA}, referring to the simplex, which is the multivariate counterpart of the Bayes space. More precisely, given a data set $X_1, \ldots, X_n \in \mathcal{B}^2(I)$, the problem of finding the first principal function $\zeta_1$ is defined as
\begin{align} \label{eq:fpca_bayes}
    \underset{\zeta \in  \mathcal{B}^2(I)}{\text{max} }\frac{1}{n} \sum_{i=1}^n \langle X_i\ominus\hat{\mu}_{B^2,X}, \zeta \rangle_{B^2}^2 ~~~ \text{subject to~} ||\zeta||_B^2 = 1,
\end{align}
where $\hat{\mu}_{B^2,X}=\frac{1}{n}\odot \bigoplus_{i=1}^n X_i$ is the sample Bayes-mean. Subsequent components $\{\zeta_i \}_{i\geq2}$ are then found by the same objective function under the additional orthogonality constraint $\langle \zeta_i , \zeta_j \rangle_{B^2} = 0$ for $j < i$. The principal functions can be found solving the eigenvalue problem 
\begin{align} \label{eq:bayes_eigenproblem}
    \hat{C}_{B^2,X} \left(\zeta_j\right) = \lambda_j \odot \zeta_j, ~ j \geq 1
\end{align}
where $\hat{C}_{B^2,X} (Y) = \frac{1}{n} \odot \bigoplus_{i=1}^n \langle X_i\ominus\hat{\mu}_{B^2,X}, Y \rangle_{B^2} \odot (X_i\ominus\hat{\mu}_{B^2,X})$ is the sample covariance operator in $\mathcal{B}^2(I)$ and $\lambda_j, j \geq 1$, are its eigenvalues. As discussed in \cite{hron2016}, this problem can be solved in $L^2(I)$ by using the clr transformation; see also Lemma \ref{lemma:bayes_mean_cov}. 

However, as previously emphasized, the eigendecomposition \eqref{eq:bayes_eigenproblem} and consequently also the optimization problem \eqref{eq:fpca_bayes}, are sensitive to the presence of outlying curves in the sample. Therefore, in order to achieve a robust estimation of the functional PCs, the underlying covariance will be based on a sub-sample consisting of the most central data points. The centrality of each observation will here be quantified by an adaptation of a regularized functional Mahalanobis distance for densities. For more insight into the notion of centrality/outlyingness for densities, see Remark \ref{rem:outlier}.

\begin{remark}\label{rem:outlier}
    In contrast to traditional FDA settings, identifying atypical curves in Bayes spaces requires rethinking outlier concepts. Because relative functions are compositional, local deviations affect the entire curve, and scaling, a key factor in conventional magnitude outliers, has no effect. Thus, outliers in this context reflect structural (shape) anomalies, such as unexpected redistribution of mass or signal across the domain. We detect such curves using the proposed concept of robust, regularized MD, flagging outliers based on the chosen quantile of its limiting distribution. We emphasize, however, that the goal of the present work is not to develop a method for outlier detection or to provide a taxonomy of outliers in relative functional data. Rather, we propose a novel approach to constructing robust PCA methods for relative functional data.
\end{remark}

\section{Notion of Mahalanobis distance for densities}\label{sec:standardization}

In recent years, various attempts to extend the concept of multivariate Mahalanobis distance to $L^2$ spaces have been proposed. The primary challenge in this functional extension is the non-invertibility of the covariance operator in infinite-dimensional cases. For instance, for a fixed $p$, \cite{galeano2015} defines the $p$-truncated Mahalanobis semidistance of $Y\in L^2(I)$ as a Mahalanobis distance of the corresponding optimal rank-$p$ projection; observe that it is indeed only a semidistance, as it lacks identifiability (zero truncated Mahalanobis distance does not necessarily imply a.s. zero process). Additionally, as argued in \cite{berrendero2020}, the partial sums corresponding to the Mahalanobis distance of optimal rank-$p$ projection diverge with $p$ and thus can depend heavily on the choice of $p$. \cite{ghiglietti2017statistical} proposed a modification of the approach presented in \cite{galeano2015} to deal with the convergence issues. \cite{berrendero2020}  introduced the concept of $\alpha$-Mahalanobis distance, motivated by the Reproducing Kernel Hilbert Space (RKHS) associated with the underlying stochastic process that generates the data. 

In a traditional multivariate setting, the Mahalanobis distance  $M(\textbf{x},\textbf{y};\boldsymbol{\Sigma})$ between two random vectors $\textbf{x},\textbf{y}\in\mathbb{R}^p$ with respect to a regular covariance matrix $\boldsymbol{\Sigma}\in\mathbb{R}^{p\times p}$ can be thought of as a Euclidean distance between the two vectors, previously standardized (whitened) by the common covariance $M^2(\textbf{x},\textbf{y};\boldsymbol{\Sigma})=(\textbf{x}-\textbf{y})'\boldsymbol{\Sigma}^{-1}(\textbf{x}-\textbf{y})=\|\boldsymbol{\Sigma}^{-1/2}\textbf{x}-\boldsymbol{\Sigma}^{-1/2}\textbf{y} \|^2.$ In that manner, the corresponding Mahalanobis norm of a random vector, i.e. the Mahalanobis distance of that random vector from its mean with respect to its covariance, equals the Euclidean norm of that random vector when standardized by its mean and the covariance. That is, $M(\textbf{x};\boldsymbol{\mu},\boldsymbol{\Sigma}):=M(\textbf{x},\boldsymbol{\mu};\boldsymbol{\Sigma})=\|\textbf{x}_{\mathrm{st}}\|$, where $\textbf{x}_{\mathrm{st}}:=\boldsymbol{\Sigma}^{-1/2}(\textbf{x}-\boldsymbol{\mu})$ is a solution to the linear problem $\boldsymbol{\Sigma}^{1/2}\textbf{x}_{\mathrm{st}}=\textbf{x}-\boldsymbol{\mu}$; for more details on whitening of multivariate data see e.g. \citep{Friedman1987EFDA,Oja2016, kessy2018optimal}. 

In infinite-dimensional spaces like $L^2(I)$, the covariance is not invertible, implying that this latter linear problem is ill-posed. One of the recent proposals on how to address this problem was given in \cite{oguamalam2024,berrendero2020}, using Tikhonov regularization. More specifically, given the regularization parameter $\alpha>0$, \cite{oguamalam2024} define the $\alpha$-standardization of a random function $Y\in L^2(I)$ with mean ${\mu}$ and covariance operator $C$ as a solution to the optimization problem
\begin{equation}\label{eq:eq_Tikhonov_L2}
   Y_\mathrm{st}^\alpha:=\argmin_{Z\in L^2(I)}\left\{\|C^{1/2}Z-(Y-{\mu})\|^2+\alpha\|Z\|^2 \right\}. 
\end{equation}

\cite{oguamalam2024}  show that the $\alpha$-Mahalanobis distance is the Euclidean norm of the solution to \eqref{eq:eq_Tikhonov_L2}. Regularization in \eqref{eq:eq_Tikhonov_L2} is also known as  $L^2$-regularization due to penalizing the $L^2$-norm of the solution and is perhaps the most widely used penalization form. However, it is often advantageous to consider regularization using a more general regularization operator $L:L^2(I)\to L^2(I)$; see e.g. \cite{morigi2007,WANG2012}. 

Here, we extend the optimization problem in \eqref{eq:eq_Tikhonov_L2} to Bayes spaces, allowing for a broader class of regularization operators under general conditions.

\subsection{Regularized standardization of density data}\label{sec:temp_rem}

Standardization (whitening) is generally defined with respect to the mean and the covariance of the random object being standardized. For the sake of readability, we first discuss the regularized standardization, assuming the mean and the covariance are known.  

Let, therefore, $X \in \mathcal{B}^2(I)$, with mean $\mu_{B^2}$ and covariance $C_{B^2,X}$ as defined in Section~\ref{sec:preliminaries}. Without loss of generality, assume that $X$ has a constant mean (the neutral element for addition in $\mathcal{B}^2(I)$ is the density of the uniform distribution over $I$, i.e., a constant function).  

Given the closed, densely defined operator $L:\mathcal{B}^2(I)\to\mathcal{B}^2(I)$ satisfying $\mathcal{N}(C_{B^2,X}^{1/2}) \cap \mathcal{N}(L) = \{0\}$, where $\mathcal{N}(\cdot)$ is the nullspace of the corresponding operator, we define the regularized standardization of $X\in \mathcal{B}^2(I)$ with respect to $L$ as the solution to the optimization problem
\begin{align}\label{eq:standardization_L}
X_{\mathrm{st}}^{\alpha,L} =_B 
    \argmin_{Y\in \mathcal{B}^2(I)} \left\{\|{C}_{B^2,X}^{1/2}Y\ominus X\|_{\mathcal{B}}^2 + \alpha\|L Y\|_{\mathcal{B}}^2   \right\},
\end{align}
where 
 $\displaystyle C_{B^2,X}^{1/2}Y=\bigoplus_{i=1}^\infty \lambda_i^{1/2} \langle \zeta_i,Y \rangle_{B^2} \odot \zeta_i,$ for $(\lambda_i,\zeta_i),\, i \geq 1$ being the $i$th eigenpair of ${C}_{B^2,X}$. Observe that $X_{\mathrm{st}}^{\alpha,L}$ is an approximate solution to $C_{B^2,X}^{1/2}X_{\mathrm{st}}^{\alpha,L} = X$, where $\alpha$ determines that level of approximation. \cite{locker1980regularization} (Theorem 3.1) guarantees the existence and uniqueness of the solution to \eqref{eq:standardization_L}.

For \( L = I \) (the identity operator), the optimization problem \eqref{eq:standardization_L} can be seen as the Bayes space counterpart of the \( \alpha \)-standardization in \eqref{eq:eq_Tikhonov_L2}. Proposition \ref{prop:L_2_B2_equivalence} shows the equivalence between \eqref{eq:standardization_L} in \( \mathcal{B}^2(I) \) and \eqref{eq:eq_Tikhonov_L2} in \( L^2(I) \) when applied to the \(\clr\)-transformed density.

\begin{proposition}\label{prop:L_2_B2_equivalence}
    For regularization parameter $\alpha>0$ 
    $$
    X_{\mathrm{st}}^{{\alpha},{I}} =_B
    \argmin_{Y\in \mathcal{B}^2(I)} \left\{\|{C}_{B^2,X}^{1/2}Y\ominus X\|_{\mathcal{B}}^2 + \alpha\|Y\|_{\mathcal{B}}^2   \right\}
    $$
    if and only if
    $$
     \clr(X_{\mathrm{st}}^{{\alpha},I})=\argmin_{Z\in L^2(I)}\left\{\|C_{\clr,X}^{1/2}Z-\clr(X)\|^2+\alpha\|Z\|^2 \right\}, 
    $$
    where $C_{B^2,X}$ and $C_{\clr, X}$ are the Bayes space covariance of $X$ and covariance of $\clr(X)$, respectively.  
\end{proposition}
Corollary \ref{cor:L_2 standardization explicit form}
in the Supplement \ref{supp:aux_results} gives the explicit solution to \eqref{eq:standardization_L} for $L=I$.

\paragraph{Choosing the regularization operator.}

Choosing regularization operators to be orthogonal projectors, as discussed in the work of \citet{morigi2007}, can be particularly advantageous when we have prior knowledge of certain features or structures within the data. A key motivation for using orthogonal projection regularization operators is that they allow us to directly control the components of the solution within the range and null space of $L$. By projecting onto a subspace that aligns with known properties of the data, we can tailor the regularization to preserve essential characteristics. The following proposition formalizes these insights and provides further motivation for selecting the appropriate regularization strategy by extending the findings of \cite{morigi2007} (Theorem 2.3) to optimization \eqref{eq:standardization_L}.
\begin{proposition}\label{prop:prop2}
    Using the notation and assumptions from Proposition \ref{prop:L_2_B2_equivalence} let $X_{\mathrm{st}}^{{\alpha},{L}}$ be the unique solution of the minimization problem \eqref{eq:standardization_L}. Consider now the perturbation of the original ill-posed problem $C_{B^2,X}^{1/2}Y = X$ modifying the right hand side as 
    $$
    C_{B^2,X}^{1/2}Y = X\oplus C_{B^2,X}^{1/2}\Tilde{X}
    $$ 
for some $\tilde{X} \in \mathcal{N}(L)$. Then the unique solution of the associated Tikhonov minimization problem
\[
    \argmin_{Y\in B^2(I)}\left\{\|C_{B^2,X}^{1/2}Y\ominus (X\oplus C_{B^2,X}^{1/2}\Tilde{X})\|_{B^2}^2+\alpha\|L Y\|_{B^2}^2 \right\}
\]
is given by $X_\mathrm{1,st}^{{\alpha},{L}}= X_\mathrm{st}^{{\alpha},{L}} \oplus \Tilde{X}$.
\end{proposition} 
If we think of $C_{B^2,X}^{1/2}X_{\mathrm{st}}^{\alpha,L}$ as a regularized approximation of $X$ (see introductory discussion in Section \ref{sec:temp_rem} and \cite{berrendero2020}), we get that $C_{B^2,X}^{1/2}X_\mathrm{1,st}^{{\alpha},{L}}=C_{B^2,X}^{1/2}X_\mathrm{st}^{{\alpha},{L}}\oplus C_{B^2,X}^{1/2}\Tilde{X}$, i.e. the perturbed part was recovered exactly. In other words, Proposition \ref{prop:prop2} suggests that the regularization operator should be chosen so that the known features of the desired solution to \eqref{eq:standardization_L} can be represented using elements in its null space. 

For further illustration, we exemplify the result of Proposition \ref{prop:prop2} by considering a regularization operator based on simplicial functional principal components. For this purpose, observe that $X$ admits a Karhunen-Loève representation
$$
X=\left(\bigoplus_{i=1}^k\langle X,\zeta_i\rangle_{B^2} \odot\zeta_i\right)\oplus\left(\bigoplus_{i>k}\langle X,\zeta_i\rangle_{B^2}\odot \zeta_i\right),
$$
where for $k\geq 0$, such that $\lambda_k>\lambda_{k+1}$, $\displaystyle X_{1:k}:=\bigoplus_{i=1}^k\langle X,\zeta_i\rangle_{B^2} \odot\zeta_i$ is the best rank-k approximation of $X$, and  $X_{(k+1):\infty}:=\bigoplus_{i>k}\langle X,\zeta_i\rangle_{B^2}\odot \zeta_i$ the residual. Additionally, it is often argued that the projection of the data onto the first few principal components is an effective way to retain the most significant features of the original dataset \citep{ramsay2005,hron2016}. Consider, therefore, the projection of the data onto the span of the first $k$ eigenfunctions of $C_{B^2,X}$ to be the data feature we would like to preserve.  As $C_{B^2,X}^{1/2}X_{1:k}\in \mathrm{span}\{\zeta_1,\dots,\zeta_k\}$, by taking $L$ to be the orthogonal projection onto the complement of $\mathrm{span}\{\zeta_1,\dots,\zeta_k\}$, the regularized approximation $C_{B^2,X}^{1/2}X_{\mathrm{st}}^{\alpha,L}$ of $X$ is of the form  
$C_{B^2,X}^{1/2}X_{\mathrm{st}}^{\alpha,L}=X_{1:k}\oplus C_{B^2,X}^{1/2}X_{2,\mathrm{st}}^{\alpha,L}$, where $X_{2,\mathrm{st}}^{\alpha,L}\in\mathrm{span}\{\zeta_{k+1},\zeta_{k+2},\dots\}$ is the 
$L^2$-regularized standardization of $X_{(k+1):\infty}$.

Thus, denoting ${W}(k)$ a projection onto the complement of the space spanned by the first $k\geq 0$ eigenfunctions of $C_{B^2,X}$, hereinafter we consider standardization of the form 
\begin{align}\label{eq:standardization_B^2_final}
X_{\mathrm{st},C_{B^2,X}}^{\alpha,k}=\argmin_{Y\in \mathcal{B}^2(I)} \left\{\|{C}_{B^2,X}^{1/2}Y\ominus X\|_{B^2}^2 + \alpha\|W(k) Y\|_{B^2}^2 \right\},
\end{align}
where we add the additional subscript $C_{B^2,X}$ to emphasize that the regularized standardization is performed with respect to $C_{B^2,X}$, and superscript $k$ is used instead of $W(k)$ for simplicity. The following proposition gives the exact form of the solution $X_{\mathrm{st},C_{B^2,X}}^{\alpha,k}$ and also shows that $\clr(X_{\mathrm{st},C_{B^2,X}}^{\alpha,k})$  is a solution to the $L^2$-equivalent of optimization problem \eqref{eq:standardization_B^2_final}.

\begin{proposition}\label{prop:L2_B2_equivalence_projector_opertor}
    Let $X\in\mathcal{B}^2(I)$ be a constant mean random density with Bayes-covariance $C_{B^2,X}$ and let $(\lambda_i,\zeta_i)$ be its $i$th eigenpair; $C_{B^2,X}\zeta_i=\lambda_i\odot\zeta_i$, $i\geq 1$ . For $k\geq 0$, such that $\lambda_k>\lambda_{k+1}$ let $W(k)$ be the projector onto the complement of the space spanned by the first $k$ eigenfunctions $\zeta_1,\dots,\zeta_k$ of $C_{B^2,X}$. Then there exists a unique solution $X_{\mathrm{st},C_{B^2,X}}^{{\alpha},k}$ of \eqref{eq:standardization_B^2_final}. Additionally, 
    $$
    X_{\mathrm{st},C_{B^2,X}}^{{\alpha},k} =_B
    \argmin_{Y\in \mathcal{B}^2(I)} \left\{\|{C}_{B^2,X}^{1/2}Y\ominus X\|_{\mathcal{B}}^2 + \alpha\|W(k)Y\|_{\mathcal{B}}^2   \right\}
    $$
    if and only if 
    $$
     \clr(X_{\mathrm{st},C_{B^2,X}}^{{\alpha},k})=\argmin_{Z\in L^2(I)}\left\{\|C_{\clr,X}^{1/2}Z-\clr(X)\|^2+\alpha\|V(k)Z\|^2 \right\}, 
    $$
    where $V(k)$ is the projector of the orthogonal complement onto the space spanned by the first $k$ eigenfunctions of the covariance $C_{\clr, X}$ of $\clr(X)$. In that case, 
    $$
    X_{\mathrm{st},C_{B^2,X}}^{{\alpha},k} =_B
    \sum_{i=1}^k {\frac{1}{\lambda_i^{1/2}}}\langle X,\zeta_i\rangle_{B^2}\zeta_i\bigoplus \sum_{i=k+1}^\infty {\frac{\lambda_i^{1/2}}{\lambda_i + \alpha}}\langle X,\zeta_i\rangle_{B^2}\zeta_i.
    $$
\end{proposition}

The technical assumption $\lambda_k>\lambda_{k+1}$ ensures well-definiteness of the projector onto the space spanned by the first $k$ eigenfunctions and poses no restriction in practice. Specifically, those PCs having the same variance are considered to be \textit{equally important}, and should then all or none be regularized.

\subsection{Regularized Mahalanobis distance for density data}\label{subsec:aMHD}

Having introduced the regularized standardization of data in Bayes space, we are ready to define the corresponding regularized Mahalanobis distance for probability density functions.

\begin{definition}[Regularized Mahalanobis distance
between two densities]\label{def:Mah_distance_XY}
    {Given a constant $\alpha> 0$, and the number of non-regularized components $k\geq 0$, the squared RDMD between $X,Y\in \mathcal{B}^2(I)$, with respect to the covariance operator $C$ is given by}
\begin{align*}
    M^2_{\alpha,k,B^2}(X,Y; C) =&~ || X_{\mathrm{st},C}^{{\alpha},k} \ominus  Y_{\mathrm{st},C}^{{\alpha},k} ||_{B^2}^2,
\end{align*}
where $X_{\mathrm{st},C}^{{\alpha},k},\,Y_{\mathrm{st},C}^{{\alpha},k}$ are the solutions to $$
 \min_{Z\in \mathcal{B}^2(I)} \left\{\|{C}^{1/2}Z\ominus X\|_{\mathcal{B}}^2 + \alpha\|W(k)Z\|_{\mathcal{B}}^2   \right\}\quad\text{and}\quad\min_{Z\in \mathcal{B}^2(I)} \left\{\|{C}^{1/2}Z\ominus Y\|_{\mathcal{B}}^2 + \alpha\|W(k)Z\|_{\mathcal{B}}^2   \right\},
$$
respectively.
\end{definition}
The RDMD of random density $X$ is, as in the multivariate setting, defined as the RDMD~\ref{def:Mah_distance_XY} between $X$ and its mean $\mu_{B^2,X}$, with respect to its covariance $C_{B^2,X}$. For the sake of further referencing, the RDMD of $X\in \mathcal{B}^2(I)$ is formally defined as follows.

\begin{definition}[Regularized Mahalanobis distance for density data]\label{def:Mah_distance_X}
    {Given a constant $\alpha> 0$, and the number of non-regularized components $k\geq 0$, the squared RDMD of the random density $X\in \mathcal{B}^2(I)$ with mean $\mu_{B^2,X}$ and covariance $C_{B^2,X}$ is given by}
\begin{align*}
    M^2_{\alpha,k,B^2}(X) =&~ M^2_{\alpha,k,B^2}(X,\mu_{B^2,X};C_{B^2,X}),
\end{align*}
where $M^2_{\alpha,k,B^2}(X,\mu_{B^2,X};C_{B^2,X})$ is defined in Definition~\ref{def:Mah_distance_XY}.
respectively.
\end{definition}

One can show that the distribution of the squared RDMD under the assumption of Gaussianity is an infinite weighted sum of independent $\chi^2(1)$ variables. For the sake of conciseness, the formal result is given in Proposition \ref{prop:Mah_distribution} in Supplement \ref{supp:aux_results}.

We conclude this section with a discussion of the connection to the functional MD introduced in \cite{galeano2015} and \cite{berrendero2020}.  

\begin{proposition}\label{prop:connection_to_other_Mah_dist}
     Let $X \in \mathcal{B}^2(I)$ with w.l.o.g. constant mean $\mu_{B^2,X}\equiv_{B^2} 1$ and $\clr(X)\in L^2(I)$ be the corresponding clr transformation with covariance $C_{\clr(X)}=\sum_{i=1}^\infty \lambda_i\xi_i\otimes\xi_i$, where $(\lambda_i,\xi_i)$ is the $i$th eigenpair of $C_{\clr(X)}$, such that $\lambda_i>0$ for $i\leq k$. Let further $M^2(\clr(X);k)=\sum_{i=1}^k\lambda_i^{-1}\langle\clr(X), \xi_i\rangle^2$, and $M_\alpha^2(\clr(X))=\sum_{i=1}^\infty \frac{\lambda_i}{(\lambda_i+\alpha)^2}\langle\clr(X), \xi_i\rangle^2$ be the k-truncated Mahalanobis distance \citep[Definition 2.1]{galeano2015} and $\alpha$-Mahalanobis distance \citep[Definition 1]{berrendero2020}, respectively. The following identities then hold:
     \begin{itemize}
         \item[i)] $M_{\alpha,0,B^2}^2(X)=M_\alpha^2(\clr(X))$,
         \item[ii)]  $\displaystyle\lim_{\alpha\to\infty}M_{\alpha,k,B^2}^2(X)=M^2(\clr(X);k)$. 
     \end{itemize}
\end{proposition}

Proposition \ref{prop:connection_to_other_Mah_dist} shows that functional MD introduced by \cite{galeano2015,berrendero2020} can be regarded as extreme cases of the RDMD from Definition \ref{def:Mah_distance_X}.

\section{Robust density PCA (RDPCA)}\label{sec:RDPCA}

\subsection{Robust Bayes covariance estimation}
Given the data sample $X_1,\dots,X_n$ and a subset $H\subseteq\{1,\dots,n\}$, $|H|=h\in[n/2,n]$ we define the trimmed Bayes mean and covariance estimators calculated using an $H$-subset of the data as
\begin{equation} \label{eq:trimmed_mean_covariance}
    \hat{\mu}_{B^2,H}=\frac{1}{h}\odot\bigoplus_{i\in H}X_i,\quad \hat{C}_{B^2,H,c_H}=c_H\frac{1}{h}\odot\bigoplus_{i\in H}(X_i\ominus\hat{\mu}_{B^2,H})\otimes_{B^2} (X_i\ominus\hat{\mu}_{B^2,H}).
\end{equation}
The constant \( c_H \) is a scaling factor to adjust for trimming and is determined under the assumption of Gaussianity.  For that purpose, we use the corollary of Proposition \ref{prop:Mah_distribution}.
\begin{corollary}\label{cor:cor_1}
Let $X\in \mathcal{B}^2(I)$ be a random density such that $\clr(X)\in L^2(I)$ is a Gaussian random process, and let $\hat{C}_{B^2,X}$ and $\hat{\mu}_{B^2,X}$ be strongly consistent estimators of the covariance operator $C_{B^2,X}$ and the constant mean $\mu_{B^2,X}\equiv_{B^2} 1$, respectively. Then, for $\alpha>0$ and $k\in \mathbb{N}_0$, such that $\lambda_k>\lambda_{k+1}$, $M_{\alpha, k,B^2}^2(X, \hat{\mu}_{B^2,X}; \hat{C}_{B^2,X})$ converges in distribution to 
   $$
  M^2_{\alpha,k,B^2}(X)\sim \begin{cases}
       y +  \sum_{i=k+1}^\infty \frac{\lambda_i^2}{(\lambda_i+\alpha)^2} \eta_i,\text{ for } k\geq 1,\\
      \sum_{i=1}^\infty\frac{\lambda_i^2}{(\lambda_i+\alpha)^2} \eta_i, \text{ for } k=0,
   \end{cases}
   $$
where $\eta_i\sim\chi^2(1)$, $i\geq 1$ are mutually independent and also independent of $y \sim \chi^2(k)$, and  $\lambda_i$, $i\geq 1$ is the $i$th eigenpair of $C_{B^2,X}$.
\end{corollary}

For more details on the necessity of scaling the trimmed covariance, see, e.g., \cite{boudt2020minimum,oguamalam2024}.  Using the notion of RDMD as a centrality measure, we choose the trimming subset $H_0$ as a solution to
\begin{align}\label{eq:functional_CMT_true}
H_0 &= \argmin_{\{H \subset \{1, \dots, n\}; |H| = h\}} \frac{1}{h} \sum_{i \in H} M_{\alpha, k,B^2}^2(X_i, \mu_{B^2,X}; C_{B^2,X}),
\end{align}
where $\alpha>0$ and $k\in\mathbb{N}_0$ are a priori chosen regularization parameters. 

Since the true mean and covariance operator are unknown,  we replace them in \eqref{eq:functional_CMT_true} with the current robust estimates. This approach leads to an implicit equation for obtaining the optimal subset:
\begin{align}\label{eq:fixed_point_alpha_2}
H_0 &= \argmin_{\{H \subset \{1, \dots, n\}; |H| = h\}} \frac{1}{h} \sum_{i \in H} 
M_{\alpha,k,B^2}^2(X_i, \hat{\mu}_{B^2,H_0}; \hat{C}_{B^2,H_0,c_{H_0}}).
\end{align}
\cite[Lemma 3.1]{oguamalam2024} guarantees that the equation \eqref{eq:fixed_point_alpha_2} has a solution.

Finally, Corollary~\ref{cor:cor_1} implies that $c_{H_0}$ can be estimated by matching the sample median of $\{M_{\alpha,k,B^2}^2(X_i, \hat{\mu}_{B^2,H_0}; \hat{C}_{B^2,H_0,c_{H_0}}),\, i=1,\dots,n\}$ with the estimate of the median of the corresponding limiting distribution given in Corollary \ref{cor:cor_1}, where the eigenvalues $\lambda_i,\, i\geq 1$ of $C_{B^2,X}$ are again replaced by their robust estimates; 
\begin{equation*}\label{eq:k}
c_{H_0}=\mathrm{med}\left\{M_{\alpha,k,B^2}^2(X_i, \hat{\mu}_{B^2,H_0}; \hat{C}_{B^2,H_0,c_{H_0}}),\, i=1,\ldots,n\right\}/\mathrm{med}\left\{  y +  \sum_{i=k+1}^\infty \frac{\hat{\lambda}_i^2}{(\hat{\lambda}_i+\alpha)^2} \eta_i\right\},
\end{equation*}
where $\eta_i\sim \chi^2(1)$, $i\geq 1$, are mutually independent and are independent of $y\sim\chi^2(k)$ for $k\geq 1$. We note that, for the purpose of PCA, the scaling is redundant. We include it, however, for completeness.

\begin{remark}\label{rem:Gaussianity}
Corollary \ref{cor:cor_1} can be used to detect potential outliers by comparing the calculated RDMD to an appropriate quantile from the limiting reference distribution; see Remark \ref{rem:outlier}. However, we emphasize how RDMD and the underlying robust estimation do not rely on Gaussianity.  In practice, RDMD values can highlight atypical observations, with non-parametric cutoffs, like, e.g., empirical quantiles or boxplots, used without distributional assumptions.  The Gaussianity is invoked only to enable an automated,  theoretically grounded cutoff for outlier detection, which may be useful in applications such as re-weighting in robust covariance estimation, as e.g. in \cite{rousseeuw1999, berrendero2020, oguamalam2024}. \end{remark}

While general consistency results under a non-contaminated model like those of \cite{CROUX1999,butler1993asymptotics} for the MCD are beyond the scope of this work, we provide a partial result under a clean model, where the trimming proportion grows sublinearly with the sample size. In this setting, we show that the robust estimators \eqref{eq:trimmed_mean_covariance} and the associated principal functions converge to the population counterparts. These results are given in Lemma \ref{sup_lemma:consistency} in Supplement Section \ref{supp:aux_results}, along with simulation studies that empirically support this convergence in broader scenarios; see Figures \ref{fig:additionalResults_consistency_sim1} and \ref{fig:additionalResults_consistency_sim2}.

\subsection{Algorithm}\label{subsec:alg}

We solve \eqref{eq:fixed_point_alpha_2} iteratively, replacing the RDMD in the $j$th iteration  with its estimate using the $(j-1)$st optimal subset:
\begin{align}\label{eq:sample_MRCT_mah_iterative}
H_j=\argmin_{\{H\subset\{1,\dots,n\};|H|=h\}}\frac{1}{h}\sum_{i\in H}\hat{M}_{\alpha,k,B^2}^2(X_i,\hat{\mu}_{B^2,H_{j-1}};\hat{C}_{B^2,H_{j-1},c_{H_{j-1}}}),
\end{align}
where $\hat{\mu}_{B^2,H_{j-1}}$ and $\hat{C}_{B^2,H_{j-1},c_{H_{j-1}}}$ are the trimmed sample mean and covariance operator based on the subset $H_{j-1}$, respectively. Thus, in the $j$th step of the procedure, we identify $h$ observations with the smallest regularized Mahalanobis distances to the robust mean estimate $\hat{\mu}_{B^2,H_{j-1}}$, w.r.t. the current robust covariance estimate $\hat{C}_{B^2,H_{j-1},c_{H_{j-1}}}$. However, in practice, we don't directly observe functions, but rather discrete realizations of the underlying functional observations, or datasets from which such functions are estimated, which is often the case with probability density functions. Let therefore 
\begin{equation}\label{eq:observed_discrete data}
X_i(t): t \in \{t_1, \dots, t_p\},\, i = 1, \dots, n,
\end{equation}
represent such a discretely observed sample. When the grid is sufficiently dense, the functional inner products required in computing the RDMD can be accurately approximated by integral sums; see, e.g.,~\cite{ostebee2002}. Accordingly, both \( X_i \) and related quantities such as \( \hat{\mu}_{B^2,H} \) are treated as vectors, and the sample covariance operator $\hat{C}_{B^2,H,c_H}$ is replaced by a matrix of pointwise covariances. The pseudocode for obtaining the robust pointwise estimators of the mean, covariance, and principal functions is given in Algorithm \ref{alg:alg1} in Supplement Section \ref{supp:aux_results}. After obtaining robust estimates on the observed grid, one may interpolate these estimates to obtain a full representation over the domain, though this step lies outside the scope of this work.

Alternatively, one can first convert discretized observations into a functional representation by projecting them onto a suitable basis. This enables evaluating each function at any point \( t \in I \), not just at the sampled time points. Using a fixed number $M \in \mathbb{N}$ of chosen basis functions $\phi_1(t), \ldots, \phi_M(t) : I \rightarrow\R^+, j = 1,\ldots,M,$ each observation can be approximated as:
\begin{equation} \label{eq:x_in_bases}
    X_i^M(t) = \bigoplus_{j=1}^M C_{i,j}\odot \phi_j(t), \quad i = 1, \dots, n,
\end{equation}
where \( [C_{i,j}] \in \mathbb{R}^{n \times M} \) is the matrix of basis coefficients, and the number of basis functions \( M \) is typically chosen to strike a balance between accurately capturing the observed data and enforcing smoothness to mitigate noise. While this representation is exact for functions of finite rank, it can also serve as an additional smoothing technique when the basis and the number of components are appropriately chosen.  
An (orthonormal) basis of our choice is the \textit{compositional spline basis} introduced in \citet{machalova2021},  designed specifically for the representation of the relative functional data, thus satisfying the unit integral constraint.  \\
Proposition~\ref{prop:Mah_distance_in_a_basis} shows that the RDMD can be computed \emph{exactly} in this framework, using only the matrix of basis coefficients $[C_{i,j}]$. The corresponding routine for robust covariance and SPCA estimation is given in Algorithm \ref{alg:alg1_in_basis}.
\begin{proposition}\label{prop:Mah_distance_in_a_basis}
    Let $\boldsymbol{\Phi}(t)=(\phi_1(t),\dots,\phi_M
    (t))'$, $t\in I$ be an orthonormal basis, and let $X_i, i=1,\dots,n$ admit a rank-M  representation as in~\eqref{eq:x_in_bases}, where $\textbf{C}=[C_{i,j}]\in\mathbb{R}^{n\times M}$ is the  the matrix of basis coefficients. Then, for   $H\subseteq\{1,\dots,n\}$, $|H|=h,$ and parameters $\alpha>0$, $k\in\mathbb{N}_0$ 
        $      \hat{M}^2_{\alpha,H,B^2}(X_i,\hat{\mu}_{B^2,H};\hat{C}_{B^2,H,c_H}) =(\textbf{e}_i-\frac{1}{h}\mathbf{1}_{H})'\textbf{C}\textbf{C}_{H}(\textbf{C}_{H}+\alpha/c_H\textbf{W}(k))^{-2}\textbf{C}\,'(\textbf{e}_i-\frac{1}{h}\mathbf{1}_{H}), 
     $
where  $\textbf{C}_{H}=\frac{1}{h}\textbf{C}\,'(\textbf{I}_{n,H}-\frac{1}{h}\textbf{J}_{n,H})\textbf{C}\in\mathbb{R}^{M\times M}
$, $\mathbf{1}_{H}\in\mathbb{R}^{n}$ is such that $(\mathbf{1}_{H})_i=1$ if $i\in H$ and $0$ otherwise, $\textbf{J}_{n,H}=\mathbf{1}_{H}\mathbf{1}_{H}'$, and $\textbf{I}_{n,H}=\mathrm{diag}(\mathbf{1}_{n,H})\in\mathbb{R}^{n\times n}$ is a diagonal matrix with $\mathbf{1}_{H}$ on its diagonal.  $\textbf{W}(k)$ is a projection matrix onto the space spanned by the span of the first $k$ eigenvectors of $\textbf{C}_{H}$; if $k=0$, $\textbf{W}(k)=\textbf{I}_M.$
\end{proposition}
\begin{remark}
    Our framework assumes some level of smoothness in the underlying functional data, an assumption embedded via regularization. In practice, this means that the methodology can also be justifiably applied to data observed on a dense, regular grid without requiring basis expansion. In fact, pre-smoothing or excessive preprocessing may mask outlying behavior, potentially interfering with outlier detection~\citep{hubert2015}. Therefore, both discretized and functional representations are supported within our framework—the choice between them should depend on the nature of the data analyst's preference.
\end{remark}

\begin{algorithm}[H] 
\footnotesize
\caption{RDPCA for data expressed in a fixed basis}\label{alg:alg1_in_basis} 
\SetKwInOut{Input}{Input}
        \Input{Orthonormal basis $\boldsymbol{\Phi}=(\phi_1,\dots,\phi_M) \in \mathcal{B}^2$, sample ${X}_1,\dots, {X}_n \in \mathcal{B}^2$, subset size $h$, initial subset $H_1\subset\{1,\dots,n\}$ with $|H_1| = h$, regularization parameter $\alpha>0$, projection dimension $k \leq n$, and tolerance level $\varepsilon_k>0$;}
        \BlankLine
        $Y_i \leftarrow \mathrm{clr}(X_i), i = 1,\ldots,n,~ \tilde{\phi}_j \leftarrow \mathrm{clr}(\phi_j), j = 1,\ldots,M$ and set $\tilde{\boldsymbol{\Phi}} = (\tilde{\phi}_1,\ldots,\tilde{\phi}_M)$;\\
	Express $Y_i$ in basis $\tilde{\boldsymbol{\Phi}}$ as $Y_i= \sum_{j=1}^M C_{i,j}\tilde{\phi}_j$, and collect $\boldsymbol{C} = [C_{i,j}] \in\mathbb{R}^{n\times M}$;\\
	\SetKwRepeat{Do}{do}{while}
	\Do{$H_0\neq H_1$}{
	$H_0\leftarrow H_1$;\\
	$\boldsymbol{C}_{H_0}=\frac{1}{h}\boldsymbol{C}'(\textbf{I}_{n,H_0}-\frac{1}{h}\textbf{J}_{n,H_0})\boldsymbol{C}=\sum_{i=1}^{M}{\hat\lambda}_{i,H_0}\textbf{u}_i\textbf{u}_i'$;\\
    Set initial scaling parameter $c_{1,H_0}\leftarrow 1$;\\
	\Do{$(c_{1,H_0}-c_{0,H_0})^2\geq\varepsilon_k$}{
	    $c_{0,H_0}\leftarrow c_{1,H_0}$;\\
        Set $\tilde{V}(k) \leftarrow I - \sum_{j=1}^k \hat{\lambda}_{j,H_0} \boldsymbol{u}_j \boldsymbol{u}_j'$;\\
     $d_{i,H_0,c_0}^2 \leftarrow (\textbf{e}_i-\frac{1}{h}\mathbf{1}_{H_0})'\boldsymbol{C}\boldsymbol{C}_{H_0}(\boldsymbol{C}_{H_0}+\alpha/c_{0,H_0}\tilde{V}(k))^{-2}\boldsymbol{C}'(\textbf{e}_i-\frac{1}{h}\mathbf{1}_{H_0})$;\\
     Generate $\eta_{k+1},\ldots,\eta_n \sim \chi^2(1)$, $y \sim \chi^2(k)$;\\
    $ c_{1,H_0}\leftarrow{\mathrm{med}\left\{d_{i,H_0,c_0}^2:i=1,\dots,n\right\}}/{\mathrm{med}\left\{y + \sum_{i=k+1}^M\frac{\hat{\lambda}_{i,H_0}^2}{(\hat{\lambda}_{i,H_0}+\alpha/c_{0,H_0})^2}\eta_i\right\}}$;\\}
    Order $d^2_{(i_1,H_0,c_1)}\leq\dots\leq d^2_{(i_n,H_0,c_1)}$;\\
    Set $H_1\leftarrow\{i_1,\dots,i_h\}$;\\
    }
    Solve symmetric eigenproblem $c_{1,H_1} \boldsymbol{C}_{H_1}(u_j) = \hat{\lambda}_j u_j;$\\
    Recalculate $\xi_j = u_j' \tilde{\boldsymbol{\Phi}}$  and set $\hat{\xi}_{j,H_\mathrm{opt}} \leftarrow \xi_j, \hat{\zeta}_{j,H_\mathrm{opt}} \leftarrow \mathrm{clr}^{-1}(\xi_j)$ and $\hat{\lambda}_{j,H_\mathrm{opt}} \leftarrow \hat{\lambda}$;\\
    \SetKwInOut{Output}{Output}
	\Output{Optimal subset $H_\mathrm{opt} := H_1$, scaling factor $c_{H_0}$, squared robust RDMD $(c_{H_0}^{-1}d_{1,H_0,c_1}^2,\dots c_{H_0}^{-1}d_{n,H_0,c_1}^2)$, robust eigenvalues $\{\hat{\lambda}_{j,H_\mathrm{opt}}\}_{j \geq 1}$ and corresponding robust FPCs $\{ \hat{\zeta}_{j,H_\mathrm{opt}} \}_{j \geq 1}, \{ \hat{\xi}_{j,H_\mathrm{opt}} \}_{j \geq 1}$  in Bayes and clr space, respectively.}
\end{algorithm}

\paragraph{Selecting the regularization parameters} \label{subsec:regParamSelection}

The regularized Mahalanobis distance depends on the tuning parameters \(\alpha>0\) and $k\in\mathbb{N}_0$. The choice of \(\alpha\) balances the smoothness of the regularized data with how well it fits the observed data. 
If \(\alpha\) is too large, over-regularization will result in poor fit, with large residuals \(\|\hat{C}_{B^2,H_\mathrm{opt}}^{1/2} (X_{i,\mathrm{st},\hat{C}_{B^2,H_\mathrm{opt}}}^{{\alpha},k}) \ominus X_i)\|_{B^2}^2\),  \(i = 1, \ldots, n\). Conversely, if \(\alpha\) is too small, the solution will overfit data errors, leading to large values of \(\|\hat{W}(k)X_{i,\mathrm{st},\hat{C}_{B^2,H_\mathrm{opt}}}^{{\alpha},k}\|^2\). To determine \(\alpha\), we adapt the approach from \cite{oguamalam2024}, iteratively adjusting \(\alpha\) so that the Bayes covariance of the standardized data has eigenvalues close to zero (for the latter part of the spectrum) or some fixed constant (for the signal part), thus maximizing the signal-to-noise ratio.

On the other hand, we can think of $k$ parameter that prevents important signal features from being over-smoothed. \cite{hron2016} argues how for densities from the $k_0$ - exponential family (with $k_0$ parameters), the number of principal components belonging to strictly positive eigenvalues is precisely $k_0$. E.g., for the Gaussian family with the known mean $k_0=1$; see \cite{hron2016} for more details. Therefore, in Section \ref{sec:RDPCA_simus} we fix $k=1$.
In more complex situations where the underlying distribution is unknown, selecting $k$ such that the first $k$ principal functions contain at least $90 \%$ of variability seems reasonable. This rule of thumb will be applied in Section \ref{sec:RDPCA_examples}.

Finally, in the supplement, we provide some ideas about the choice of the subset size $h$.

% ---------------------------------------------------------------
% Simulations --------------------------------------------------
% ---------------------------------------------------------------

\section{Simulations}\label{sec:RDPCA_simus}

Within two simulation studies, we evaluate the accuracy of the proposed robust approach in estimation of the covariance and principal components under contamination with a fixed proportion $c  \in \{0,0.05,0.1,0.2\}$ of the data is replaced by outliers. Outliers are introduced via Huber's $\varepsilon$-contamination model \cite{Huber1964} as $X\sim (1-c)F_{reg} + c F_{out}$, where $F_{reg}$, $F_{out}$ denote the distribution of the regular and outlying observations, respectively.

To mitigate the effect of the data processing on the results, in the simulations we use the raw data sampled on a dense grid and robustly estimate the covariance and PCs using Algorithm~\ref{alg:alg1} from the supplement, with the regularization parameter $\alpha$ automatically chosen according to the procedure discussed in Section \ref{subsec:alg}. Furthermore, we fix $h = \lfloor 0.75n \rfloor$ and $k = 1$.

We assess both the accuracy and robustness of the covariance operator and principal component estimation, and compare the results with those of the non-robust SFPCA method. To measure the performance of RDPCA, we consider two statistics: the integrated square error (ISE), approximating the norm of the difference between estimates of the true and sub-sampled covariances, and the mean cosine between the first five pairs of eigenfunctions based on the same estimates. For more details, see Supplement Section \ref{supp:simulations}. For completeness, we also examine the performance of several outlier detection methods applied after the clr transformation, as commonly done in the functional data analysis literature \citep{arribas-gil2014, berrendero2020,oguamalam2024}. Detailed results are provided in Supplementary Section \ref{sec:add_simul_results_supp}.

% ---------------------------------------------------------------
% Simulation 1 --------------------------------------------------
% ---------------------------------------------------------------

\subsection{Simulation 1} \label{sec:RDPCA_sim1}

In the first simulation, in each of $100$ replications, we consider a data set of size $n=500$ consisting of $n-\lfloor cn\rfloor$ regular- and $\lfloor cn\rfloor$ contaminated curves, $c\in\{0,\,0.05,\,
0.1,\,0.2\}$. Each curve is estimated using kernel density estimation (KDE) on a generated sample and evaluated at an equidistant grid of $p = 50$ time points. The construction of a regular/outlying sample is as follows:
\begin{align*}
    X_\mathrm{reg} =~ \mathrm{KDE}\left(Y_1,\ldots,Y_{N}\right),~ X_\mathrm{out} =~ \mathrm{KDE}\left(Y_1,\ldots,Y_{N},\tilde{Y}_1,\ldots,\tilde{Y}_{\lfloor 0.1N \rfloor }\right)
\end{align*}
where $Y_i \sim N(0,1),~ \tilde{Y}_i \sim U([q_1;q_2] \cup [-q_2;-q_1]), N = 250$ and $q_1,q_2$ being the $0.1\%$ and $0.5\%$ quantile of the standard normal distribution, respectively. KDE was calculated using the \texttt{R}-package \texttt{kde1d}~\citep{kde1d}. 

\begin{figure}[!h]
    \centering
    \includegraphics[width=.9\linewidth]{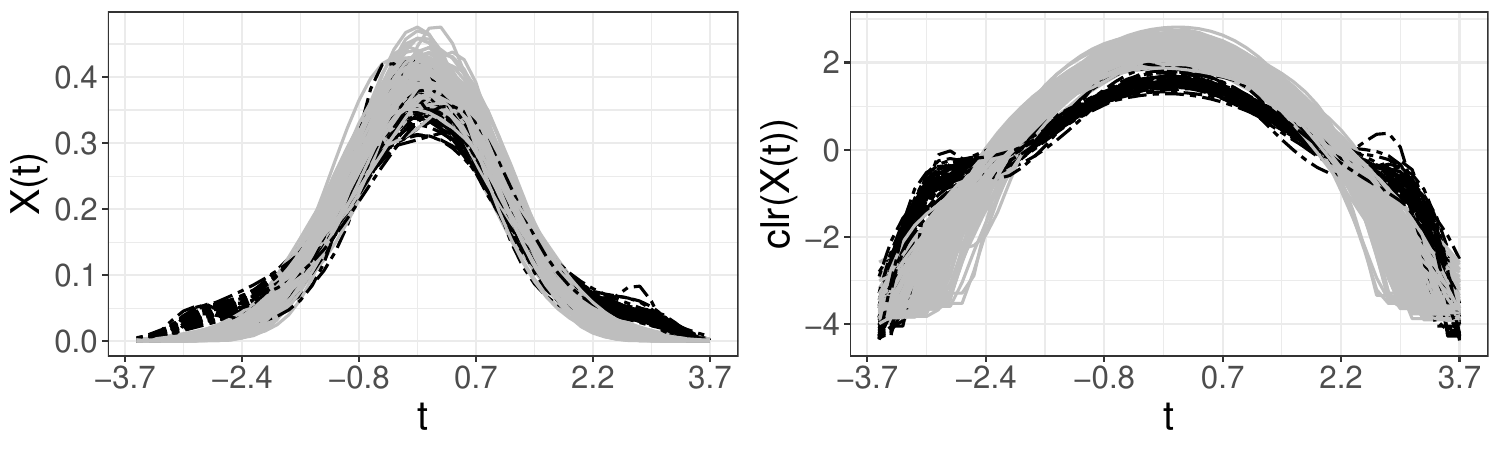}
    \caption{\small Visualization of density functions (left) and clr transformed counterparts (right). Solid curves represent the main processes, while the dashed ones indicate the outliers.
    }
    \label{fig:dataModel2}
\end{figure}

By the construction of these densities, a fixed proportion $c$ of the data is contaminated within the tails of the distribution. Figure \ref{fig:dataModel2} presents an example of this model, displaying both the densities and their transformed counterparts, for a contamination rate of $c = 0.2$. In the supplement, we extend this simulation with two additional models, where the densities are constructed in a similar fashion for different distributions of $Y_i$ and $\tilde{Y}_i$.

\begin{figure}[!h]
    \centering
    \includegraphics[width=.9\linewidth]{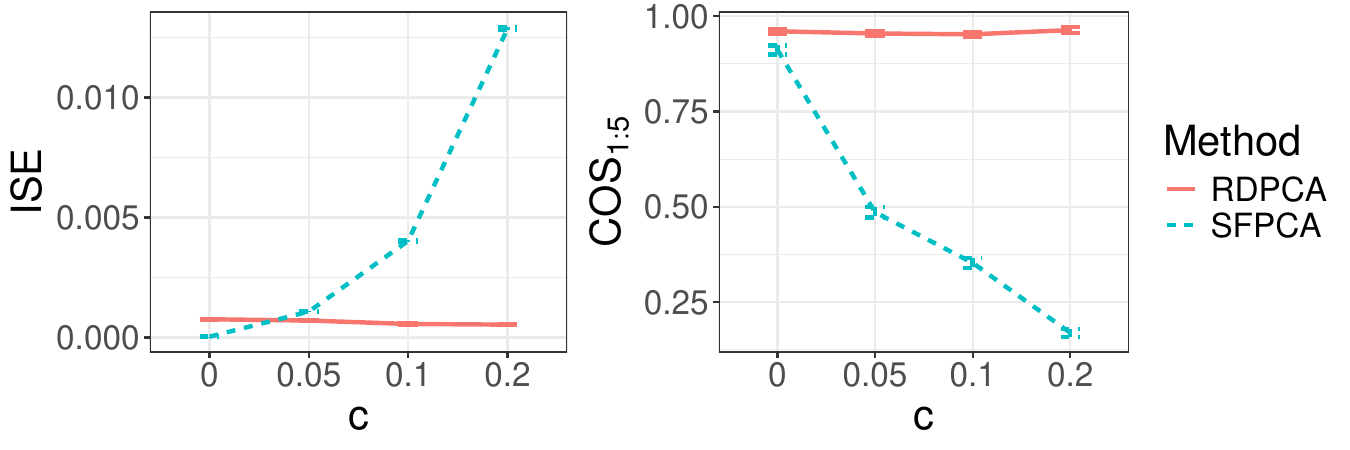}
    \caption{\small Mean ($\pm$ standard error) ISE between the estimated and true covariances (left) and cosine similarity between the first five pairs of estimated and true eigenfunctions (right), obtained using RDPCA (solid) and SFPCA (dashed).}
    \label{fig:CovmeasuresLinesSim1Mod2}
\end{figure}

Figure \ref{fig:CovmeasuresLinesSim1Mod2} depicts the accuracy of the estimation of principal components w.r.t. the contamination rate $c$. Both the ISE and the mean cosine between the first five pairs of principal functions suggest that RDPCA significantly outperforms SFPCA in the presence of outliers. Interestingly, the right panel of Figure \ref{fig:CovmeasuresLinesSim1Mod2} indicates that RDPCA outperforms SFPCA in estimating the principal functions even in the non-contaminated setting.

\begin{figure}[h]
    \centering
    \includegraphics%[height=1.7in]
    [width=.9\linewidth]{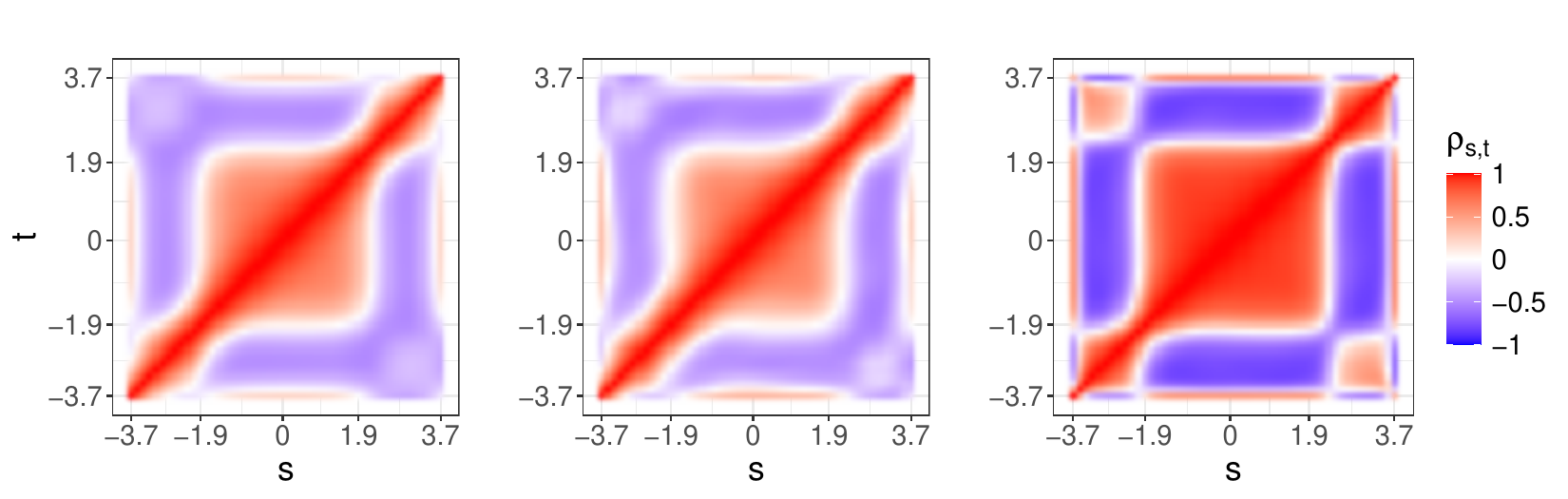}
    \caption{\small True (left), RDPCA with $\alpha=0.32$ (middle), and SFPCA (right) correlation function, for the example shown in Figure \ref{fig:dataModel2}.
    }
    \label{fig:CorrSim1}
\end{figure}

\begin{figure}[!h]
    \centering
    \includegraphics[width=.7\linewidth]{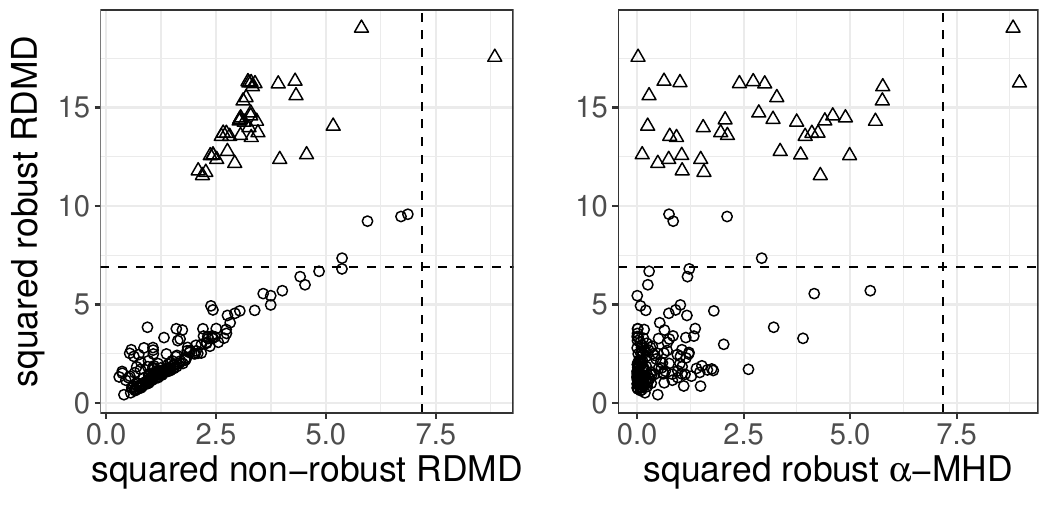}
    \caption{\small Distance-distance plot of squared robust vs. non-robust RDMD (left) and robust $\alpha$-Mahalanobis \citep{berrendero2020} (right) distances for example of Figure \ref{fig:dataModel2}. The regularization parameter is $\alpha = 0.32$. Dashed lines indicate the corresponding cutoff values under Gaussianity. Circles and triangles correspond to true regular and outlying observations, respectively.
    }
    \label{fig:ddplotSim1}
\end{figure}

Figure \ref{fig:CorrSim1} illustrates how the robust correlation function nicely aligns with the correlation function based on a Monte Carlo estimate of the true underlying model. On the contrary, the non-robust correlation reveals a significantly different structure. In Figure \ref{fig:ddplotSim1} we compare the squared RDMD computed using RDPCA to the squared RDMD obtained using SPCA and the squared robust $\alpha$-Mahalanobis \citep{berrendero2020} distance, ignoring the compositional structure of the densities. The dashed lines indicate the $95\%$ quantile of the limiting distribution of these distances assuming Gaussianity and estimated by Monte Carlo simulation; see Corollary \ref{cor:cor_1}. A density is considered outlying if its squared Mahalanobis distance exceeds this cutoff. The circles represent the true regular observations, whereas the triangles correspond to the outliers. According to the cutoff values (dashed lines), only RDPCA can determine outliers in a meaningful way. In the supplement, we added some additional results related to this comparison.

% ---------------------------------------------------------------
% Simulation 2 --------------------------------------------------
% ---------------------------------------------------------------

\subsection{Simulation 2} \label{sec:RDPCA_sim2}

In the second simulation, we generate data sets of size $n = 500$ according to
\begin{align}\label{eq:model_simu2}
    \clr(X_\mathrm{reg}(t)) =&~ s_1 \xi_1 + s_2 \xi_2 + s_3 \xi_3 + s_4 \xi_4 ,\\
    \clr(X_\mathrm{out}(t)) =&~ s_1 \xi_1 + s_2 \xi_2 + s_3 \xi_3 + s_4 \xi_4 + s_5\xi_5,
\end{align}
where $\xi_1 = \sqrt{2}\mathrm{sin}(2\pi t), \xi_2 = \sqrt{2}\mathrm{cos}(2\pi t), \xi_3 = \sqrt{2}\mathrm{sin}(4\pi t), \xi_4 = \sqrt{2}\mathrm{cos}(4\pi t)$ are the eigenfunctions of the covariance operator of the clr transformed densities. The scores $s_1, s_2, s_3, s_4$ are drawn, in one setting, from a normal distribution, $s_i \sim N(0,\lambda_i)$ or, in another setting, from a multivariate t-distribution; $(s_1,s_2,s_3,s_4) \sim t_5(\boldsymbol{0},\mathrm{diag}(\lambda_1,\lambda_2,\lambda_3,\lambda_4))$ and $(s_1,s_2,s_3,s_4,s_5) \sim t_5(\boldsymbol{0},\mathrm{diag}(\lambda_1,\lambda_2,\lambda_3,\lambda_4,\lambda_5))$. In both cases we set $(\lambda_1,\lambda_2,\lambda_3,\lambda_4,\lambda_5) = (2,1,1/2,1/4,4)$. The resulting process is then evaluated at an equidistant grid of $p = 100$ time points.  Visualization of the data is given in the supplement in Figure \ref{fig:dataClrSim2Mod12}.

\begin{figure}[!h]
    \centering
    \includegraphics[width=.95\linewidth]{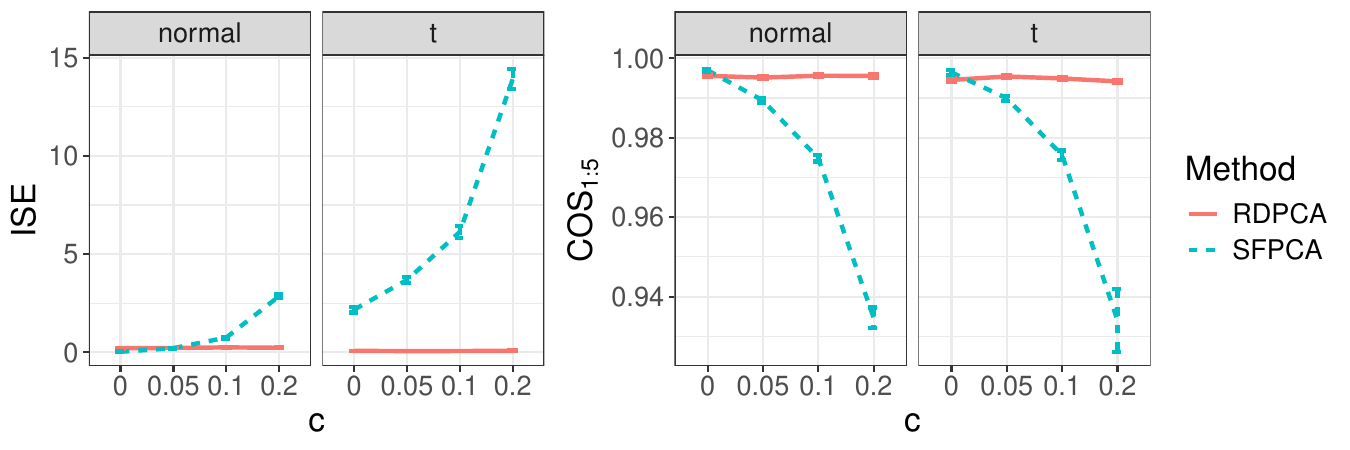}
    \caption{\small Mean ($\pm$ standard error) ISE between the estimated and true covariances (left) and cosine similarity between the first five pairs of estimated and true eigenfunctions (right), obtained using RDPCA (solid) and SFPCA (dashed). The notation “normal" and “t" refer to the distribution of the scores in the underlying model as given at the beginning of the section.
    }
    \label{fig:CovmeasuresLinesSim2Mod3} 
\end{figure}

Each setting is replicated 100 times, and the accuracy of RDPCA is evaluated by the ISE of the difference between the estimated and the true covariance, as well as the mean cosine between the first four pairs of eigenfunctions based on the estimated and the true covariance. The results are shown in Figure \ref{fig:CovmeasuresLinesSim2Mod3}, and again demonstrate that the robust procedure outperforms SFPCA in the presence of outliers. Interestingly, for $t$-distributed scores, even in the setting with no outliers, the robust procedure yields better results.

\begin{figure}[!h]
    \centering
    \includegraphics[width=.9\linewidth]{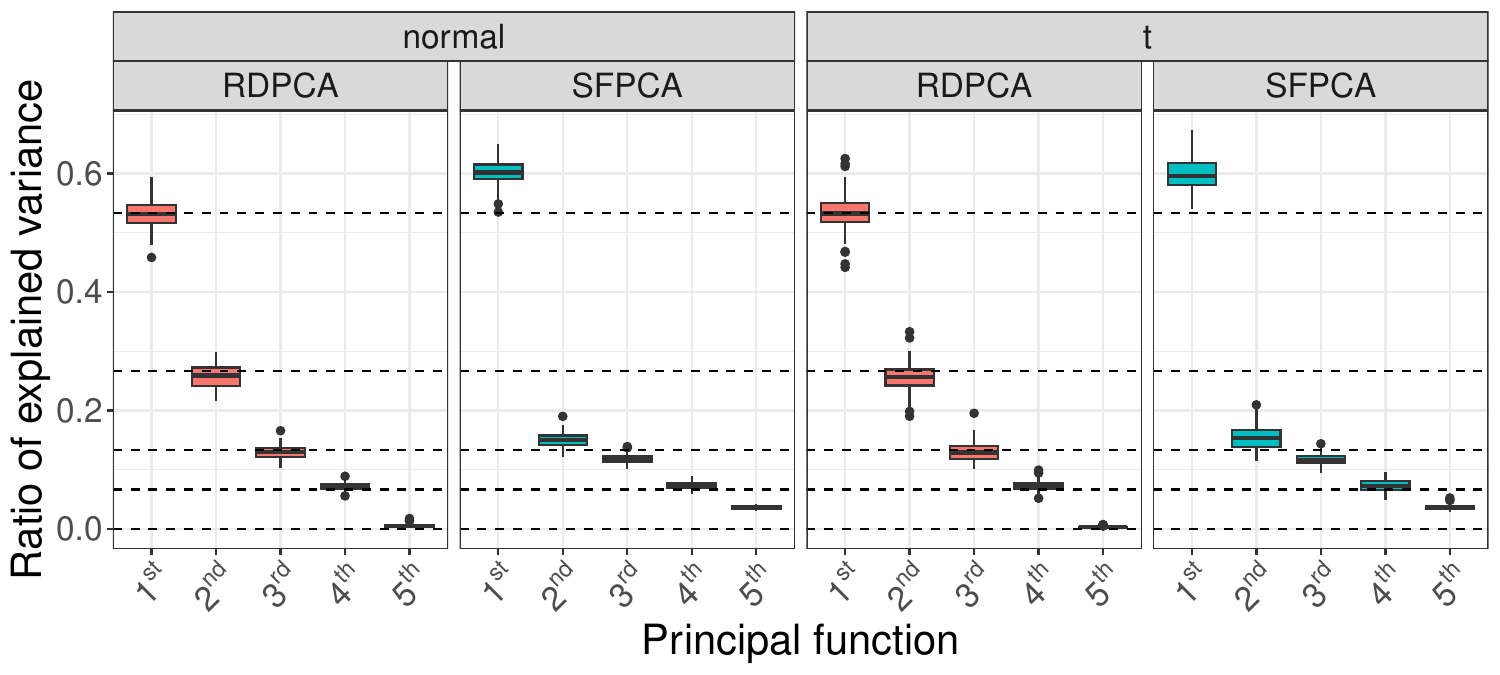}
    \caption{\small Estimated ratio of explained variance for the first five principal functions over different score distributions at $c = 0.2$. The dashed reference lines depict the true proportion. The notation “normal" and “t" refer to the distribution of the scores in the underlying model as described at the beginning of this section.
    }
    \label{fig:explainedVarSim2Mod3}
\end{figure}

Finally, we compare the estimated ratios of explained variance from the robust and non-robust methods to the true values. As shown in Figure \ref{fig:explainedVarSim2Mod3}, the robust method provides estimates that closely match the true ratios.

% -------------------------------------------------------------------------------------------------
% --- Real data example ---------------------------------------------------------------------------
% -------------------------------------------------------------------------------------------------
\section{Real-data example}\label{sec:RDPCA_examples}

To illustrate the practical performance of RDPCA, we apply it to two real-data examples. The computations were carried out on a machine equipped with an AMD Ryzen 9 5900HX core with Radeon Graphics (16 cores, 3.30 GHz) and 32 GB of RAM, running Windows 11 Pro. The implementation was written in \texttt{R} (version 4.4.2). Calculations connected to RDPCA were run using 15 parallel cores and required approximately 5 and 1 minute(s) for the first and second example, respectively.

In addition to robust PCA, during the examples, we also examine atypical curves according to Algorithm \ref{alg:alg1_in_basis}. For this, we estimate the $95\%$-quantile of the limiting distribution of the RDMD under the assumption of Gaussianity; see Corollary \ref{cor:cor_1}. A density is then considered outlying if its squared RDMD exceeds this cutoff. 

% ----------------------------
% --- Example 1 --------------
% ----------------------------
\subsection{EPXMA spectra} \label{sec:RDPCA_epxma}

The considered dataset consists of electron probe X-ray microanalysis (EPXMA) spectra of various glass vessel types. This type of analysis is commonly used to determine the chemical composition of glasses by measuring X-rays at different wavelengths emitted from a sample after being hit by a beam of electrons. As is commonly done in the analysis of spectral data, we analyze EPXMA spectra as relative functional data.

The dataset contains 180 such observations divided into four distinct groups corresponding to different glass vessel types. The largest group, referred to as “sodic", comprises 145 spectra. The remaining groups, designated as “potassic", “potasso-calcic", and “calcic", are represented by $15, 10,$ and $10$ spectra, respectively. It is important to note that the latter 38 spectra within the sodic group were acquired under varying experimental conditions. Each observation is characterized by measurements across 1920 wavelengths. For the purposes of this analysis, we focus on wavelengths 51--750 as these encompass the primary variability observed in the data and to avoid measurement artifacts; for more details, see \cite{lemberge2000}. This results in $n = 180,\, p = 700$.

\begin{figure}[!h]
    \centering
    \includegraphics[width=\linewidth]{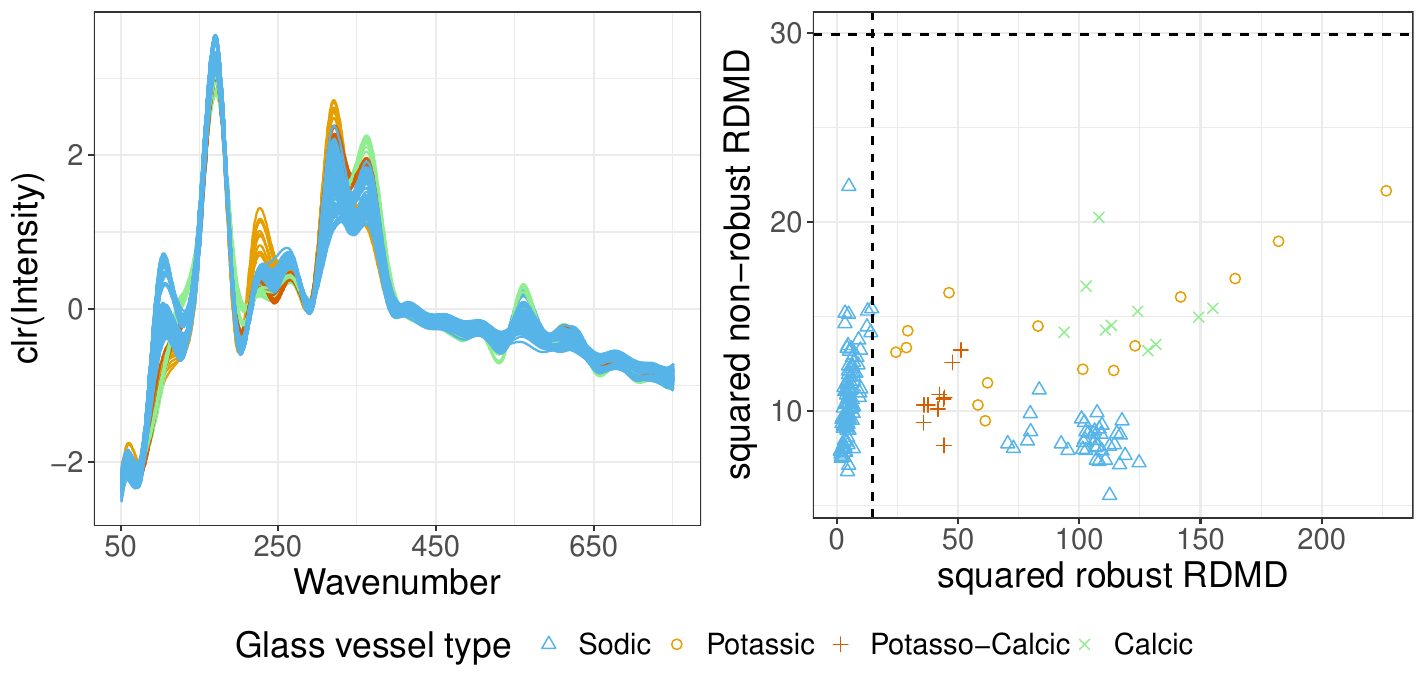}
    \caption{\small Left: clr transformed spectra colored according to glass vessel type. Right: Distance-distance plot of the squared robust vs. non-robust RDMD. Dashed lines indicate the theoretical cutoff values under Gaussianity.}
    \label{fig:epxma}
\end{figure}

We analyze the data using Algorithm \ref{alg:alg1_in_basis}, where the data is expressed using $31$ compositional splines. The left plot of Figure \ref{fig:epxma} shows the smoothed trajectories of the clr transformed observations colored according to their respective glass vessel type. For the analysis, we consider the first 107 observations from the sodic group as regular observations. We use $h = \lceil 0.5n \rceil$ and $k = 4$, and automatically select  $\alpha = 0.03$; see Section \ref{subsec:alg}.

The right plot of Figure \ref{fig:epxma} displays the squared robust versus the non-robust RDMD. The non-robust approach detects virtually no outlying observations, while the robust procedure can clearly differentiate between the majority of “sodic" and the remaining observations. 

\begin{figure}[!h]
    \centering
    \includegraphics[width=0.8\linewidth]{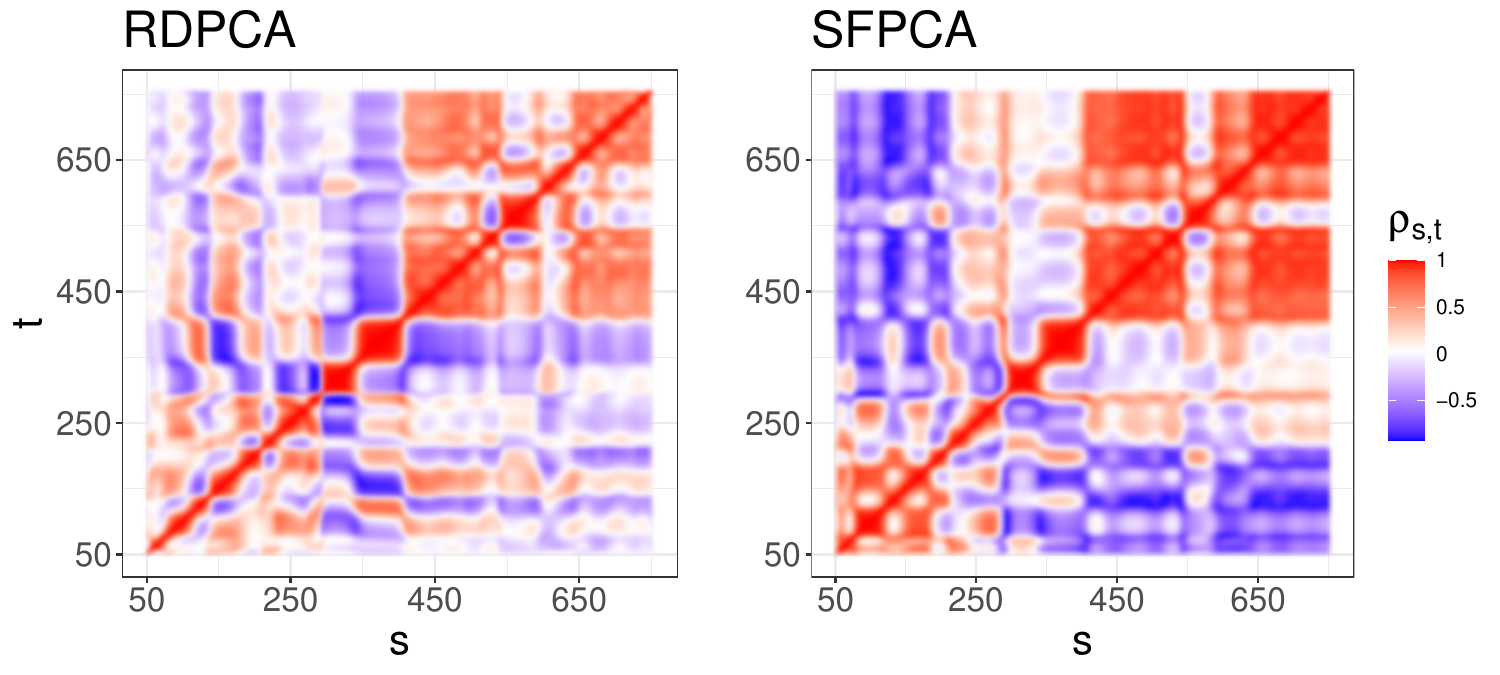}
    \caption{\small Correlation function of EPXMA data based on RDPCA and SFPCA.
    }
    \label{fig:epxma.corr}
\end{figure}

Figure \ref{fig:epxma.corr} showcases the consequent influence onto the covariance, where we compare smoothed robust to non-robust correlation function. The non-robust correlation shows its bias towards the outlying curves at parts of the domain where significant differences arise.

% ----------------------------
% --- Example 2 --------------
% ----------------------------
\subsection{Fertility data} \label{sec:RDPCA_fertility}

The second data set consists of age-specific fertility rates (ASFR) across different countries. ASFR represents the number of women giving birth per $1,000$ within specific age groups, providing a detailed view of how fertility changes with age. The resulting curves are analyzed as density data because we are interested in the underlying distribution across the different age groups rather than the magnitude of it. We examine data from the Human Fertility Database,  available at  \href{https://www.humanfertility.org/Home/Index}{https://www.humanfertility.org/Home/Index}, focusing on the years $1956$, $1990$, and $2019$, to gain insight into the evolution of fertility dynamics. 

\begin{figure}[!h]
    \centering
    \includegraphics[width=.9\linewidth]{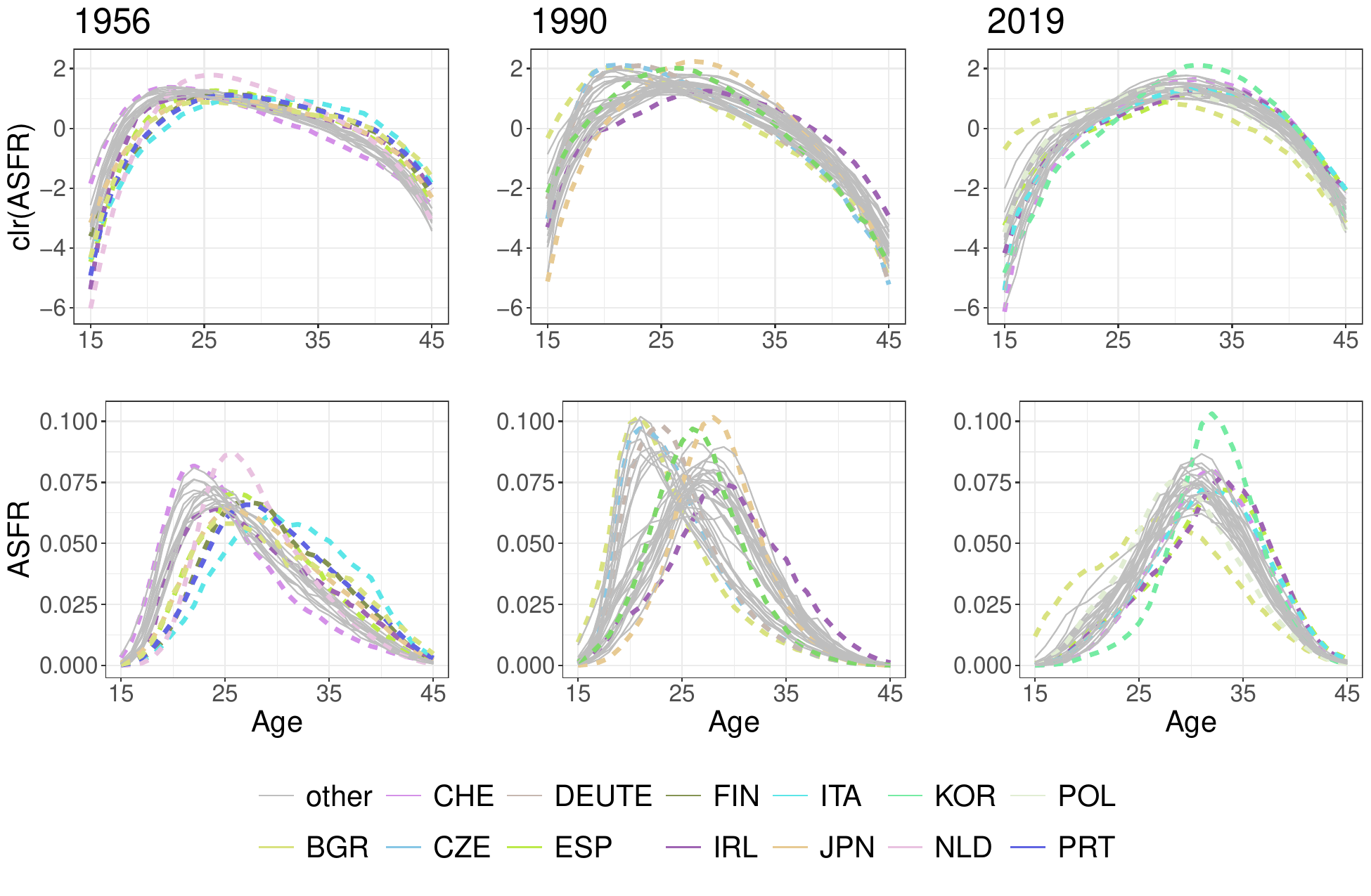}
    \caption{\small Fertility densities (lower row) and clr transformed densities (upper row) corresponding to years 1956, 1990, and 2019. Curves identified as outlying by Algorithm \ref{alg:alg1_in_basis} are displayed as dashed lines.
    }
    \label{fig:fertilityOutliers}
\end{figure}

In this example, we apply Algorithm \ref{alg:alg1_in_basis} with 21 compositional splines as basis functions. The data is visualized in Figure \ref{fig:fertilityOutliers}. Outlying countries, according to their squared RDMD, are represented by dashed curves. During the early stages of this period, peak fertility was primarily achieved by young women in their early 20s. Large families were common, reflected in a rather slow decrease in the majority of the fertility curves. During the following decades, mostly influenced by the introduction of reliable contraceptives and a change in the mindset about life quality, later first pregnancies became more and more common. Throughout Europe, this phenomenon started earlier in the West than in the East, as can be seen by the fertility curves corresponding to the year $1990.$ Since then, the peak fertility has steadily increased and sits now at around 30 years. For more details, we refer to the insightful article by \cite{beaujouan2020} and references therein.

While the variability in the Bayes space seems to be present within the peaks, the upper row of Figure \ref{fig:fertilityOutliers} shows that the main variability within the clr space is significantly higher in the tails of the distribution; see also \cite{hron2016}. 

\begin{figure}[!h]
    \centering
    \includegraphics[width=.9\linewidth]{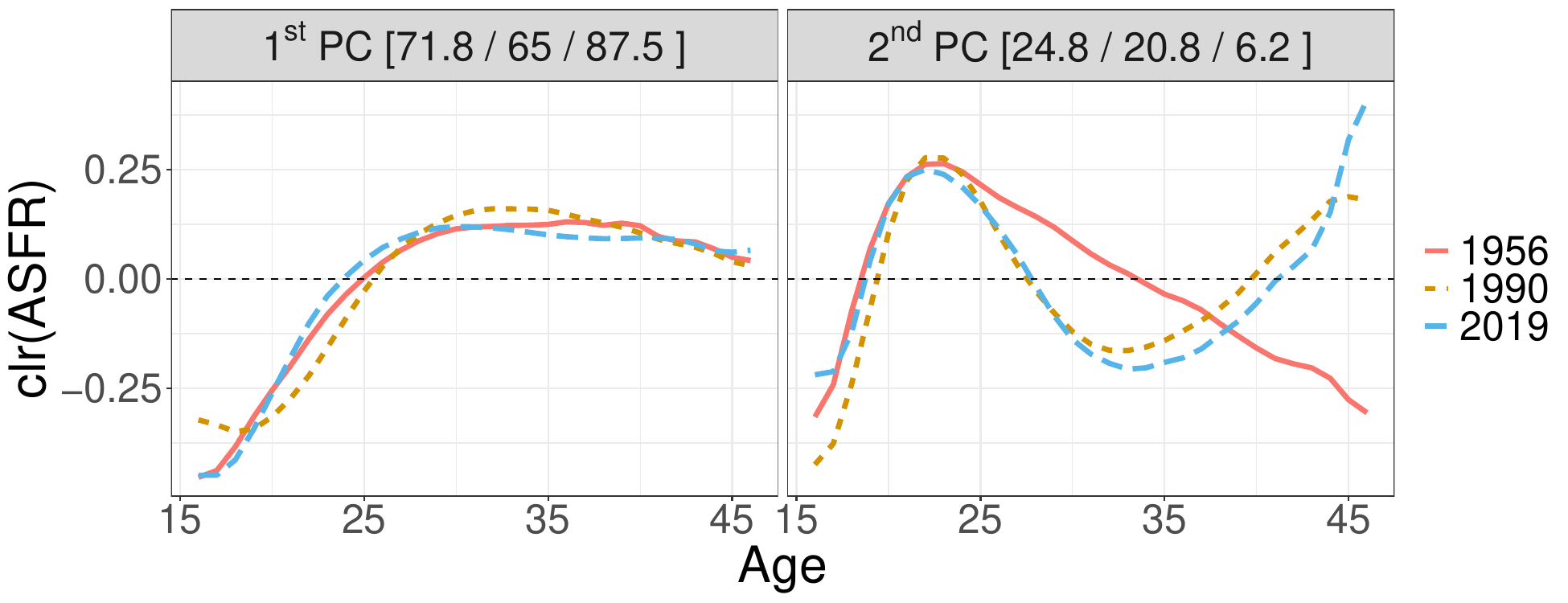}
    \caption{\small Comparison between the first two robust principal functions of the fertility data for years $1956, 1990, 2019$. Corresponding percentage of explained variance in brackets (left to right).
    }
    \label{fig:fertilityEigenfunction.clr}
\end{figure}

In Figure \ref{fig:fertilityEigenfunction.clr}, the first two principal functions are displayed. In the brackets, we included the corresponding percentage of explained variance. For each year, the first component mainly captures the variability in the tails of the densities. Here, the principal functions focus on the left tail and the following ages until the early 20s. The second principal function contains significantly less information, mainly explaining the right tail and the peak fertility.

\begin{figure}[!h]
    \centering
    \includegraphics[width=.9\linewidth]{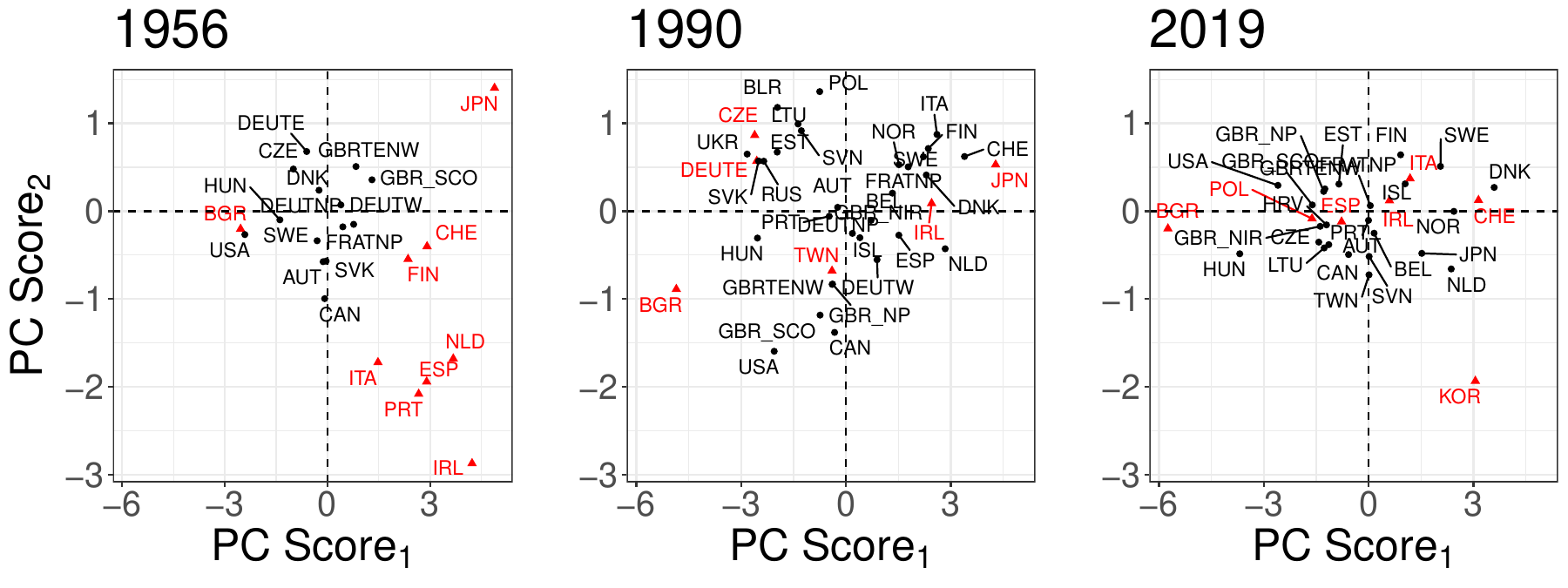}
    \caption{\small Robust PC scores for the first two eigenfunctions for the years $1956, 1990, 2019$. Circles correspond to regular observations, while the triangles indicate outlying countries.
    }
    \label{fig:fertilityScores.clr.12}
\end{figure}

Following the construction of robust principal components, we look at the corresponding estimated principal component scores $\hat{s}_j$ associated with each country. In Figure \ref{fig:fertilityScores.clr.12}, the first two robust scores for each country are visualized. 
For 1956, the outliers, as indicated by the triangles, all lay at the border of the data cloud. For Japan, the first and second scores are exceedingly high, indicating not only relatively low fertility in younger age groups but also proportionally very high peak fertility. For the outliers of 1990 and 2019, we obtain a similar picture dominated by mostly extreme scores. For 1990, this mainly reflects countries that have an unusually high or low fertility for younger and older age groups. Similarly, in 2019, these extreme values explain high fertility rates at younger ages as well as late peak fertility. 

\section{Conclusions and discussion}
\label{sec:concl}

Despite the attention that the Bayes space methodology for density data analysis has attracted over the last decades, little focus has been paid to robust frameworks necessary for meaningful analysis in the presence of anomalies. Therefore, this paper introduces robust density PCA (RDPCA) as a methodology to robustly estimate the covariance operator and functional principal components of densely sampled univariate density data. Within this newly proposed method, we extend the notion of a functional Mahalanobis distance \citep{berrendero2020, galeano2015} to the Bayes space. The resulting regularized density Mahalanobis distance (RDMD) leads to a robust covariance estimation based on a subsample of those curves that best fit the underlying distribution. Based on this estimate, robust PCs for density function can be determined as summarized in Algorithm \ref{alg:alg1}. If the data can be expressed in a fixed basis, we provide an equivalent approach in Algorithm \ref{alg:alg1_in_basis}. During simulations, we demonstrated the efficiency of RDPCA as well as the advantages it has against a non-robust approach in the presence of outliers. Furthermore, if outliers arise within the tails of the densities, the clr transformation can amplify these deviations. Especially in these scenarios, we observed that common robust procedures that do not respect the nature of the data would fail.

One of our initial assumptions was that the data had been observed at a dense grid. In the case where the data is only available at a sparse grid, further extensions would require smoothing by an appropriate basis, e.g., CB-splines \citep{machalova2021} specifically developed for Bayes spaces. Naturally, the next step from univariate data would be to consider multivariate densities. During multivariate functional data analysis, each observation contains the recording of several “functional" variables. In this setting, not only is the covariance between different time points considered, but also the relation between individual variables. If the second dimension is also continuously observed, the observations are so-called random surfaces. As these statistical fields show increasingly practical relevance (\cite{berrendero2011}; \cite{górecki2018}; \cite{dai2018}; \cite{masak2023} and references therein), expanding RDPCA to these areas seems worthwhile. Calculating the RDMD involved regularizing by a suitable operator that smooths out unwanted noise components while keeping the relevant signal within the (uncontaminated) data unaffected. Another class of meaningful operators that utilize the functional nature of the data are differential operators, often used in Tikhonov regularization. One has to keep in mind that these operations have yet to be defined for Bayes spaces. Next to PCA, the regularized Mahalanobis distance could be used for concepts like linear or quadratic discriminant analysis for the classification of density data. As these methods rely on similarity measures involving several covariance operators, the robust classification of densities based on the RDMD would be suitable.

\section*{Supplementary material}

\textbf{Supplement:} The supplement contains auxiliary results, further developing ideas on RDPCA. Next, it provides additional simulation analysis corresponding to Section \ref{sec:RDPCA_sim1} and \ref{sec:RDPCA_sim2}, and a further analysis of the fertility data from Section \ref{sec:RDPCA_fertility}. The supplement finishes with the proofs of theoretical results.

\section*{Acknowledgement}

The authors acknowledge support from the Austrian Science Fund (FWF), project number I 5799-N, and the Czech Science Foundation, project numbers 22-15684L and 25-15447S. The present research has also been partially supported by MUR, grant Dipartimento di Eccellenza 2023-2027.

\section*{Disclosure statement}
The authors report there are no competing interests to declare.

\bibliographystyle{chicago}
\bibliography{references}  %%% Uncomment this line and comment out the ``thebibliography'' section below to use the external .bib file (using bibtex) .

\begin{thebibliography}{}

\bibitem[\protect\citeauthoryear{Arribas-Gil and Romo}{Arribas-Gil and Romo}{2014}]{arribas-gil2014}
Arribas-Gil, A. and J.~Romo (2014, 03).
\newblock {Shape outlier detection and visualization for functional data: the {O}utliergram}.
\newblock {\em Biostatistics\/}~{\em 15\/}(4), 603--619.

\bibitem[\protect\citeauthoryear{Beaujouan}{Beaujouan}{2020}]{beaujouan2020}
Beaujouan, E. (2020).
\newblock {L}atest-late fertility? {D}ecline and resurgence of late parenthood across the low-fertility countries.
\newblock {\em Population and Development Review\/}~{\em 46\/}(2), 219--247.

\bibitem[\protect\citeauthoryear{Berrendero, Bueno-Larraz, and Cuevas}{Berrendero et~al.}{2020}]{berrendero2020}
Berrendero, J.~R., B.~Bueno-Larraz, and A.~Cuevas (2020).
\newblock {O}n {M}ahalanobis distance in functional settings.
\newblock {\em Journal of Machine Learning Research\/}~{\em 21\/}(9), 1--33.

\bibitem[\protect\citeauthoryear{Berrendero, Justel, and Svarc}{Berrendero et~al.}{2011}]{berrendero2011}
Berrendero, J.~R., A.~Justel, and M.~Svarc (2011).
\newblock Principal components for multivariate functional data.
\newblock {\em Computational Statistics \& Data Analysis\/}~{\em 55\/}(9), 2619--2634.

\bibitem[\protect\citeauthoryear{Boudt, Rousseeuw, Vanduffel, and Verdonck}{Boudt et~al.}{2020}]{boudt2020minimum}
Boudt, K., P.~J. Rousseeuw, S.~Vanduffel, and T.~Verdonck (2020).
\newblock The minimum regularized covariance determinant estimator.
\newblock {\em Statistics and Computing\/}~{\em 30\/}(1), 113--128.

\bibitem[\protect\citeauthoryear{Butler, Davies, and Jhun}{Butler et~al.}{1993}]{butler1993asymptotics}
Butler, R., P.~Davies, and M.~Jhun (1993).
\newblock Asymptotics for the minimum covariance determinant estimator.
\newblock {\em The Annals of Statistics\/}, 1385--1400.

\bibitem[\protect\citeauthoryear{Chakraborty and Chaudhuri}{Chakraborty and Chaudhuri}{2014}]{Chakraborty2014}
Chakraborty, A. and P.~Chaudhuri (2014).
\newblock {The spatial distribution in infinite dimensional spaces and related quantiles and depths}.
\newblock {\em The Annals of Statistics\/}~{\em 42\/}(3), 1203 -- 1231.

\bibitem[\protect\citeauthoryear{Croux and Haesbroeck}{Croux and Haesbroeck}{1999}]{CROUX1999}
Croux, C. and G.~Haesbroeck (1999).
\newblock Influence function and efficiency of the minimum covariance determinant scatter matrix estimator.
\newblock {\em Journal of Multivariate Analysis\/}~{\em 71\/}(2), 161--190.

\bibitem[\protect\citeauthoryear{Dai and Genton}{Dai and Genton}{2018}]{dai2018}
Dai, W. and M.~G. Genton (2018).
\newblock Multivariate functional data visualization and outlier detection.
\newblock {\em Journal of Computational and Graphical Statistics\/}~{\em 27\/}(4), 923--934.

\bibitem[\protect\citeauthoryear{Delicado}{Delicado}{2011}]{delicado2011}
Delicado, P. (2011).
\newblock Dimensionality reduction when data are density functions.
\newblock {\em Computational Statistics \& Data Analysis\/}~{\em 55\/}(1), 401--420.

\bibitem[\protect\citeauthoryear{Egozcue, D{\'i}az-Barrero, and Pawlowsky-Glahn}{Egozcue et~al.}{2006}]{egozcue2006}
Egozcue, J.~J., J.~L. D{\'i}az-Barrero, and V.~Pawlowsky-Glahn (2006).
\newblock Hilbert space of probability density functions based on {A}itchison geometry.
\newblock {\em Acta Mathematica Sinica\/}~{\em 22\/}(4), 1175--1182.

\bibitem[\protect\citeauthoryear{Febrero, Galeano, and González-Manteiga}{Febrero et~al.}{2008}]{febrero2008}
Febrero, M., P.~Galeano, and W.~González-Manteiga (2008).
\newblock {Outlier detection in functional data by depth measures, with application to identify abnormal NOx levels}.
\newblock {\em Environmetrics\/}~{\em 19\/}(4), 331--345.

\bibitem[\protect\citeauthoryear{Fraiman and Muniz}{Fraiman and Muniz}{2001}]{fraiman2001}
Fraiman, R. and G.~Muniz (2001, Dec).
\newblock Trimmed means for functional data.
\newblock {\em Test\/}~{\em 10\/}(2), 419--440.

\bibitem[\protect\citeauthoryear{Friedman}{Friedman}{1987}]{Friedman1987EFDA}
Friedman, J.~H. (1987).
\newblock {E}xploratory {P}rojection {P}ursuit.
\newblock {\em Journal of the American Statistical Association\/}~{\em 82\/}(397), 249--266.

\bibitem[\protect\citeauthoryear{Galeano, Joseph, and Lillo}{Galeano et~al.}{2015}]{galeano2015}
Galeano, P., E.~Joseph, and R.~E. Lillo (2015).
\newblock {T}he {M}ahalanobis {D}istance for {F}unctional {D}ata {W}ith {A}pplications to {C}lassification.
\newblock {\em Technometrics\/}~{\em 57\/}(2), 281--291.

\bibitem[\protect\citeauthoryear{Ghiglietti, Ieva, and Paganoni}{Ghiglietti et~al.}{2017}]{ghiglietti2017statistical}
Ghiglietti, A., F.~Ieva, and A.~M. Paganoni (2017).
\newblock {S}tatistical inference for stochastic processes: two-sample hypothesis tests.
\newblock {\em Journal of Statistical Planning and Inference\/}~{\em 180}, 49--68.

\bibitem[\protect\citeauthoryear{G{\'o}recki, Krzy{\'{s}}ko, Waszak, and Wo{\l}y{\'{n}}ski}{G{\'o}recki et~al.}{2018}]{górecki2018}
G{\'o}recki, T., M.~Krzy{\'{s}}ko, {\L}.~Waszak, and W.~Wo{\l}y{\'{n}}ski (2018, Mar).
\newblock Selected statistical methods of data analysis for multivariate functional data.
\newblock {\em Statistical Papers\/}~{\em 59\/}(1), 153--182.

\bibitem[\protect\citeauthoryear{Horv\'{a}th and Kokoszka}{Horv\'{a}th and Kokoszka}{2012}]{HorvathKokoszka2012}
Horv\'{a}th, L. and P.~Kokoszka (2012).
\newblock {\em Inference for Functional Data with Applications}.
\newblock Springer Series in Statistics. Springer.

\bibitem[\protect\citeauthoryear{Hron, Menafoglio, Templ, Hrůzová, and Filzmoser}{Hron et~al.}{2016}]{hron2016}
Hron, K., A.~Menafoglio, M.~Templ, K.~Hrůzová, and P.~Filzmoser (2016).
\newblock {S}implicial principal component analysis for density functions in {B}ayes spaces.
\newblock {\em Computational Statistics \& Data Analysis\/}~{\em 94}, 330--350.

\bibitem[\protect\citeauthoryear{Huber}{Huber}{1964}]{Huber1964}
Huber, P.~J. (1964).
\newblock {Robust Estimation of a Location Parameter}.
\newblock {\em The Annals of Mathematical Statistics\/}~{\em 35\/}(1), 73 -- 101.

\bibitem[\protect\citeauthoryear{Hubert, Rousseeuw, and Segaert}{Hubert et~al.}{2015}]{hubert2015}
Hubert, M., P.~J. Rousseeuw, and P.~Segaert (2015, Jul).
\newblock Multivariate functional outlier detection.
\newblock {\em Statistical Methods {\&} Applications\/}~{\em 24\/}(2), 177--202.

\bibitem[\protect\citeauthoryear{Kessy, Lewin, and Strimmer}{Kessy et~al.}{2018}]{kessy2018optimal}
Kessy, A., A.~Lewin, and K.~Strimmer (2018).
\newblock Optimal whitening and decorrelation.
\newblock {\em The American Statistician\/}~{\em 72\/}(4), 309--314.

\bibitem[\protect\citeauthoryear{Kokoszka and Reimherr}{Kokoszka and Reimherr}{2017}]{kokoszka2017}
Kokoszka, P. and M.~Reimherr (2017).
\newblock {\em Introduction to Functional Data Analysis}.
\newblock Chapman \& Hall / CRC numerical analysis and scientific computing. CRC Press.

\bibitem[\protect\citeauthoryear{Lemberge, Raedt, Janssens, Wei, and Van~Espen}{Lemberge et~al.}{2000}]{lemberge2000}
Lemberge, P., I.~Raedt, K.~Janssens, F.~Wei, and P.~Van~Espen (2000).
\newblock {Q}uantitative analysis of 16-17th century archaeological glass vessels using {PLS} regression of {EPXMA} and $\mu$-{XRF} data.
\newblock {\em Journal of Chemometrics\/}~{\em 14}, 751--763.

\bibitem[\protect\citeauthoryear{Locker and Prenter}{Locker and Prenter}{1980}]{locker1980regularization}
Locker, J. and P.~Prenter (1980).
\newblock {R}egularization with differential operators. {I}. {G}eneral theory.
\newblock {\em Journal of Mathematical analysis and applications\/}~{\em 74\/}(2), 504--529.

\bibitem[\protect\citeauthoryear{López-Pintado and Romo}{López-Pintado and Romo}{2009}]{pintado2009}
López-Pintado, S. and J.~Romo (2009).
\newblock On the concept of depth for functional data.
\newblock {\em Journal of the American Statistical Association\/}~{\em 104\/}(486), 718--734.

\bibitem[\protect\citeauthoryear{Machalov{\'a}, Talsk{\'a}, Hron, and G{\'a}ba}{Machalov{\'a} et~al.}{2021}]{machalova2021}
Machalov{\'a}, J., R.~Talsk{\'a}, K.~Hron, and A.~G{\'a}ba (2021, Jun).
\newblock Compositional splines for representation of density functions.
\newblock {\em Computational Statistics\/}~{\em 36\/}(2), 1031--1064.

\bibitem[\protect\citeauthoryear{Masak and Panaretos}{Masak and Panaretos}{2023}]{masak2023}
Masak, T. and V.~M. Panaretos (2023).
\newblock Random surface covariance estimation by shifted partial tracing.
\newblock {\em Journal of the American Statistical Association\/}~{\em 118\/}(544), 2562--2574.

\bibitem[\protect\citeauthoryear{Menafoglio, Grasso, Secchi, and Colosimo}{Menafoglio et~al.}{2018}]{menafoglio2018}
Menafoglio, A., M.~Grasso, P.~Secchi, and B.~M. Colosimo (2018).
\newblock Profile monitoring of probability density functions via simplicial functional pca with application to image data.
\newblock {\em Technometrics\/}~{\em 60\/}(4), 497--510.

\bibitem[\protect\citeauthoryear{Menafoglio, Guadagnini, Guadagnini, and Secchi}{Menafoglio et~al.}{2021}]{Menafoglio2021}
Menafoglio, A., L.~Guadagnini, A.~Guadagnini, and P.~Secchi (2021).
\newblock Object oriented spatial analysis of natural concentration levels of chemical species in regional-scale aquifers.
\newblock {\em Spatial Statistics\/}~{\em 43}, 100494.

\bibitem[\protect\citeauthoryear{Morigi, Reichel, and Sgallari}{Morigi et~al.}{2007}]{morigi2007}
Morigi, S., L.~Reichel, and F.~Sgallari (2007).
\newblock {O}rthogonal projection regularization operators.
\newblock {\em Numerical Algorithms\/}~{\em 44\/}(2), 99--114.

\bibitem[\protect\citeauthoryear{Nagler}{Nagler}{2024}]{kde1d}
Nagler, T. (2024).
\newblock {\em kde1d: {U}nivariate {K}ernel {D}ensity {E}stimation}.
\newblock R package version 1.0.7, \url{https://cran.r-project.org/web/packages/kde1d/index.html}.

\bibitem[\protect\citeauthoryear{Oguamalam, Radojičić, and Filzmoser}{Oguamalam et~al.}{2024}]{oguamalam2024}
Oguamalam, J., U.~Radojičić, and P.~Filzmoser (2024).
\newblock {M}inimum regularized covariance trace estimator and outlier detection for functional data.
\newblock {\em Technometrics\/}~{\em 66\/}(4), 588--599.

\bibitem[\protect\citeauthoryear{Oja, Sirkiä, and Eriksson}{Oja et~al.}{2016}]{Oja2016}
Oja, H., S.~Sirkiä, and J.~Eriksson (2016).
\newblock {S}catter matrices and independent component analysis.
\newblock {\em Austrian Journal of Statistics\/}~{\em 35}, 175–189.

\bibitem[\protect\citeauthoryear{Ostebee and Zorn}{Ostebee and Zorn}{2002}]{ostebee2002}
Ostebee, A. and P.~Zorn (2002).
\newblock {\em {C}alculus from {G}raphical, {N}umerical, and {S}ymbolic {P}oints of {V}iew}.
\newblock Number Bd. 2 in Calculus from Graphical, Numerical, and Symbolic Points of View. Harcourt College Publishers.

\bibitem[\protect\citeauthoryear{Pawlowsky-Glahn, Egozcue, and Tolosana-Delgado}{Pawlowsky-Glahn et~al.}{2015}]{pawlowsky15}
Pawlowsky-Glahn, V., J.~Egozcue, and R.~Tolosana-Delgado (2015).
\newblock {\em {M}odeling and analysis of compositional data}.
\newblock Wiley, Chichester.

\bibitem[\protect\citeauthoryear{Petersen, Zhang, and Kokoszka}{Petersen et~al.}{2022}]{petersen2022}
Petersen, A., C.~Zhang, and P.~Kokoszka (2022).
\newblock {M}odeling probability density functions as data objects.
\newblock {\em Econometrics and Statistics\/}~{\em 21}, 159--178.

\bibitem[\protect\citeauthoryear{Ramsay and Silverman}{Ramsay and Silverman}{1997}]{ramsay2005}
Ramsay, J.~O. and B.~W. Silverman (1997).
\newblock {\em Principal components analysis for functional data}.
\newblock New York, NY: Springer New York.

\bibitem[\protect\citeauthoryear{Ramsay and Silverman}{Ramsay and Silverman}{2005}]{silverman2005}
Ramsay, J.~O. and B.~W. Silverman (2005).
\newblock {\em {Functional Data Analysis}}.
\newblock Springer.

\bibitem[\protect\citeauthoryear{Rousseeuw and Driessen}{Rousseeuw and Driessen}{1999}]{rousseeuw1999}
Rousseeuw, P.~J. and K.~V. Driessen (1999).
\newblock {A} fast algorithm for the minimum covariance determinant estimator.
\newblock {\em Technometrics\/}~{\em 41\/}(3), 212--223.

\bibitem[\protect\citeauthoryear{Secchi, Stamm, and Vantini}{Secchi et~al.}{2013}]{Stamm2013}
Secchi, P., A.~Stamm, and S.~Vantini (2013).
\newblock {Inference for the mean of large $p$ small $n$ data: A finite-sample high-dimensional generalization of Hotelling’s theorem}.
\newblock {\em Electronic Journal of Statistics\/}~{\em 7\/}(none), 2005 -- 2031.

\bibitem[\protect\citeauthoryear{Talská, Menafoglio, Hron, Egozcue, and Palarea-Albaladejo}{Talská et~al.}{2020}]{talska2020}
Talská, R., A.~Menafoglio, K.~Hron, J.~J. Egozcue, and J.~Palarea-Albaladejo (2020).
\newblock {W}eighting the domain of probability densities in functional data analysis.
\newblock {\em Stat\/}~{\em 9\/}(1), e283.
\newblock e283 sta4.283.

\bibitem[\protect\citeauthoryear{van~den Boogaart, Egozcue, and Pawlowsky-Glahn}{van~den Boogaart et~al.}{2010}]{vanboogaart2010}
van~den Boogaart, K.-G., J.~J. Egozcue, and V.~Pawlowsky-Glahn (2010).
\newblock {B}ayes linear spaces.
\newblock {\em SORT : statistics and operations research transactions\/}~{\em 34\/}(2), 201--222.

\bibitem[\protect\citeauthoryear{van~den Boogaart, Pawlowsky-Glahn, and Egozcue}{van~den Boogaart et~al.}{2014}]{vanboogaart2014}
van~den Boogaart, K.-G., V.~Pawlowsky-Glahn, and J.~J. Egozcue (2014).
\newblock {B}ayes {H}ilbert spaces.
\newblock {\em Australian \& New Zealand Journal of Statistics\/}~{\em 56}, 171–194.

\bibitem[\protect\citeauthoryear{Wang}{Wang}{2012}]{WANG2012}
Wang, Z. (2012).
\newblock {M}ulti-parameter {T}ikhonov regularization and model function approach to the damped {M}orozov principle for choosing regularization parameters.
\newblock {\em Journal of Computational and Applied Mathematics\/}~{\em 236\/}(7), 1815--1832.

\end{thebibliography}


\begin{thebibliography}{}

\bibitem[\protect\citeauthoryear{Arribas-Gil and Romo}{Arribas-Gil and Romo}{2014}]{arribas-gil2014}
Arribas-Gil, A. and J.~Romo (2014, 03).
\newblock {Shape outlier detection and visualization for functional data: the {O}utliergram}.
\newblock {\em Biostatistics\/}~{\em 15\/}(4), 603--619.

\bibitem[\protect\citeauthoryear{Berrendero, Bueno-Larraz, and Cuevas}{Berrendero et~al.}{2020}]{berrendero2020}
Berrendero, J.~R., B.~Bueno-Larraz, and A.~Cuevas (2020).
\newblock {O}n {M}ahalanobis distance in functional settings.
\newblock {\em Journal of Machine Learning Research\/}~{\em 21\/}(9), 1--33.

\bibitem[\protect\citeauthoryear{Boudt, Rousseeuw, Vanduffel, and Verdonck}{Boudt et~al.}{2020}]{rousseeuw2020}
Boudt, K., P.~J. Rousseeuw, S.~Vanduffel, and T.~Verdonck (2020).
\newblock The {M}inimum {R}egularized {C}ovariance {D}eterminant estimator.
\newblock {\em Statistics and Computing\/}~{\em 30\/}(1), 113--128.

\bibitem[\protect\citeauthoryear{Febrero, Galeano, and González-Manteiga}{Febrero et~al.}{2008}]{febrero2008}
Febrero, M., P.~Galeano, and W.~González-Manteiga (2008).
\newblock {Outlier detection in functional data by depth measures, with application to identify abnormal NOx levels}.
\newblock {\em Environmetrics\/}~{\em 19\/}(4), 331--345.

\bibitem[\protect\citeauthoryear{Fraiman and Muniz}{Fraiman and Muniz}{2001}]{fraiman2001}
Fraiman, R. and G.~Muniz (2001, Dec).
\newblock Trimmed means for functional data.
\newblock {\em Test\/}~{\em 10\/}(2), 419--440.

\bibitem[\protect\citeauthoryear{Hall and Hosseini-Nasab}{Hall and Hosseini-Nasab}{2006}]{hall2006}
Hall, P. and M.~Hosseini-Nasab (2006).
\newblock On properties of functional principal components analysis.
\newblock {\em Journal of the Royal Statistical Society Series B: Statistical Methodology\/}~{\em 68\/}(1), 109--126.

\bibitem[\protect\citeauthoryear{Hildebrandt}{Hildebrandt}{1912}]{Hildebrandt1912}
Hildebrandt, T.~H. (1912).
\newblock Necessary and sufficient conditions for the interchange of limit and summation in the case of sequences of infinite series of a certain type.
\newblock {\em Annals of Mathematics\/}~{\em 14\/}(1/4), 81--83.

\bibitem[\protect\citeauthoryear{Hormann and Kokoszka}{Hormann and Kokoszka}{2011}]{Hormann2011ConsistencyOT}
Hormann, S. and P.~Kokoszka (2011).
\newblock Consistency of the mean and the principal components of spatially distributed functional data.
\newblock {\em Bernoulli\/}~{\em 19}, 1535--1558.

\bibitem[\protect\citeauthoryear{Hyndman and {Shahid Ullah}}{Hyndman and {Shahid Ullah}}{2007}]{HYNDMAN2007}
Hyndman, R.~J. and M.~{Shahid Ullah} (2007).
\newblock Robust forecasting of mortality and fertility rates: A functional data approach.
\newblock {\em Computational Statistics \& Data Analysis\/}~{\em 51\/}(10), 4942--4956.

\bibitem[\protect\citeauthoryear{Kokoszka and Reimherr}{Kokoszka and Reimherr}{2017}]{kokoszka2017}
Kokoszka, P. and M.~Reimherr (2017).
\newblock {\em Introduction to Functional Data Analysis}.
\newblock Chapman \& Hall / CRC numerical analysis and scientific computing. CRC Press.

\bibitem[\protect\citeauthoryear{Machalov{\'a}, Talsk{\'a}, Hron, and G{\'a}ba}{Machalov{\'a} et~al.}{2021}]{machalová2021}
Machalov{\'a}, J., R.~Talsk{\'a}, K.~Hron, and A.~G{\'a}ba (2021, Jun).
\newblock Compositional splines for representation of density functions.
\newblock {\em Computational Statistics\/}~{\em 36\/}(2), 1031--1064.

\bibitem[\protect\citeauthoryear{Oguamalam, Radojičić, and Filzmoser}{Oguamalam et~al.}{2024}]{oguamalam2024}
Oguamalam, J., U.~Radojičić, and P.~Filzmoser (2024).
\newblock {M}inimum regularized covariance trace estimator and outlier detection for functional data.
\newblock {\em Technometrics\/}~{\em 66\/}(4), 588--599.

\bibitem[\protect\citeauthoryear{Rousseeuw}{Rousseeuw}{1984}]{rousseeuw1984}
Rousseeuw, P.~J. (1984).
\newblock Least median of squares regression.
\newblock {\em Journal of the American Statistical Association\/}~{\em 79\/}(388), 871--880.

\bibitem[\protect\citeauthoryear{Sun and Genton}{Sun and Genton}{2011}]{sun2011}
Sun, Y. and M.~G. Genton (2011).
\newblock Functional boxplots.
\newblock {\em Journal of Computational and Graphical Statistics\/}~{\em 20\/}(2), 316--334.

\end{thebibliography}





% \end{thebibliography}

\end{document}

\maketitle

% -------------------
% --- Supplement ----
% -------------------

This supplementary material provides additional content supporting the article \textit{Robust functional PCA for relative data}. In Section \ref{supp:aux_results}, we present auxiliary theoretical considerations that extend discussions from Section \ref{sec:RDPCA} of the main part. Section \ref{supp:simulations} contains further simulation studies, both complementing and extending the experiments presented in the main part. Section \ref{supp:realdata} includes additional findings for the real data example introduced in Section \ref{sec:RDPCA_fertility} of the main text. Finally, in Section \ref{supp:proofs}, we provide detailed proofs of all theoretical results stated in the main paper and the supplement.

\section{Auxiliary results} \label{supp:aux_results}

\subsection{RDPCA for densely observed data}

Algorithm \ref{alg:alg1} provides the pseudocode for estimating robust FPCs, their corresponding eigenvalues, as well as robust mean and covariance when the underlying data has been sampled on a dense grid.

\begin{algorithm}[h!]
\singlespacing
\small
\SetAlgoLined
\caption{RDPCA}\label{alg:alg1}
\SetKwInOut{Input}{Input}

        \Input{sample $X_1,\ldots,X_n \in \mathcal{B}^2$ observed at $p$ time points $t_1,\dots,t_p$, subset size $h$, initial subset $H_1\subset\{1,\dots,n\}$ with $|H_1|=h$, regularization parameter $\alpha > 0$, projection dimension $k \leq n$, and the tolerance level $\varepsilon_k>0$;}
        \BlankLine
	$Y_i\leftarrow\clr(X_i)$, $i=1,\dots,n$;
	\SetKwRepeat{Do}{do}{while}
	\Do{$H_0\neq H_1$}{
	$H_0\leftarrow H_1$;\\
    Sample mean: 
    $\bar{Y}_{H_0}\leftarrow \frac{1}{h}\sum_{i\in H_0}Y_i$;\\
Non-scaled covariance: 
 $\hat{C}_{\clr,H_0} \leftarrow \frac{1}{h} \sum_{i \in H_0} \left( Y_i - \bar{Y}_{H_0} \right) \otimes \left( Y_i - \bar{Y}_{H_0} \right)$;\\
Calculate spectral decomposition  $\hat{C}_{\clr,H_0}=\sum_{i\geq 1}\hat{\lambda}_{i,H_0}\hat{\xi}_{i,H_0}\otimes\hat{\xi}_{i,H_0}$;\\
	Set initial scaling parameter $c_{1,H_0}\leftarrow 1$;\\
    \Do{$(c_{1,H_0}-c_{0,H_0})^2\geq\varepsilon_k$}{
	    $c_{0,H_0}\leftarrow c_{1,H_0}$;\\  
     Set $\hat{V}(k)\leftarrow I-\sum_{i=1}^k
    \hat{\xi}_{i,H_0}\otimes\hat{\xi}_{i,H_0}$;\\
    Standardization: 
$Y_{i,\mathrm{st},H_0}^{\alpha,k}=\left(\hat{C}_{\clr,H_0}+\alpha / c_{0,H_0} \hat{V}(k)\right)^{-1}\hat{C}_{\clr,H_0}^{1/2}(Y_i - \bar{Y}_{H_0})$;\\
$d_{i,H_0,c_0}^2\leftarrow\|Y_{i,\mathrm{st},H_0}^{\alpha,k} \|^2$;\\
    Generate $\eta_{k+1},\dots,\eta_n \sim \chi^2(1),\quad y \sim \chi^2(k)$;\\
    $\displaystyle c_{1,H_0}\leftarrow\frac{\mathrm{med}\left\{d_{i,H_0,c_0}^2:i=1,\dots,n\right\}}{\mathrm{med}\left\{y +  \sum_{i=k+1}^\infty \frac{\lambda_i^2}{(\lambda_i + \alpha)^2}\eta_i\right\}}$;\\
	}
    Order $d^2_{(i_1,H_0,c_1)}\leq\dots\leq d^2_{(i_n,H_0,c_1)}$;\\
    Set $H_1\leftarrow\{i_1,\dots,i_h\}$;\\
    }
    Solve symmetric eigenproblem $ c_{1,H_1} \hat{C}_{\clr,H_1} (\xi_j) = \lambda_j \xi_j$;\\
    Set $\hat{\xi}_{j,H_\mathrm{opt}} \leftarrow \xi_j$, $\hat{\zeta}_{j,H_\mathrm{opt}} = \clr^{-1}(\xi_j)$ and $\hat{\lambda}_{j,H_\mathrm{opt}} \leftarrow \lambda_j$;\\
    
\SetKwInOut{Output}{Output}
\Output{Optimal subset $H_\text{opt} := H_1$, scaling factor $c_{H_0}$, squared robust RDMD $(c_{1,H_0}^{-1}d_{(1,H_0,c_1)}^2,\dots c_{1,H_0}^{-1}d_{(n,H_0,c_1)}^2)$, robust eigenvalues $\{\hat{\lambda}_{j,H_\mathrm{opt}}\}_{j \geq 1}$ and corresponding robust FPCs $\{ \hat{\zeta}_{j,H_\mathrm{opt}} \}_{j \geq 1}, \{ \hat{\xi}_{j,H_\mathrm{opt}} \}_{j \geq 1}$  in Bayes and clr space, respectively.}
\end{algorithm}

\subsection{Selecting the subset size $h$} \label{supp:choice_of_h}

\begin{figure}[htbp]
    \centering
\includegraphics[width=\linewidth]{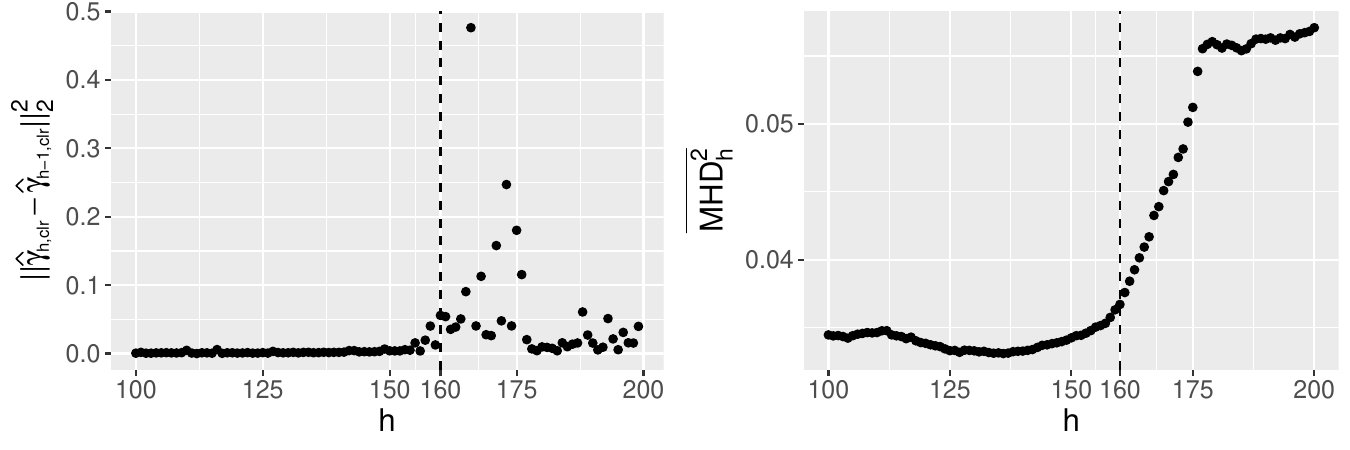}
    \caption{\small Illustration of rule of thumb for selecting the subset size $h$ for Model 1.1 $(n = 200, p = 50, c = 0.2)$ of the simulation study in Section~\ref{sec:RDPCA_sim1}. Left: squared norm of difference of estimated covariance functions for consecutive values of $h$. The term $\hat{\gamma}_{h,\mathrm{clr}}$ refers to the sub-sampled covariance function calculated at a subset of size $h$. Right: trajectory of the objective of Problem \ref{eq:fixed_point_alpha_2} as a function of subset size $h$. The term $\overline{\mathrm{MHD}^2_h}$ refers to the mean squared RDMD calculated at a subset of size $h$. For both approaches, an increase at around $h=160$ is noticeable, caused by the first outliers being included in the subset.
    }
    \label{fig:objective}
\end{figure}

On the one hand, the optimal subset should serve as a good fit to the majority of the data by containing at least half of the observations, i.e. $h \geq n/2$. On the other hand, if we assume that our data set contains a certain fraction $\lceil c \cdot n\rceil$ of outlying observations, the optimal subset should not be chosen too large, i.e., $h \leq n - \lceil c \cdot n\rceil$. This trade-off between robustness and efficiency is discussed in \cite{rousseeuw1984} in more detail. 
In practice, the actual number of contaminated samples within a sample is unknown beforehand. Therefore, \cite{rousseeuw2020} introduced a data-driven approach where they look at differences in the norm or the objective of the underlying estimator based on subsequent values of $h$. A significant change in the slope of the consequent plot can indicate the presence of outliers. In Figure \ref{fig:objective}, we give an illustrative example of the proposed methods.
%\vspace{-0.05cm}

\subsection{Additional theoretical results}

\begin{corollary}\label{cor:L_2 standardization explicit form}
     Let 
    $\displaystyle X_{\mathrm{st}}^{{\alpha},{I}} =_B
    \argmin_{Y\in \mathcal{B}^2(I)} \left\{\|{C}_{B^2,X}^{1/2}Y\ominus X\|_{\mathcal{B}}^2 + \alpha\|Y\|_{\mathcal{B}}^2   \right\},$ for regularization parameter $\alpha>0$. 
    Then 
    \begin{align*}
X_{\mathrm{st}}^{\alpha,I} =_B \bigoplus_{i=1}^\infty \frac{\lambda_i^{1/2}}{\lambda_i + \alpha} \odot \langle \zeta_i,X \rangle_{B^2} \odot \zeta_i,
\end{align*}
where $(\lambda_i,\zeta_i)$ is an $i$-th eigenpair of $C_{B^2,X}$ satisfying $C_{B^2,X}\zeta_i=\lambda_i\odot\zeta_i$, $i\geq 1$.
\end{corollary}

\begin{proposition}\label{prop:Mah_distribution}
   Let $X\in \mathcal{B}^2(I)$ be a constant-mean random density such that $\clr(X)$ is a Gaussian random process. Then, for any $\alpha>0$ and $k\in\mathbf{N}_0$
   $$
   M^2_{\alpha,k,B^2}(X)\sim \begin{cases}
       y +  \sum_{i=k+1}^\infty \frac{\lambda_i^2}{(\lambda_i+\alpha)^2} \eta_i,\text{ for } k\geq 1,\\
\sum_{i=1}^\infty\frac{\lambda_i^2}{(\lambda_i+\alpha)^2} \eta_i, \text{ for } k=0,
   \end{cases}
   $$
where $\eta_i\sim\chi^2(1)$, $i\geq 1$ are mutually independent and also independent of $y \sim \chi^2(k)$, and  $\lambda_i$, $i\geq 1$ is the $i$th eigenvalue $C_{B^2,X}$.
\end{proposition}

\begin{lemma}\label{sup_lemma:consistency}
    Let $\mathcal{X}=\{X_1,\dots,X_{N_1}\}$ be the iid sample, $X_i\in\mathcal{B}^2(I)$ with Bayes mean $\mu_{B^2,X}$ and covariance $C_{B^2,X}$, and assume additionally that $\mathbb{E}\|X_i\|_{\mathcal{B}^2}^4<\infty$. Let further $\hat{\mu}_{B^2,H}$ and $\hat{C}_{B^2,H,c_H}$ be the robust estimators of the mean and covariance based on the optimal H-subset as in \eqref{eq:trimmed_mean_covariance} and let $h = \lfloor N - \rho N^{\delta}  \rfloor$, for some $\rho \in [0.5,1]$ and $0\leq\delta<1$ Then, the following holds:
    \begin{itemize}
        \item[i)]  \begin{equation}
        \| {C}_{B^2,X} \ominus \hat{C}_{B^2,H,1} \|_{\mathrm{HS},B^2} \overset{P}{\rightarrow} 0 \text{ as } N \rightarrow \infty.
    \end{equation}
    $\| {C}_{B^2,X} \ominus \hat{C}_{B^2,H,c_H} \|_{\mathrm{HS},B^2}^2 = \sum_{i\geq 1}\|{C}_{B^2,X}\phi_i \ominus \hat{C}_{B^2,H,c_H}\phi_i\|_{\mathcal{B}^2}^2$ refers to the Hilbert-Schmidt norm of the Bayes space, where $\phi_1,\phi_2,\dots \in\mathcal{B}^2(I)$ is an orthonormal basis of $\mathcal{B}^2(I)$.
    \item[ii)] Assuming that the eigenvalues $\lambda_i$, $i\geq 1$ are simple,  $\|s_i\psi_{B^2,i}-\hat{\psi}_{B^2,H}\| \overset{P}{\rightarrow} 0$, where $\psi_{B^2,i}$, and $\hat{\psi}_{B^2,H}$ are the $i$th eigenfunction of ${C}_{B^2,X}$ and $\hat{C}_{B^2,H,X}$, respectively, and $s_i\in\{1-,1\}$ is the sign correction. 
    \end{itemize}
\end{lemma}

\section{Simulations} \label{supp:simulations}

% ---------------------------------------------------------------
% Simulation 1 --------------------------------------------------
% ---------------------------------------------------------------
\subsection{Simulation 1} \label{supp:RDPCA_sim1}

\begin{figure}[ht]
    \centering
    \includegraphics[width=.9\linewidth]{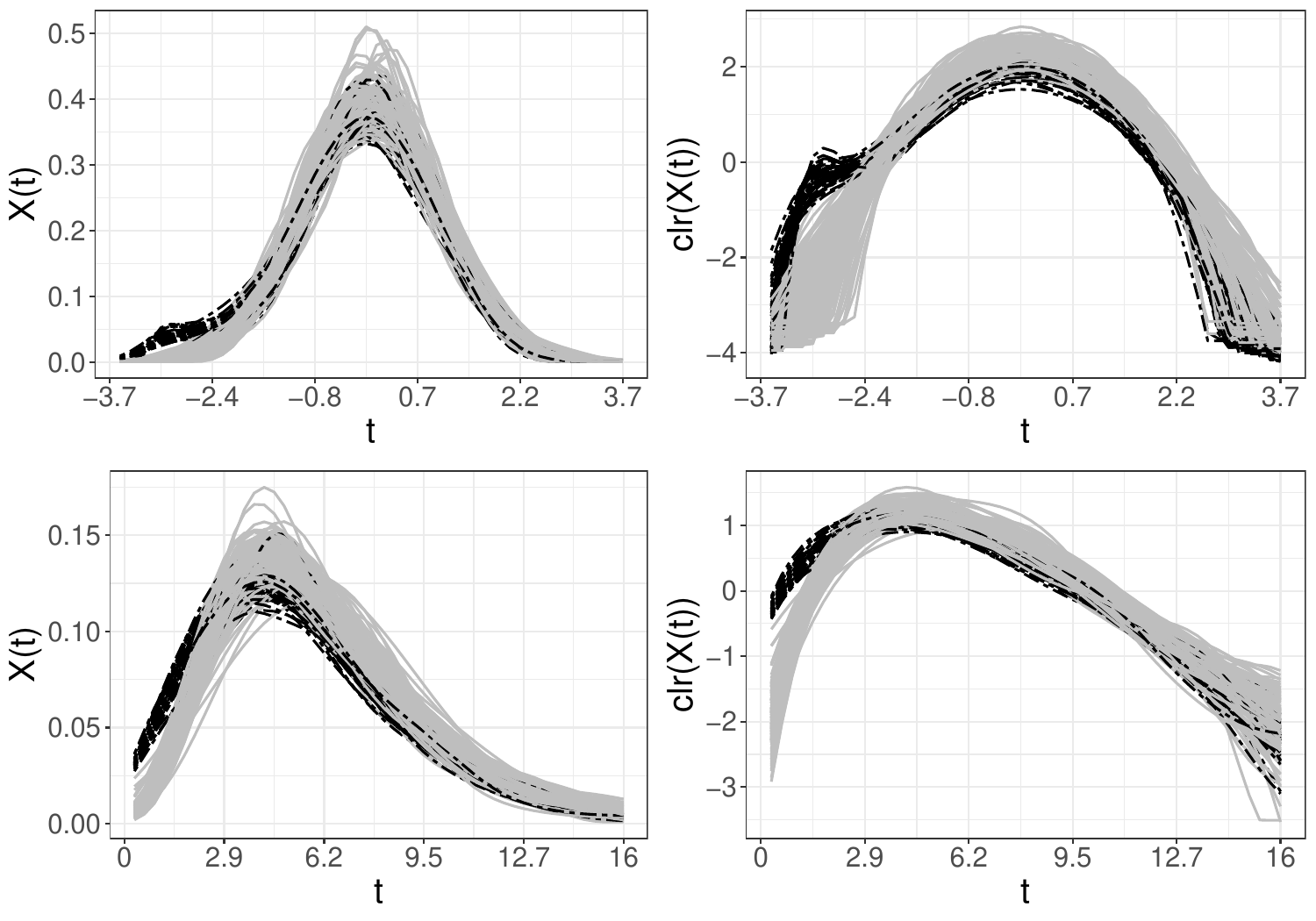}
    \caption{\small Visualization of densities (left) and clr transformed densities (right) for Model 1.2 (top row) and Model 1.3 (bottom row). Solid curves represent the main processes, while the dashed ones indicate the outliers. The sample size is $n = 200$, $p = 50$ time points, and the contamination rate is $c = 0.2$.
    }
    \label{fig:dataClr}
\end{figure}

In this section, we extend the simulation of Section \ref{sec:RDPCA_sim1} by
\begin{align*}
    \texttt{Model 1.2}: X_\mathrm{reg} =&~ \mathrm{KDE}\left(Y_1,\ldots,Y_{N}\right),~ X_\mathrm{out} = \mathrm{KDE}\left(Y_1,\ldots,Y_{N},\tilde{Y}_1,\ldots,\tilde{Y}_{\lfloor 0.05N \rfloor }\right),~ \mathrm{and} \\
    \texttt{Model 1.3}: X_\mathrm{reg} =&~ \mathrm{KDE}\left(Z_1,\ldots,Z_{N}\right),~ X_\mathrm{out} = \mathrm{KDE}\left(Z_1,\ldots,Z_{N},\tilde{Z}_1,\ldots,\tilde{Z}_{\lfloor 0.1N \rfloor }\right).
\end{align*}
For Model 1.2 we have $Y_i \sim N(0,1),~ \tilde{Y}_i \sim U([q_1;q_2]), N = 250,$ with $q_1, q_2$ being the $0.01\%$ and $0.25\%$ quantile of the standard normal distribution, while for 1.3 we set $Z_i \sim \chi^2_6,~ \tilde{Z}_i \sim U([q_3;q_4])$, and $q_3, q_4$ to the $0.1\%$ and $0.2\%$ quantiles of $\chi^2_6$. To showcase the densities and the clr transformed curves, we provide Figure \ref{fig:dataClr}. 

\begin{figure}[ht]
    \centering
    \includegraphics[width=.9\linewidth]{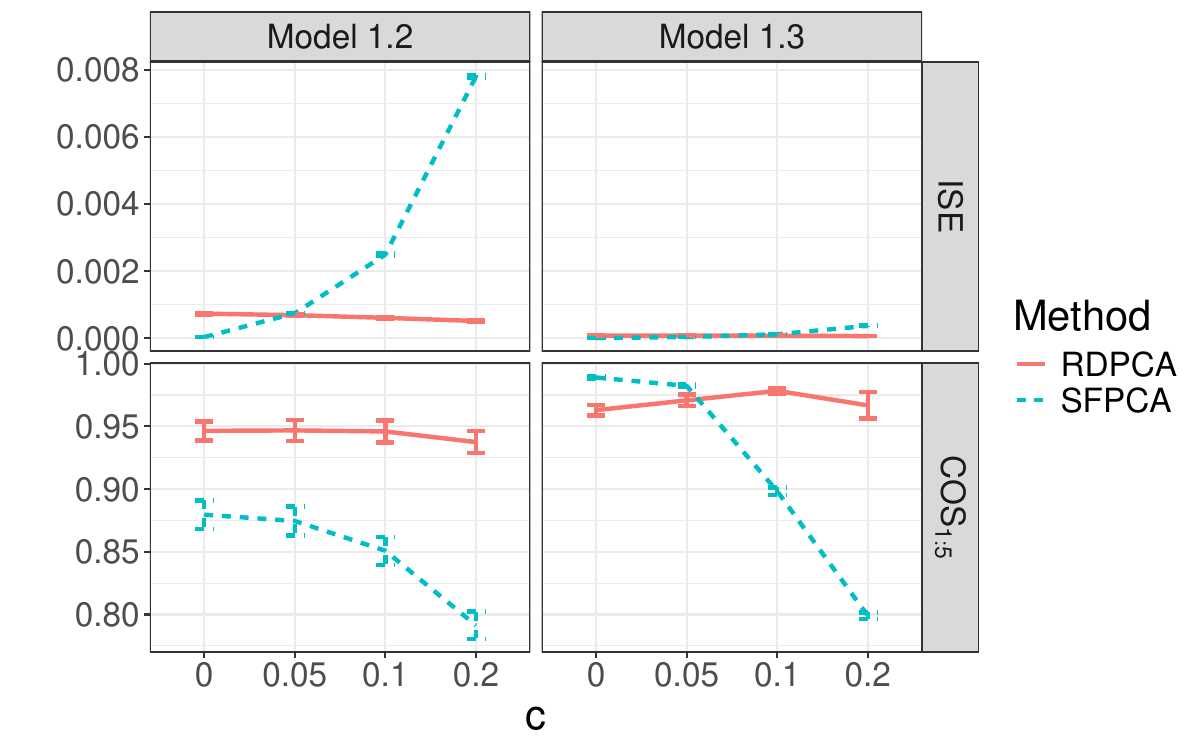}
    \caption{\small Integrated square error and mean cosine between estimated and true eigenfunctions. Displayed are the mean $+ / -$ one standard error of each measure over all iterations at each level of the corresponding parameter configuration. 
    Top row: ISE. Bottom row: Mean Cosine between the first five pairs of eigenfunctions. All measurements are calculated based on the covariance estimated on RDPCA (solid) and SFPCA (dashed), and respectively compared with a sample estimate based on a Monte Carlo simulation of the true distribution.}
    \label{fig:CovmeasuresLinesSim1}
\end{figure}

Both settings were analyzed for the same parameter configurations of $c, \alpha, h,$ and $k$ as in Section \ref{sec:RDPCA_sim1}. We evaluate the performance of RDPCA  and compare it to SFPCA using two statistical measures. The integrated square error (ISE) measures the accuracy between an estimate of the true underlying covariance and the covariance based on the robust/non-robust approach. Precisely, given estimates $\hat{v}, \hat{\gamma}$ of the covariance function, the ISE is defined as $\text{ISE} = p^{-2} \sum_{i,j = 1}^p \left( \hat{\gamma}(t_i, t_j) - \hat{v}(t_i, t_j) \right)^2$. The estimate of the true covariance is based on the sample covariance of a Monte Carlo simulation where many samples following the true distribution are generated. The corresponding results are shown in Figure \ref{fig:CovmeasuresLinesSim1}. This is followed by the mean cosine between the first five pairs of eigenfunctions based on estimated covariance and the true covariance. These values are depicted at the bottom of Figure \ref{fig:CovmeasuresLinesSim1}. Similar to the initial presentation of this simulation in Section \ref{sec:RDPCA_sim1}, both the ISE and the mean cosine indicate that RDPCA outperforms SFPCA when outliers are present.

\begin{figure}[ht]
    \centering
    \includegraphics[width=.9\linewidth]{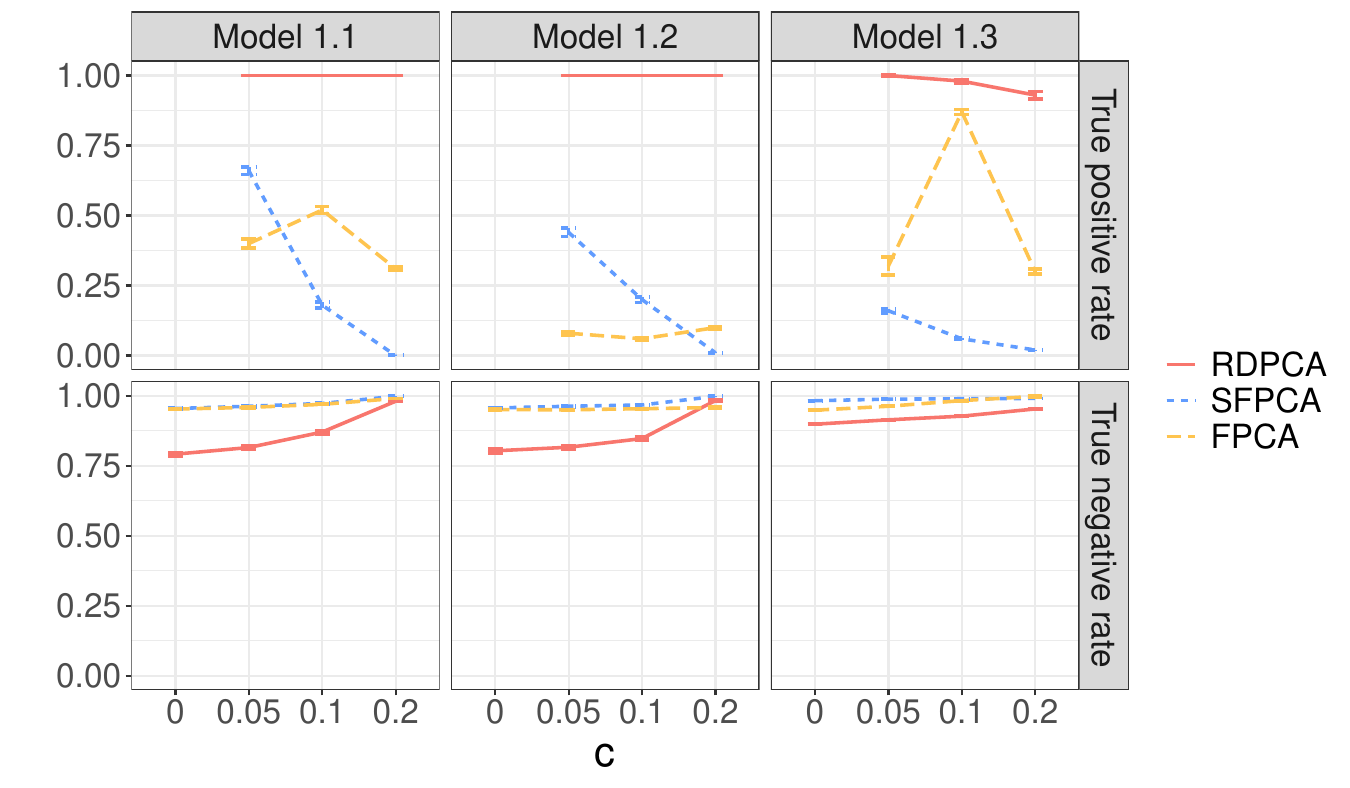}
    \caption{\small True positive and true negative rate of models of the first simulation, comparing RDPCA, SFPCA, and robust FPCA. Displayed are the mean $+ / -$ one standard error over all iterations at each level of the corresponding parameter configuration.}
    \label{fig:ErrorRatesSim1}
\end{figure}

Finally, we evaluate the advantage of the clr transformation over a non-compositional, simple functional approach. As the clr transformation focuses on relative changes, rather than absolute differences, small deviations within the tails of the distribution carry significantly more weight after transforming the data. Based on the squared RDMD and $\alpha$-Mahalanobis distances as well as the corresponding theoretical cutoff values, we can calculate true positive and true negative rates. In Figure \ref{fig:ErrorRatesSim1}, the resulting rates are presented. Here, we used a simulated cutoff value at the $95\%$-quantile of the underlying limiting distributions, see \ref{cor:cor_1} and \cite{berrendero2020}. Model 1.1 refers to the model presented in Section \ref{sec:RDPCA_sim1}. In situations where outliers differ in the tails of the domain, a direct approach would completely fail in correctly distinguishing regular and irregular observations from each other. Thus, when analyzing density data, the clr transformation can be an important and necessary tool to emphasize the relative scale property. 

\begin{figure}[ht]
    \centering
    \includegraphics[width=.9\linewidth]{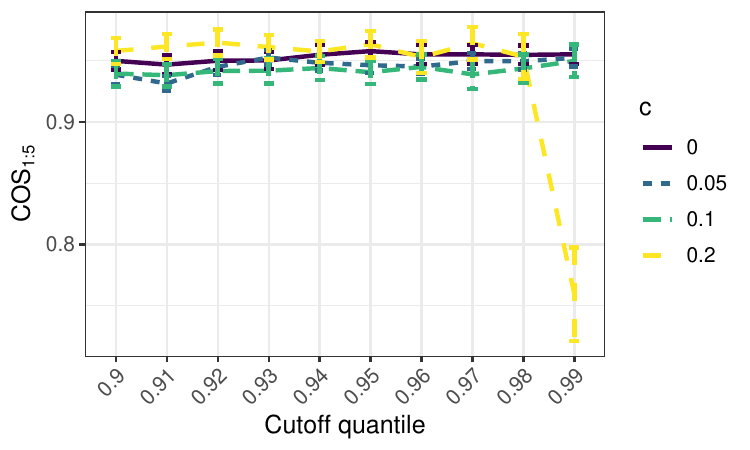}
    \caption{\small Mean cosine between estimated true and reweighted covariance based on regular observations according to different quantiles under the assumptions of Gaussianity. Calculated at Model 1.1 ($n = 500, p = 50$) over 50 repetitions for each parameter configuration. }
    \label{fig:QuantileConsistency}
\end{figure}

To improve the efficiency of the covariance estimator, we used a re-weighting step of the covariance based on the subset. The idea was to not only use observations from the optimal subset, but rather all observations whose RDMD lay below a certain cutoff and are thus considered as regular. In Figure \ref{fig:QuantileConsistency}, we show the interaction between this re-weighting step and the choice of the cutoff under the assumptions of Gaussianity (see Corollary \ref{cor:cor_1} in the main paper) for a small simulation study of Model 1.1 ($n = 500, p = 50$). This is illustrated by the mean cosine between the estimated true and the reweighted covariance. The latter is calculated based on the regular observations according to the chosen cutoff. As we see, the choice of the quantile and therefore the number of discarded regular observations, does not significantly influence the performance of the sub-sampled covariance estimator.

% ---------------------------------------------------------------
% Simulation 2 --------------------------------------------------
% ---------------------------------------------------------------
\subsection{Simulation 2} \label{supp:RDPCA_sim2}

In this extension of the simulation study conducted in Section \ref{sec:RDPCA_sim2}, we construct data by
\begin{align*}
    \texttt{Model 2.2: } & \clr(X_\mathrm{reg}(t)) = s_1 \xi_1 + s_2 \xi_2 + s_3 \xi_3 + s_4 \xi_4 ,\\
    &  \clr(X_\mathrm{out}(t)) = \tilde{s}_1 \xi_1 + \tilde{s}_2 \xi_2 + \tilde{s}_3 \xi_3 + \tilde{s}_4 \xi_4
\end{align*}
for the eigenfunctions $\xi_1(t) = \sqrt{2}\text{sin}\left(2 \pi t\right), \xi_2(t) = \sqrt{2}\text{cos}\left(2 \pi t\right), \xi_3(t) = \sqrt{2}\text{sin}\left(4 \pi t\right), \xi_4(t) = \sqrt{2}\text{cos}\left(4 \pi t\right)$ of the covariance operator of the clr transformed densities. The corresponding scores satisfy $\mathbb{E}\left[s_j\right] = 0$ while $\text{var}(s_j)$ is equal to the $j$-th eigenvalue $\lambda_j$ of the covariance operator. By Mercers Theorem \citep{kokoszka2017} the covariance function satisfies $\gamma(s,t) = \sum_{j=1}^4 s_j \xi_j(s) \xi_j(t)$. 

\begin{figure}[ht]
    \centering
    \includegraphics[width=.9\linewidth]{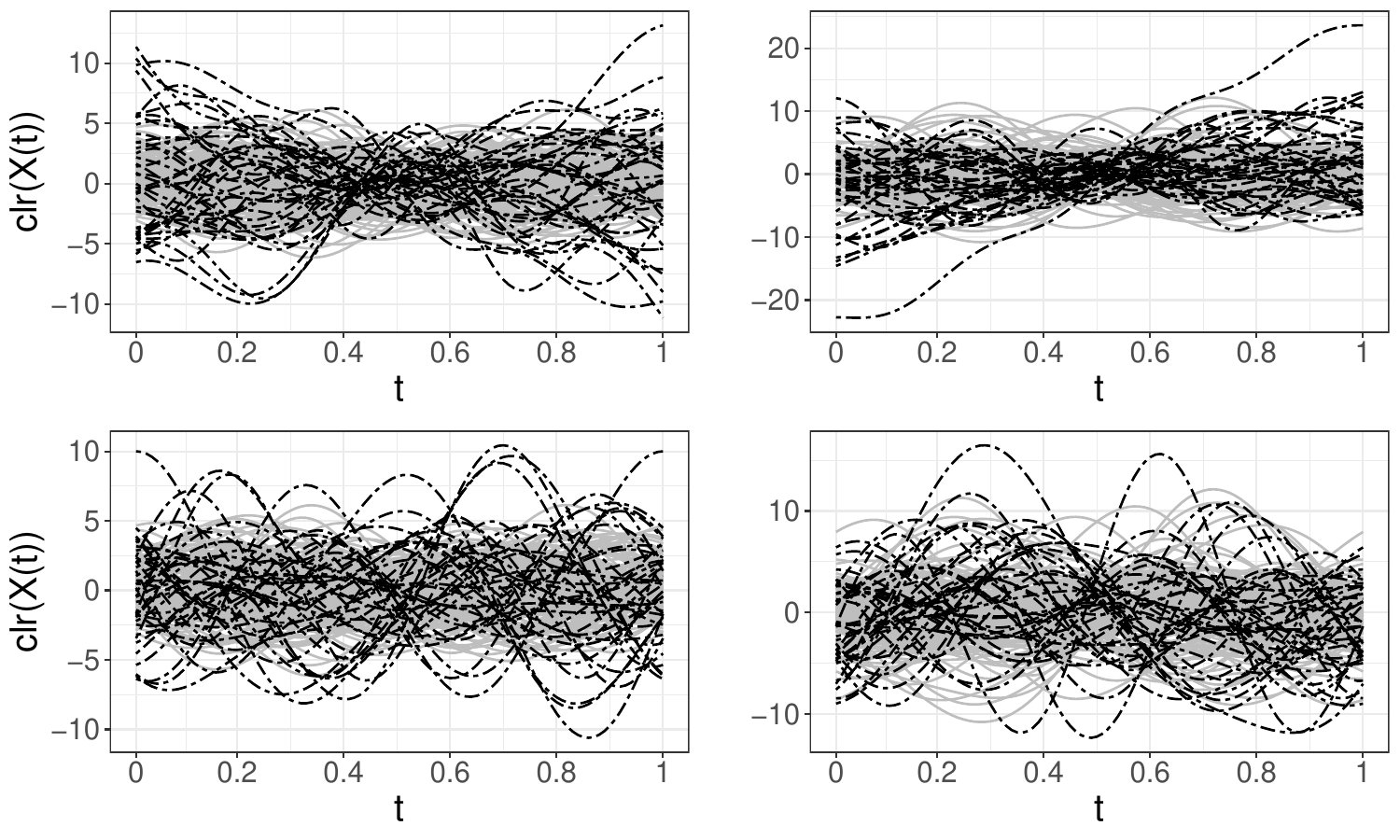}
    \caption{\small Visualization of curves for Model 2.1 (top row; introduced in the main part in Section \ref{sec:RDPCA_sim2}) and Model 2.2 (bottom row) with normal (left) and multivariate t-distributed (right) scores. Solid curves represent the main processes, dashed ones indicate outliers. Parameters: $n = 200, p = 100, c=0.2$.
    }
    \label{fig:dataClrSim2Mod12}
\end{figure}

To showcase Model 2.1 (from Section \ref{sec:RDPCA_sim2} of the main part) and Model 2.2, we provide Figure \ref{fig:dataClrSim2Mod12}. The setup of Model 2.1 was already covered in the main part. For Model 2.2, the outlying scores follow a different distribution from the scores corresponding to regular observations. As in Section \ref{sec:RDPCA_sim2}, the regular scores either individually follow an independent normal distribution, i.e. $s_i \sim N(0,\lambda_i)$ or jointly a multivariate t-distribution, i.e. $(s_1,s_2,s_3,s_4) \sim t_5(\boldsymbol{0},\mathrm{diag}(\lambda_1,\lambda_2,\lambda_3\lambda_4))$ with $(\lambda_1,\lambda_2,\lambda_3\lambda_4) = (2,1,1/2,1/4)$. Scores $\tilde{s}_i$ of the outlying curves follow the same scheme under different scaling parameters $(\tilde{\lambda}_1,\tilde{\lambda}_2,\tilde{\lambda}_3\tilde{\lambda}_4) = (4,3,5/2,9/4)$.

\begin{figure}[ht]
    \centering
    \includegraphics[width=.9\linewidth]{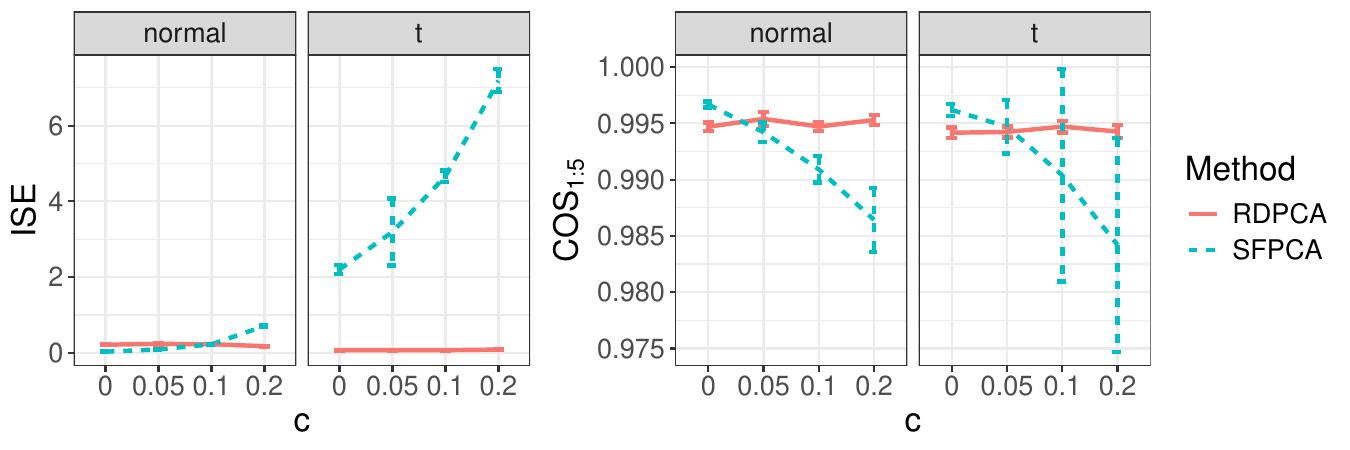}
    \caption{Integrated square error and cosine between first four pairs of estimated and true eigenfunctions. Displayed are the mean $+ / -$ one standard error of each measure over all iterations at each level of the corresponding parameter configuration. All measurements are calculated based on RDPCA and SFPCA covariance estimators and respectively compared with a true covariance function $\gamma(s,t) = \sum_{j=1}^4 s_j \xi_j(s) \xi_j(t).$ The notation “normal" and “t" refer to the distribution of the scores in the underlying model as described at the beginning of this section.}
    \label{fig:CovmeasuresLinesSim2Mod12} 
\end{figure}

As in the previous simulation, this analysis also considers varying contamination rates, i.e. $c=\{0,0.05,0.1, 0.2\}$, an automatically selected regularization parameter $\alpha$, and $n = 500, h = \lfloor 0.75n \rfloor$, $k = 1$. Each parameter configuration is iterated 100 times. Also, the performance is measured by the ISE, the mean cosine of the first four pairs of eigenfunctions, and compared to SFPCA. The corresponding results are shown in the left part of Figure \ref{fig:CovmeasuresLinesSim2Mod12}. On the right is the mean cosine based on the estimated covariance and the true covariance. Both measures suggest that the robust approach performs similarly or even better than the non-robust approach in the presence of outliers. 

\begin{figure}[ht]
    \centering
    \includegraphics[width=.9\linewidth]{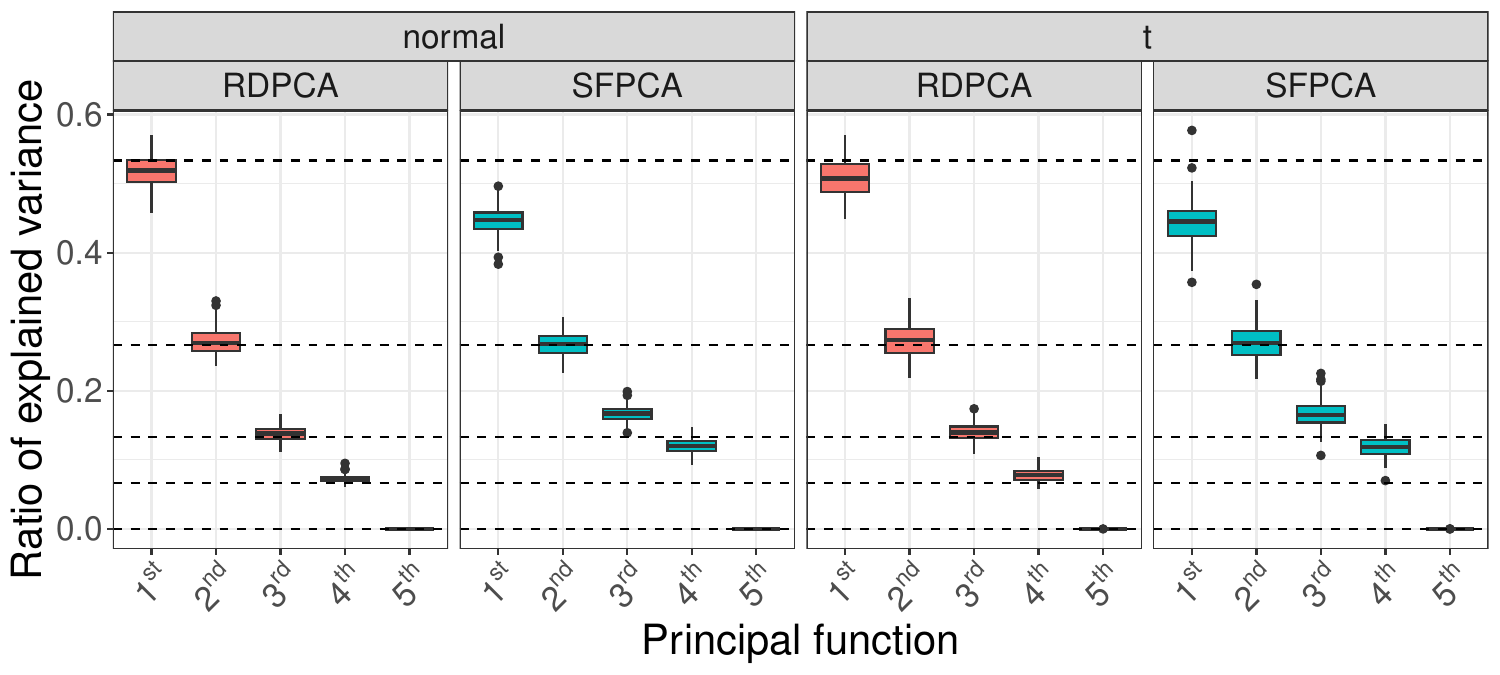}
    \caption{Estimated ratio of explained variance for the first five principal functions over different score distributions at $c = 0.2$. The dashed reference lines depict the true proportion.}
    \label{fig:explainedVarSim2Mod12}
\end{figure}

Another important feature is the accuracy of the estimation of the eigenvalues $\lambda_j$. To measure this performance, we consider the explained variance of the first five estimated eigenfunctions. These results are depicted in Figure \ref{fig:explainedVarSim2Mod12}, where we compare the ratio of explained variance to the true proportion for a contamination rate of $c = 0.2$. The results clearly show that only RDPCA can precisely identify the correct ratios for all estimated eigenfunctions. 

% ---------------------------------------------------------------------------------
% Additional simulation results  --------------------------------------------------
% ---------------------------------------------------------------------------------
\subsection{Additional simulation results} \label{sec:add_simul_results_supp}

This chapter presents an in-depth analysis of the introduced simulations, where we examine the sensitivity of RDPCA with respect to the hyperparameters $h$ and $k$. Furthermore, we investigate the large-sample behavior of Algorithm \ref{alg:alg1}. Finally, we compare RDPCA to different outlier detection methods, often considered in functional settings (\cite{arribas-gil2014}; \cite{berrendero2020}; \cite{oguamalam2024}).

\begin{figure}[!h]
    \centering
    \includegraphics[width=.9\linewidth]{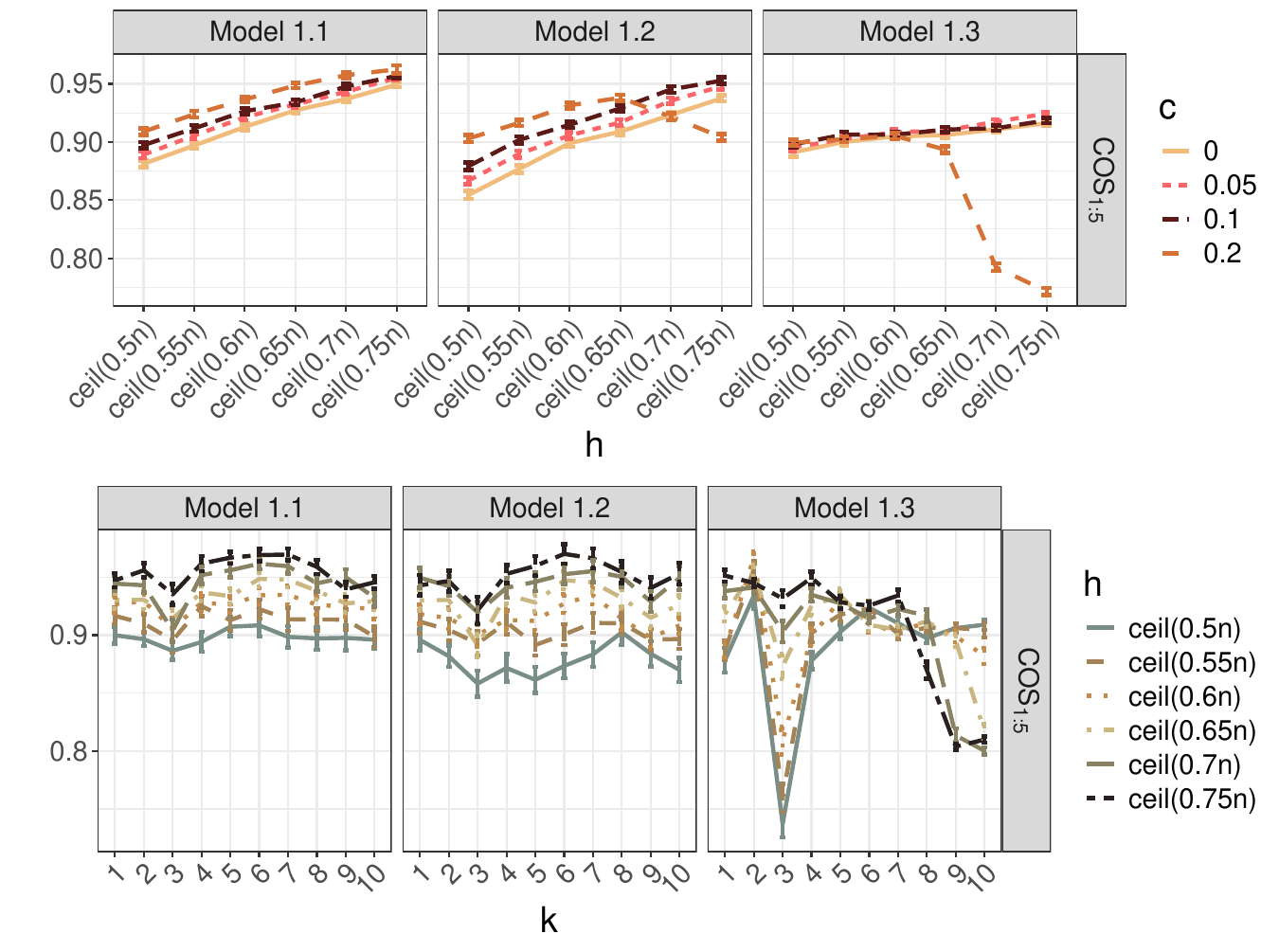}
    \caption{\small Sensitivity analysis of first simulation with respect to the subset size $h$ (top) and the projection dimension $k$ (bottom) based on raw covariance.
    }
    \label{fig:additionalResults_sensitivity_sim1}
\end{figure}

\begin{figure}[!h]
    \centering
    \includegraphics[width=.9\linewidth]{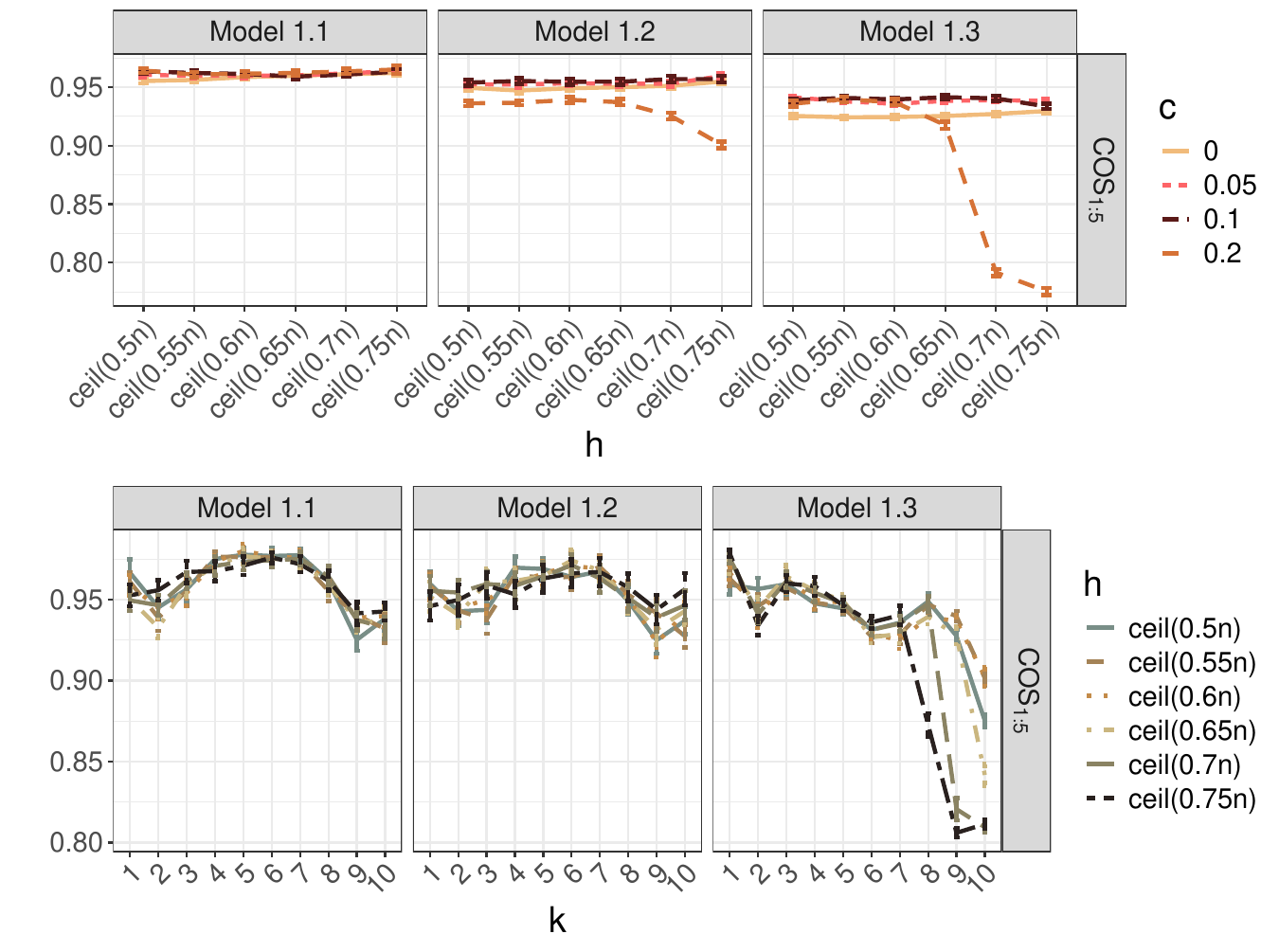}
    \caption{\small Sensitivity analysis of first simulation with respect to the subset size $h$ (top) and the projection dimension $k$ (bottom) based on weighted covariance.
    }
    \label{fig:additionalResults_sensitivity_sim1_weighted}
\end{figure}

\begin{figure}[!h]
    \centering
    \includegraphics[width=.9\linewidth]{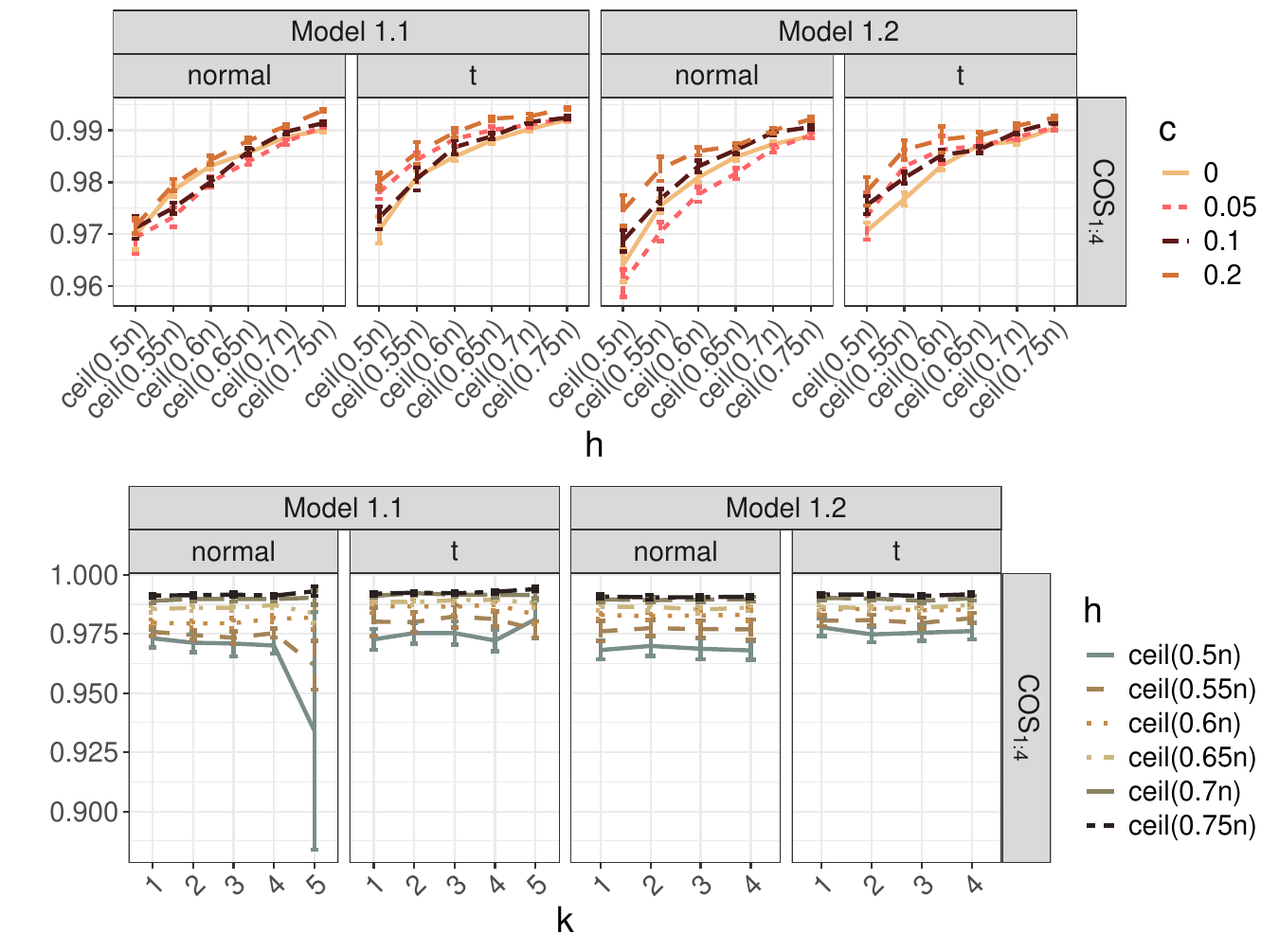}
    \caption{\small Sensitivity analysis of second simulation with respect to the subset size $h$ (top) and the projection dimension $k$ (bottom) based on raw covariance.
    }
    \label{fig:additionalResults_sensitivity_sim2}
\end{figure}

\begin{figure}[!h]
    \centering
    \includegraphics[width=.9\linewidth]{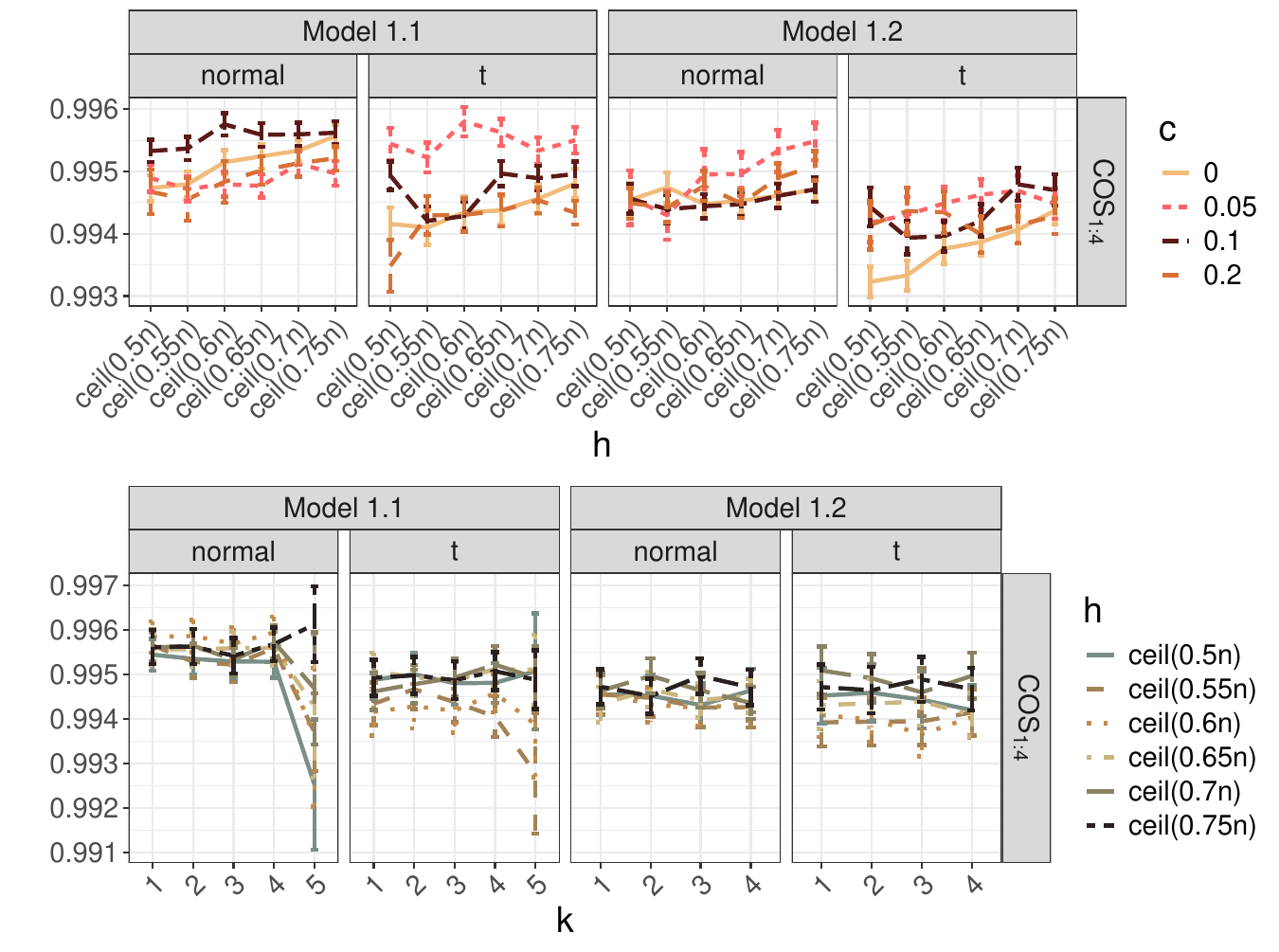}
    \caption{\small Sensitivity analysis of second simulation with respect to the subset size $h$ (top) and the projection dimension $k$ (bottom) based on weighted covariance.
    }
    \label{fig:additionalResults_sensitivity_sim2_weighted}
\end{figure}

For the hyperparameter analysis, we consider different combinations of the projection dimension $k$ and the subset size $h$, i.e., $k \in \{1,\ldots, 10\}$ and $h \in \{\lceil0.5n\rceil, \lceil0.55n\rceil, \ldots, \lceil0.75n\rceil\}.$ The corresponding results of the first simulation are presented in Figure \ref{fig:additionalResults_sensitivity_sim1} and \ref{fig:additionalResults_sensitivity_sim1_weighted}. Here, we consider a fixed sample size of $n = 500$ as well as $p = 50$. In Figure \ref{fig:additionalResults_sensitivity_sim1}, we evaluate the performance based on the mean cosine between the estimated true and sub-sampled covariance based on the optimal subset. In the top plot, we investigate the sensitivity with respect to $h$ and fix $k=1$. Mostly throughout all three models, a slight increase in performance for increasing values of $h$ is noticeable. This makes sense since we have a maximum of $ 20\%$  of outliers, and therefore, increasing $h$ should only improve the accuracy. The plot below depicts the sensitivity of $k$ for a fixed contamination rate $c = 0.2$. Also, for this parameter, a rather stable behavior is apparent. In Figure \ref{fig:additionalResults_sensitivity_sim1_weighted}, additional results based on re-weighting are provided. Using that approach seems to completely mitigate the choice of $h$ while also being relatively stable with respect to $k$.

Results of the second simulation are displayed in Figure \ref{fig:additionalResults_sensitivity_sim2} and \ref{fig:additionalResults_sensitivity_sim2_weighted} in a similar fashion. Here, we fix $n = 500, p = 100$. Again, results show that RDPCA seems to be quite stable with respect to the choice of the parameters $k$ and $h$. It has to be noted that due to the construction of the data, for higher values of $k$, the underlying projection matrix $\hat{\boldsymbol{V}}(k)$ contains the zero vector, which leads to numeric instabilities. Therefore, these values are excluded from the figure.

\begin{figure}[!h]
    \centering
    \includegraphics[width=.9\linewidth]{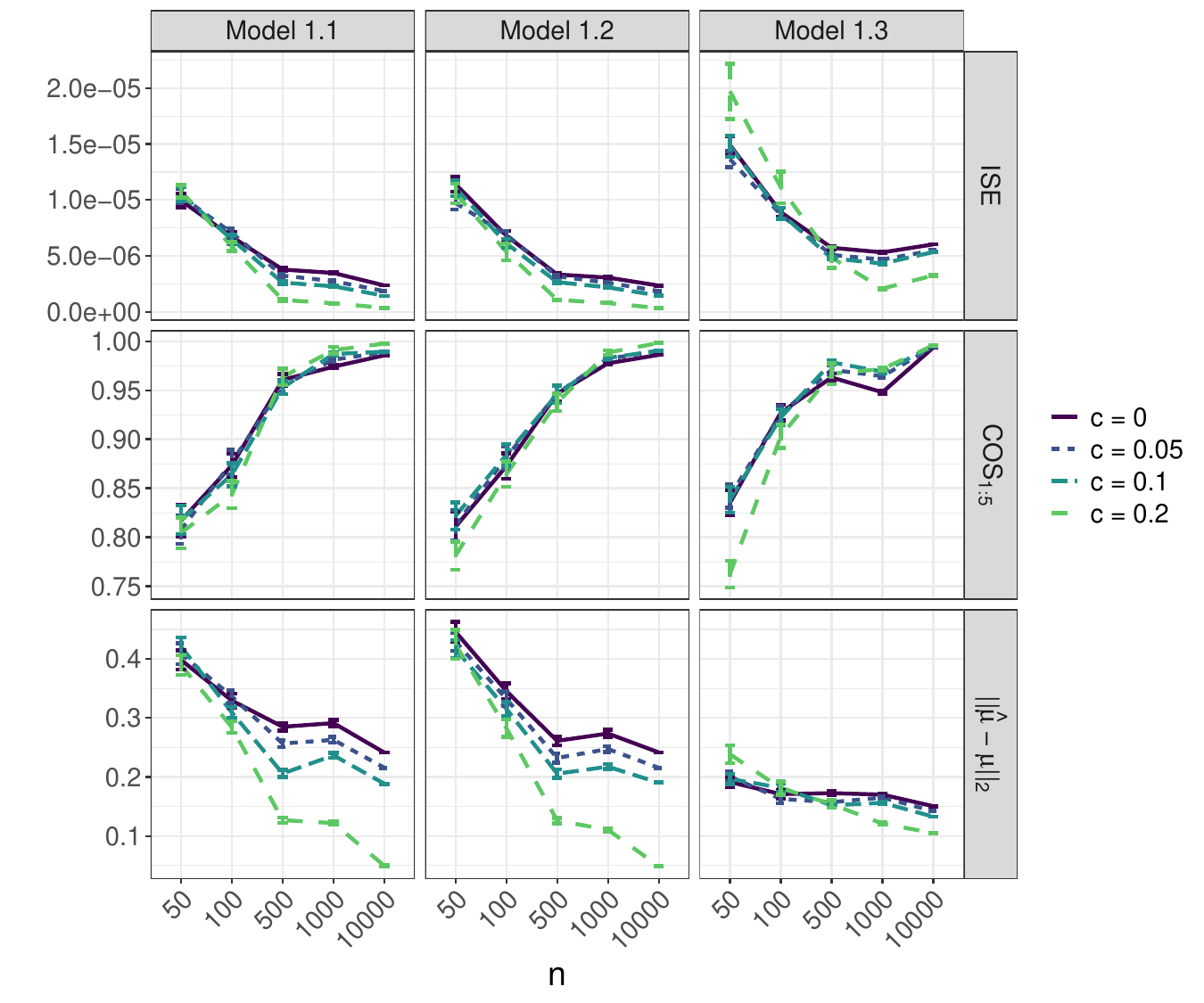}
    \caption{\small Consistency analysis of the three models of the first simulation. Displayed are the trends of ISE (top), mean cosine (middle), and mean error between estimated and true mean (bottom) for increasing sample sizes. 
    }
    \label{fig:additionalResults_consistency_sim1}
\end{figure}

\begin{figure}[!h]
    \centering
    \includegraphics[width=.9\linewidth]{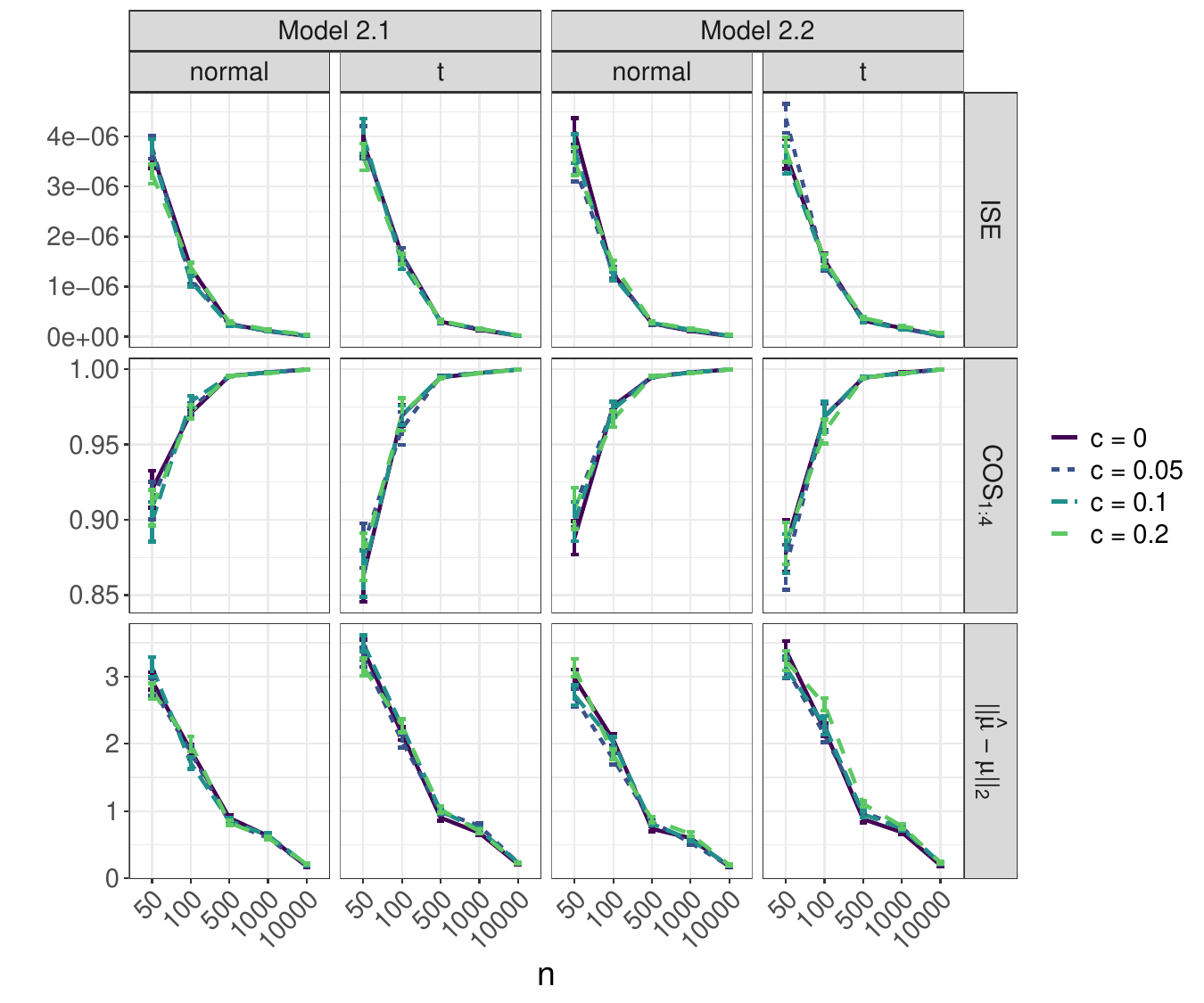}
    \caption{\small Consistency analysis of the two models of the second simulation. Displayed are the trends of ISE (top), mean cosine (middle), and mean error between estimated and true mean (bottom) for increasing sample sizes. 
    }
    \label{fig:additionalResults_consistency_sim2}
\end{figure}

Next, we further investigated the large sample behavior of RDPCA to analyze the consistency of the sub-sampled covariance with respect to the true covariance. We fix parameters $p, h, k$ and subsequently increase the sample size $n \in \{50, 100, 500, 1000, 10000 \}$. The results for the first simulation are shown in Figure \ref{fig:additionalResults_consistency_sim1}. In the top and middle row, we report the ISE and the mean cosine. The bottom row presents the $L^2$-norm between the sub-sampled mean $\hat{\mu}_{\mathrm{clr},H}$ and the true estimated mean $\mu$. The consistency analysis for the second simulation is displayed in Figure \ref{fig:additionalResults_consistency_sim2} and yields very similar results. Both plots indicate consistent results.

\begin{figure}[!h]
    \centering
    \includegraphics[width=.9\linewidth]{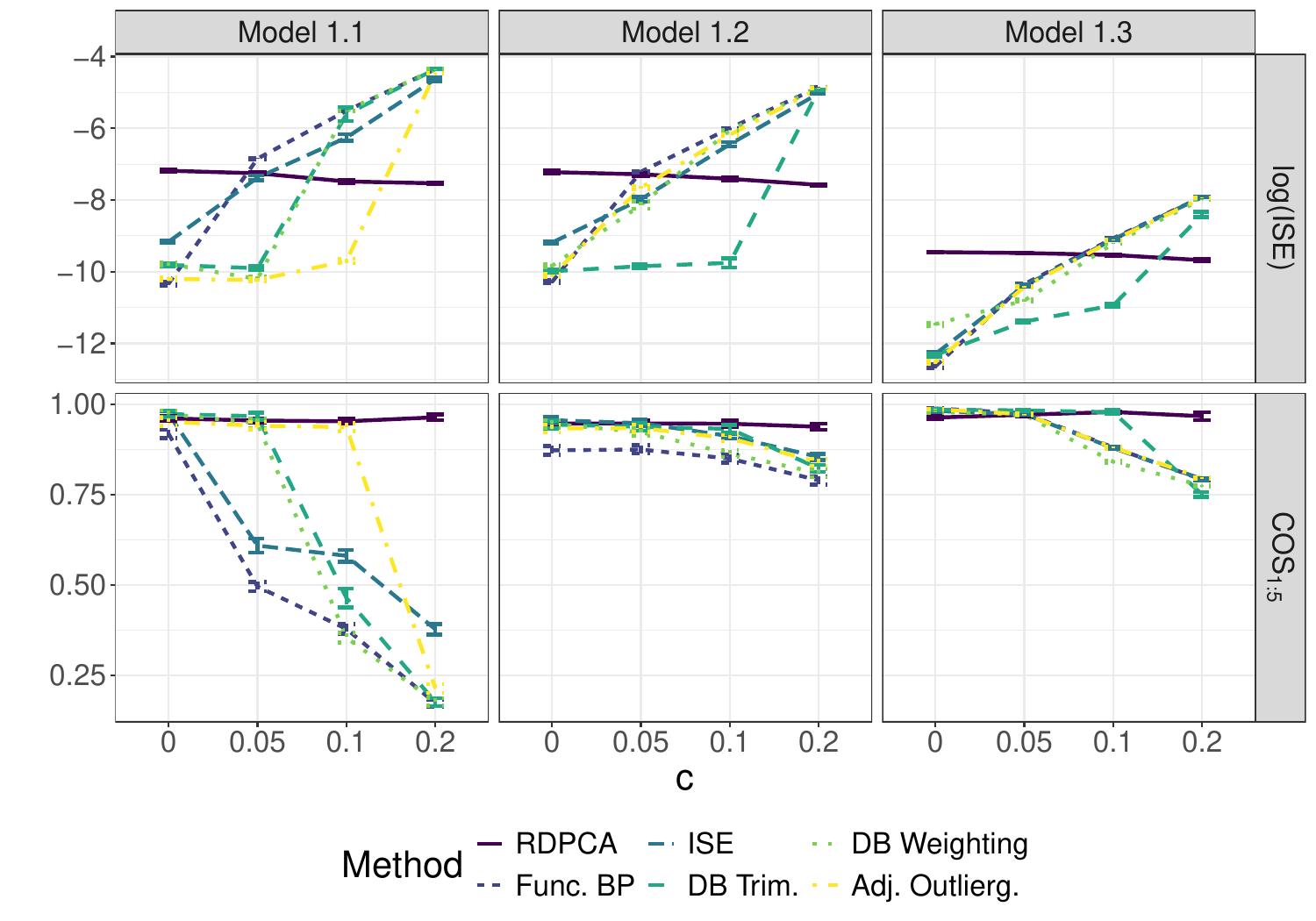}
    \caption{\small Comparison of performance of RDPCA and different functional outlier detection methods in the first simulation setting. Measured are the logarithm of ISE (top) and the mean cosine (bottom). 
    }
    \label{fig:additionalResults_comparison_sim1}
\end{figure}

\begin{figure}[!h]
    \centering
    \includegraphics[width=.9\linewidth]{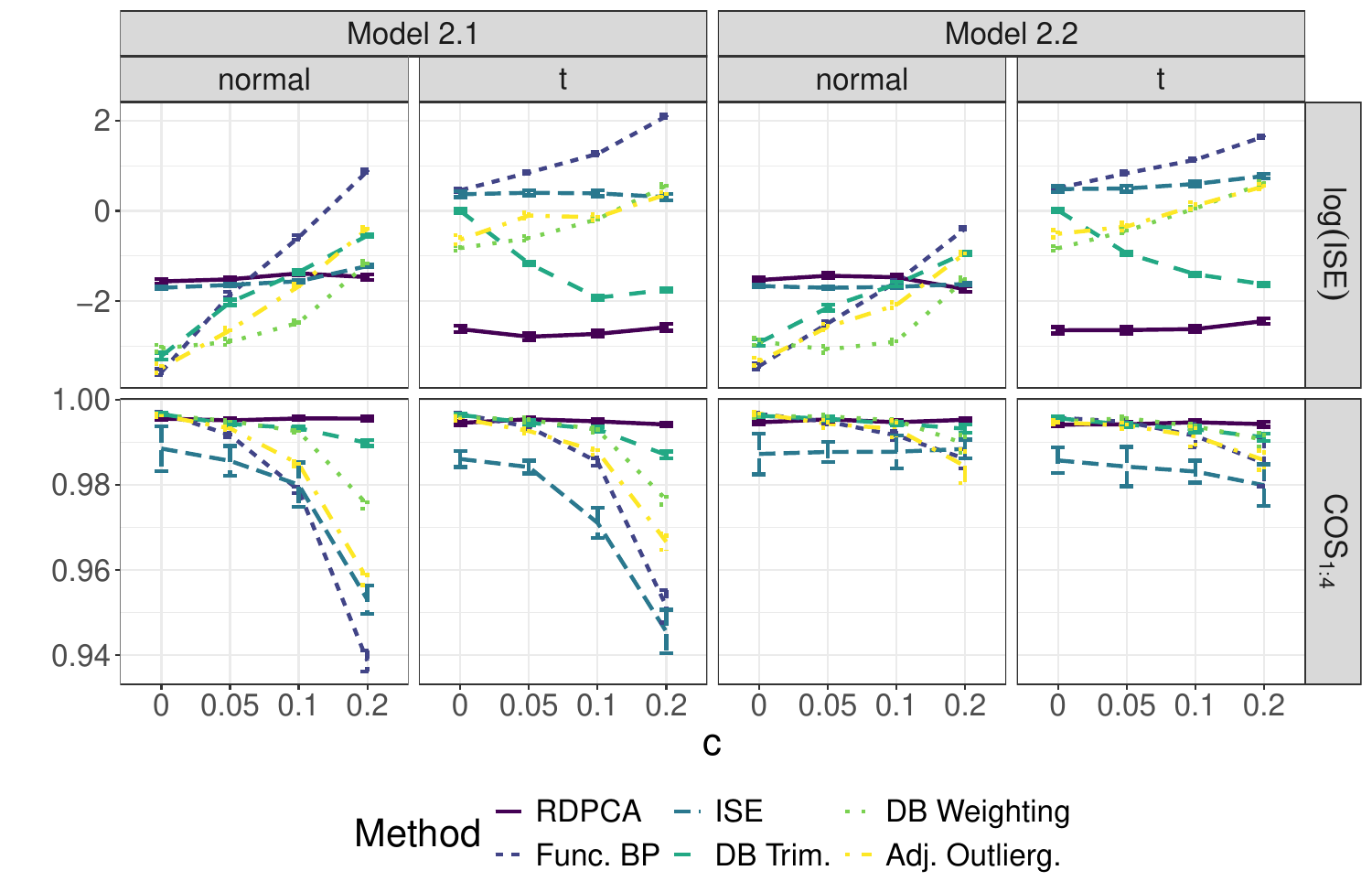}
    \caption{\small Comparison of performance of RDPCA and different functional outlier detection methods in the second simulation setting. Measured are the logarithm of the ISE (top) and the mean cosine (bottom).  
    }
    \label{fig:additionalResults_comparison_sim2}
\end{figure}

Finally, we take a look at a comparative simulation study by extending the previous results in Section \ref{sec:RDPCA_sim1} and \ref{sec:RDPCA_sim2} of the main paper as well as Section \ref{supp:RDPCA_sim1} and \ref{supp:RDPCA_sim2} in the supplement, considering additional outlier detection methods that are commonly used in functional data analysis. These procedures can be classified into three groups based on the underlying framework. \textit{Functional boxplots}  \citep{sun2011}, \textit{depth-based trimming} and \textit{depth-based weighting} \citep{febrero2008} as well as the \textit{adjusted outliergram} \citep{arribas-gil2014} are based on the concept of functional depth. These methods quantify the outlyingness of each data curve by the corresponding centrality within the data sample. More details can be found in \cite{fraiman2001} and the corresponding references. The second group considers functional principal components and can be measured by the \textit{integrated square error} introduced in \cite{HYNDMAN2007}. The third group uses the Mahalanobis distance and contains RDPCA.

Results for both simulations are in Figure \ref{fig:additionalResults_comparison_sim1} and \ref{fig:additionalResults_comparison_sim2}, respectively. We compare the ISE as well as the mean cosine. Besides RDPCA, the covariance estimates for each concept are based on the sub-sampled covariances calculated on the non-outliers given by the respective \texttt{R} function output. The results reveal comparable performance for lower contamination rates, while RDPCA clearly outperforms the other approaches in the presence of more outliers. 

Additionally, Table \ref{tab:comparison_runtimes} provides the average runtime in seconds for both comparative simulation studies. The computation times are aggregated over all configurations as the individual choices do not significantly influence the latter. Overall, the previous results underline a favorable trade-off between computational efficiency and accuracy of RDPCA.

\begin{table}[]
\small
\begin{tabular}{c|c|c|c|c|c|c}
Method             & RDPCA  & Func. BP  & ISE     & DB Trim.  & DB Weighting & Adj. Outlierg. \\ \hline
Mean runtime in s. & 115 / 111 & 228 / 765 & 36 / 29 & 427 / 508 & 415 / 488    & 8 / 18        
\end{tabular}
\caption{\small Average runtime in seconds of each Method. The first and second values correspond to the first and second simulation settings, respectively. Aggregated over all parameter configurations.}
\label{tab:comparison_runtimes}
\end{table}

\section{Real-data example} \label{supp:realdata}

\subsection{Fertility data} 

In this section, more details around the analysis of the fertility data from the Human Fertility Database (HFD) in Section \ref{sec:RDPCA_fertility} are provided. At the time of this analysis, the HFD provides age-specific fertility rates (ASFR) for 39 countries spanning from 1891 to 2023. To ensure reliable analysis, we focus on years where data for at least 20 countries is available, narrowing our time frame to 1956--2020. We further preprocess the data by concentrating on the age groups from 15 to 45 years, resulting in $p=31$ age groups. This focus is justified by biological and statistical considerations: this age range captures the majority of female reproductive years, and fertility rates outside this range are significantly lower and more susceptible to extreme variability.

For each year, we are applying RDPCA based on Algorithm \ref{alg:alg1_in_basis} to the resulting ASFR data, using 21 compositional splines \citep{machalová2021} as basis functions. The remaining parameters are $k = 3, h = \lceil 0.5n \rceil$, while $\alpha$ is selected automatically. It is important to note that not all countries have data continuously available across all years. 

\begin{figure}[ht]
    \centering
    \includegraphics[width=1\linewidth]{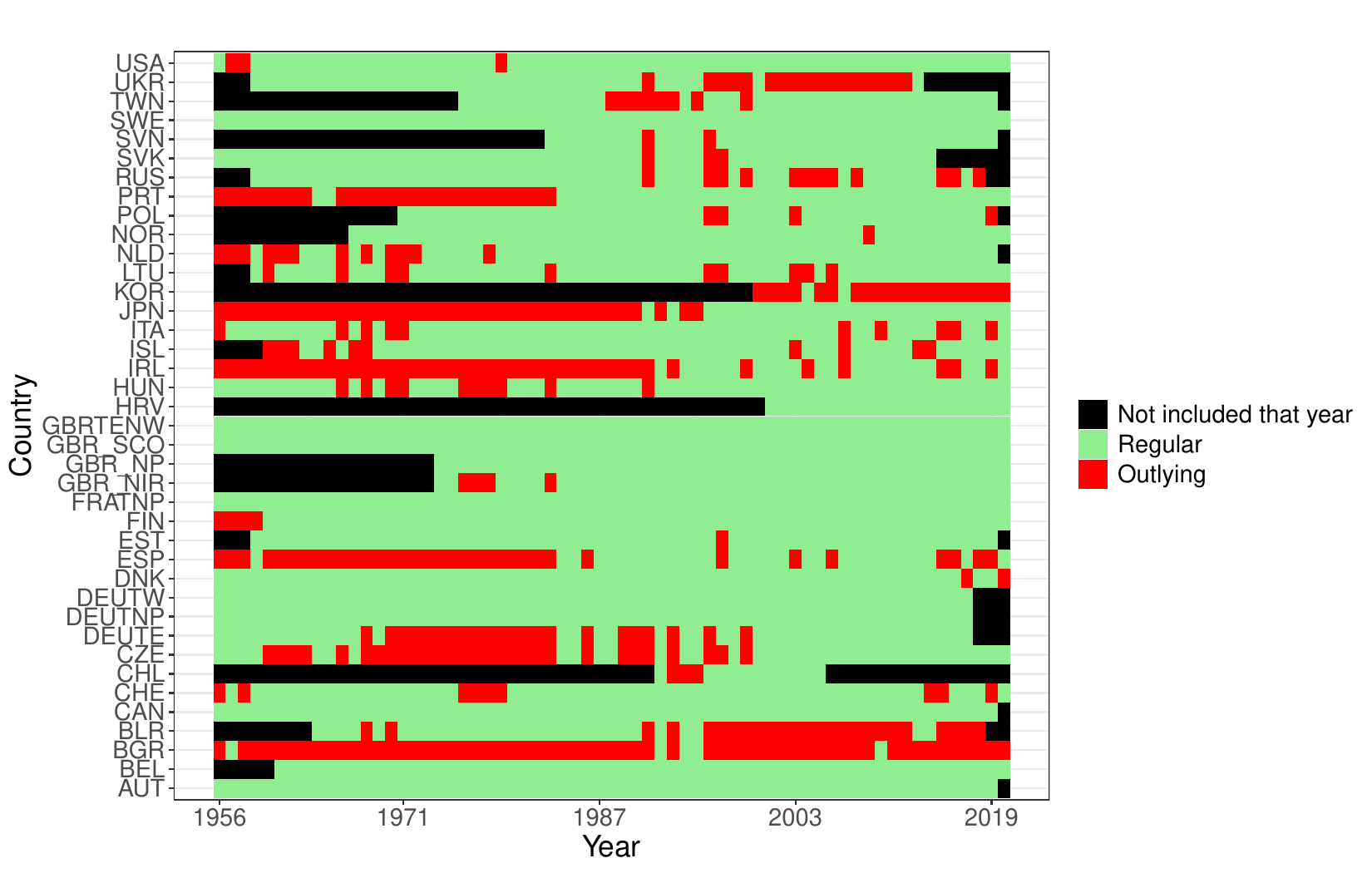}
    \caption{Yearly evolution of anomaly detection within fertility curves.
    }
    \label{fig:fertilityOutlierEvolution}
\end{figure}

In Figure \ref{fig:fertilityOutlierEvolution}, we summarize the yearly evolution of the outlying behavior between $1956$ and $2020$ over all countries. In the first half of the observed time period, we primarily get persistent outliers like Spain (ESP) and Portugal (PRT), or the Czech Republic (CZE) and Japan (JPN). For some of these countries, the reason why they behave outlying is their outstandingly late (first-) pregnancies. Amongst others, this was caused by economic uncertainties, large families, and children tending to stay longer at home before moving out. As opposed to this, rather early parenthood, influenced by the Soviet Union, was present in the Czech Republic. During the more recent years, countries like Russia (RUS), Belarus (BLR), and Ukraine (UKR) exhibit an outlying behavior. As discussed above, overall pregnancies shifted from younger to older ages due to several reasons. Within Eastern Europe, especially for the latter countries, this trend was delayed and is partially still ongoing. 

\begin{figure}[ht]
    \centering
    \includegraphics[width=.9\linewidth]{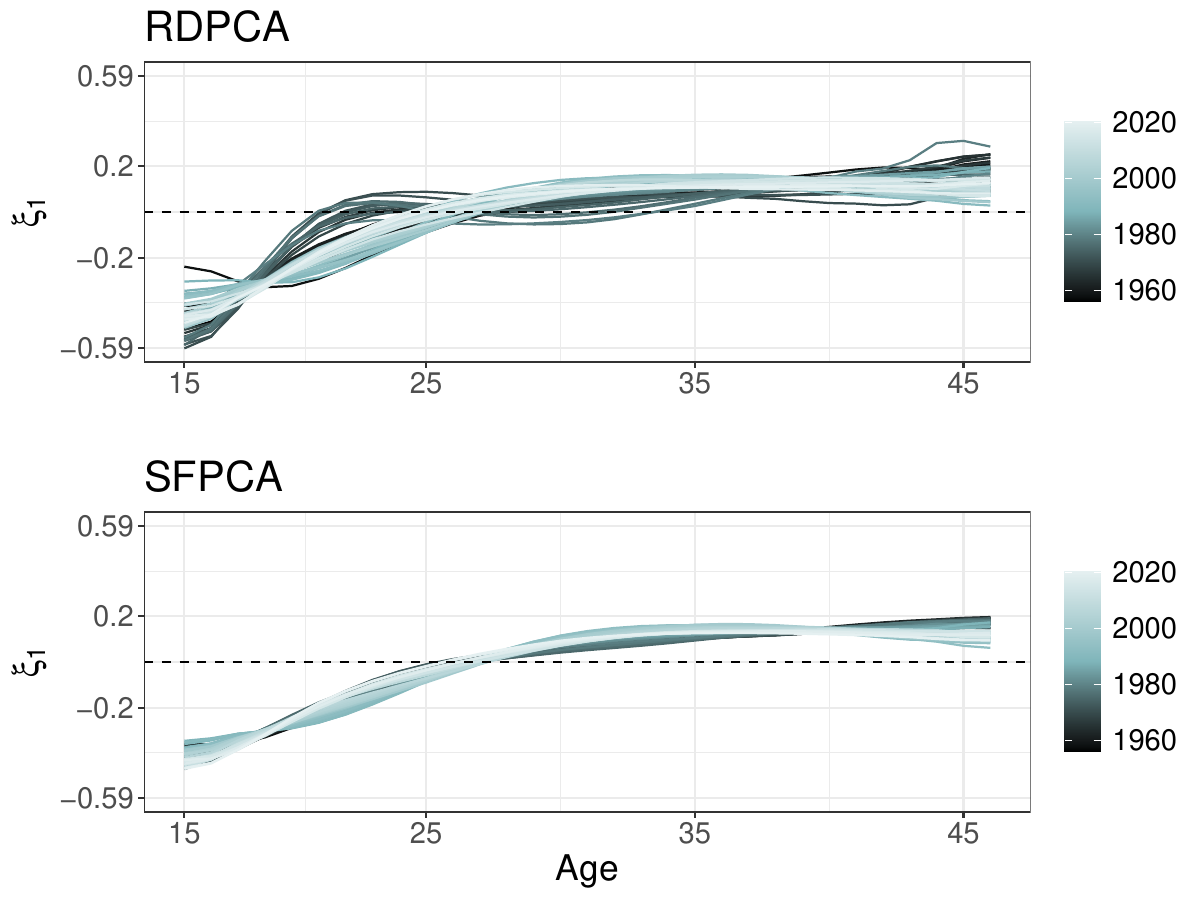}
    \caption{First principal function for fertility curves over the years 1956 until 2020.
    }
    \label{fig:fertilityPrincipals}
\end{figure}

As it was already shown in Section \ref{sec:RDPCA_fertility}, the first principal function mainly captures information contained in the left tail. A closer look at the complete period in Figure \ref{fig:fertilityPrincipals} reveals an interesting dynamic around the 90s. During that time, high fertility in Western European countries shifted earlier compared to the East. This is displayed by the shape of the principal functions that exhibit two extrema, representing the East and the West, respectively. On the other hand, the simplicial principal components fail to capture this trend at all. There is no apparent change over the consulted period.

\clearpage

% ----------------------------------------------------------
% Proofs  --------------------------------------------------
% ----------------------------------------------------------
\section{Proofs} \label{supp:proofs}

\begin{proof}[Proof of Lemma \ref{lemma:bayes_mean_cov}]
\begin{itemize}
    \item[i)] Observe first that the Bayes mean $\mu_{B^2,X}$ is connected to the “standard" mean of $\clr(X)$ through:
\begin{align}\label{eq:bayes_mean_2}
\langle \clr(\mu_{B^2,X}), \clr(f)\rangle&=\langle \mu_{B^2,X}, f\rangle_{B^2}=\mathbb{E}(\langle X,f\rangle_{B^2})\nonumber\\
&=\mathbb{E}(\langle \clr(X),\clr(f)\rangle)=\langle\mathbb{E}(\clr(X)),\clr(f)\rangle,
\end{align}
where the last equality holds since $\mathbb{E}(\clr(X))$ is the expectation of $\clr(X)$. Moreover, for a constant function $g\equiv c\in\mathbb{R}$
$$
\langle \clr(\mu_{B^2,X}), g\rangle = 0,\quad \langle\mathbb{E}(\clr(X)),g\rangle = \mathbb{E}(\langle \clr(X),g\rangle)=0.
$$
Additionally, \eqref{eq:bayes_mean_2} shows that $\clr(\mu_{B^2,X})$ and $\mathbb{E}(\clr(X))$ coincide on projections onto every $f_0\in L_0(I)$. As every function $f\in\ L^2(I)$ can be written as $f=f_0+c_f$, where $c_f=\int_I f(t)\mathrm{d}t$ and $f_0\in L_0(I)$, we deduce that  $\clr(\mu_{B^2,X})$ and $\mathbb{E}(\clr(X))$ coincide on projections onto every $f\in L(I)$, thus proving the first claim.
\item[ii)] 
The following identity connecting the Bayes covariance of $X$ and standard, $L^2$ covariance of $\clr(X)$ holds:
For $f,g\in\mathcal{B}^2(I)$
\begin{align}
\langle \clr(C_{B^2,X}f),\clr(g)\rangle&=
\langle C_{B^2,X}f,g\rangle_{B^2}=\mathbb{E}\left(\langle X\ominus\mu_{B^2,X},f\rangle_{B^2}\langle X\ominus\mu_{B^2,X},g\rangle_{B^2} \right)\nonumber\\
&=\mathbb{E}\left(\langle \clr(X)-\mathbb{E}(X),\clr(f)\rangle\langle \clr(X)-\mathbb{E}(X),\clr(g)\rangle \right)\nonumber\\
&=\langle C_{\clr(X)}\clr(f),\clr(g)\rangle\nonumber,
\end{align}
thus proving that $\clr(C_{B^2,X} f)$ and $C_{\clr(X)}\clr(f)$ coincide on projections in $L_0$. Using the same argument as for the expectation, we obtain that for every $f\in\mathcal{B}^2(I)$
$$
\clr(C_{B^2,X} f)=_{a.e.}C_{\clr(X)}\clr(f).
$$
\item[iii)] This statement is a direct consequence of (ii) and the definition of Bayes space inner product.
\end{itemize}
\end{proof}

\begin{proof}[Proof of Proposition \ref{prop:L_2_B2_equivalence}]
Using the definition of Bayes-space norm and $\clr$ transformation, we first obtain that
\begin{align}\label{eq:prop1.1}
  \clr({C}_{B^2,X}^{1/2}Y)&=\clr\left(\bigoplus_{i=1}^\infty (\lambda_i^{1/2} \langle \zeta_i,Y \rangle_{B^2}) \odot \zeta_i \right)=\sum_{i=1}^\infty \lambda_i^{1/2} \langle \zeta_i,Y \rangle_{B^2} \clr\left(\zeta_i \right)\nonumber\\
  &=\sum_{i=1}^\infty \lambda_i^{1/2} \langle \clr(\zeta_i),\clr(Y) \rangle_{L^2} \clr\left(\zeta_i \right)
\end{align}
for $(\lambda_i,\zeta_i),\, i \geq 1$ being the $i$th eigenpair of ${C}_{B^2,X}$. Additionally, Lemma \ref{lemma:bayes_mean_cov} gives that 
\begin{align}\label{eq:prop1.2}
  \sum_{i=1}^\infty \lambda_i^{1/2} \langle \clr(\zeta_i),\clr(Y) \rangle_{L^2} \clr\left(\zeta_i \right)=C_{\clr,X}^{1/2}\clr(Y).
\end{align}
\eqref{eq:prop1.1} and \eqref{eq:prop1.2}, together with the definition of the Bayes-space norm, finally give that 
\begin{equation}\label{eq:prop1.3}
\argmin_{Y\in \mathcal{B}^2(I)} \left\{\|{C}_{B^2,X}^{1/2}Y\ominus X\|_{B^2}^2 + \alpha\|Y\|_{B^2}^2  \right\}=\argmin_{Y\in \mathcal{B}^2(I)} \left\{\|C_{\clr,X}^{1/2}\clr(Y)-\clr(X)\|^2+\alpha\|\clr(Y)\|^2   \right\}.    
\end{equation}
Let now 
\begin{equation}\label{eq:prop1.4}
 X_{\mathrm{st}}^{{\alpha},{I}} =_B
    \argmin_{Y\in \mathcal{B}^2(I)} \left\{\|{C}_{B^2,X}^{1/2}Y\ominus X\|_{\mathcal{B}}^2 + \alpha\|Y\|_{\mathcal{B}}^2   \right\}.
\end{equation}
Denoting $ L_0^2(I)\subset L^2(I)$ to be the set of all $\clr$-transformed elements of $\mathcal{B}^2(I)$, i.e. the set of $L^2$-functions with zero internal,  \eqref{eq:prop1.3} then allows us to reparameterize the optimization problem \eqref{eq:prop1.4} as
$$
X_{\mathrm{st}}^{{\alpha},{I}} =_B \clr^{-1}(Y),\quad Y=
    \argmin_{Z\in  L_0^2(I)} \left\{\|{C}_{\clr,X}^{1/2}Z - \clr(X)\|^2 + \alpha\|Z\|^2   \right\}.
$$ 
We now need to show that 
$$
\argmin_{Y\in  L_0^2(I)} \left\{\|{C}_{\clr,X}^{1/2}Y - \clr(X)\|^2 + \alpha\|Y\|^2   \right\}=\argmin_{Y\in L^2(I)} \left\{\|{C}_{\clr,X}^{1/2}Y - \clr(X)\|^2 + \alpha\|Y\|^2 \right\}.
$$
We know that 
$$
\argmin_{Y\in L^2(I)} \left\{\|{C}_{\clr,X}^{1/2}Y - \clr(X)\|^2 + \alpha\|Y\|^2 \right\}=C_{\clr,X}^{1/2}(C_{\clr,X}+\alpha I )\clr(X) =: Y_{L^2};
$$
see e.g. \cite{berrendero2020} for more details. Linearity of the integral implies that $Y_{L^2}\in L_0^2(I)$, thus completing the first part of the proof. The other implication is shown analogously.  
\end{proof}

\begin{proof}[Proof of Corollary \ref{cor:L_2 standardization explicit form}]
    Proposition \ref{prop:L_2_B2_equivalence} implies that it is enough to show that 
    $$\clr\left( \bigoplus_{i=1}^\infty \frac{\lambda_i^{1/2}}{\lambda_i + \alpha} \odot \langle \zeta_i,X \rangle_{B^2} \odot \zeta_i \right)$$ solves the equivalent optimization problem in $L^2$. Properties of Bayes space norm and  operations give that 
    \begin{align*}
\clr\left( \bigoplus_{i=1}^\infty \frac{\lambda_i^{1/2}}{\lambda_i + \alpha} \odot \langle \zeta_i,X \rangle_{B^2} \odot \zeta_i \right) &= \sum_{i=1}^\infty\frac{\lambda_i^{1/2}}{\lambda_i+\alpha}\langle\clr(\zeta_i),\clr(X)\rangle\clr(\zeta_i)\\
&=C_{\clr,X}^{1/2}(C_{\clr,X}+\alpha I)^{-1}\clr(X),        
    \end{align*}
    which is an optimizer of the equivalent optimization problem in $L^2$; see the final part of the proof of Proposition \ref{prop:L_2_B2_equivalence} for more insight.
\end{proof}

\begin{proof}[Proof of Proposition \ref{prop:prop2}]
    As $\tilde{X} \in \mathcal{N}(L)$, the minimization problem in the statement of the proposition can be rewritten as 
    \begin{align*}
    \argmin_{Y\in B^2(I)}\left\{\|C_{B^2,X}^{1/2}(Y\ominus\tilde{X})\ominus X\|_{B^2}^2+\alpha\|L (Y-\tilde{X})\|_{B^2}^2 \right\},
\end{align*}
admitting a unique solution in the form $Y\ominus\tilde{X}=X_\mathrm{st}^{\boldsymbol{\alpha},{L}}$.
\end{proof}

\begin{proof}[Proof of Proposition \ref{prop:L2_B2_equivalence_projector_opertor}]
The equivalence between the two optimization problems is shown as in the proof of Proposition \eqref{prop:L_2_B2_equivalence}. We next show that the optimization problem in a Bayes space has an explicit solution. Denote first $W(-k)$ to be the orthogonal projector onto the span $\mathrm{span}(\zeta_1,\dots,\zeta_k)$ of the first $k$ eigenfunctions of $X$. Then, for every $Y\in\mathcal{B}^2(I)$, we can write $Y=W(-k)Y\oplus W(k)Y$. This further allows us to rewrite
\begin{align*}
    X_{\mathrm{st};C_{B^2,X}}^{{\alpha},k} =_B
    \argmin_{Y\in \mathcal{B}^2(I)} &\left\{\|\left({C}_{B^2,X}^{1/2}W(k)Y\ominus W(k)X\right)\oplus \left({C}_{B^2,X}^{1/2}W(-k)Y\ominus W(-k)X\right)\|_{\mathcal{B}}^2\right.\\ 
    &+ \left.\alpha\|W(k)Y\|_{\mathcal{B}}^2   \right\}.
\end{align*}
As $W(k){C}_{B^2,X}^{1/2}W(k)Y={C}_{B^2,X}^{1/2}W(k)Y$ and $W(-k){C}_{B^2,X}^{1/2}W(-k)Y={C}_{B^2,X}^{1/2}W(-k)Y$ it follows that ${C}_{B^2,X}^{1/2}W(k)Y\ominus W(k)X$ and ${C}_{B^2,X}^{1/2}W(-k)Y\ominus W(-k)X$ are orthogonal in $\mathcal{B}^2(I)$. Thus,  
\begin{align*}
    X_{\mathrm{st};C_{B^2,X}}^{{\alpha},k} =_B
    \argmin_{Y\in \mathcal{B}^2(I)} &\left\{\|{C}_{B^2,X}^{1/2}W(k)Y\ominus W(k)X\|_{\mathcal{B}}^2+ \alpha\|W(k)Y\|_{\mathcal{B}}^2   \right.\\
    &+ \left.\|{C}_{B^2,X}^{1/2}W(-k)Y\ominus W(-k)X\|_{\mathcal{B}}^2\}\right\}. 
\end{align*}
Observe now two optimization problems independently
\begin{align}
    \min_{Y_1\in \mathcal{B}^2(I)} &\left\{\|{C}_{B^2,X}^{1/2}W(k)Y_1\ominus W(k)X\|_{\mathcal{B}}^2+ \alpha\|Y_1\|_{\mathcal{B}}^2   \right\}\label{eq_two_optimzations1}\\
    \min_{Y_2\in \mathcal{B}^2(I)} &\left\{\|{C}_{B^2,X}^{1/2}W(-k)Y_2\ominus W(-k)X\|_{\mathcal{B}}^2  \right\}\label{eq_two_optimzations2}.
\end{align}
Regularization in \eqref{eq_two_optimzations1} is said to be in a standard form and admits a closed-form solution 
\begin{align*}
Y_1&=_{\mathcal{B}^2} {C}_{B^2,X}^{1/2}W(k)\left({C}_{B^2,X}^{1/2}W(k){C}_{B^2,X}^{1/2} +\alpha I\right)^{-1}W(k)X\\
&=\sum_{i=k+1}^\infty \frac{\lambda_i^{1/2}}{\lambda_i+\alpha}\langle \zeta_i,X \rangle_{B^2}\zeta_i\in \mathrm{span}(\zeta_{k+1},\zeta_{k+2}\dots),
\end{align*}
see  the proof of \cite[Proposition 2]{berrendero2020} for more details. Additionally, since $C^{1/2}_{B^2,X}W(-k)$ is invertible, \eqref{eq_two_optimzations1} has a solution 
$$
Y_2=\sum_{i=1}^k\lambda_i^{-1/2}\langle \zeta_i,X \rangle_{B^2}\zeta_i\in\mathrm{span}(\zeta_{1},\dots,\zeta_{k}).
$$
It is then straightforward to verify that $Y=Y_1\oplus Y_2$ is the solution to the original optimization problem. Similarly, one can show that 
$$
\sum_{i=1}^k\lambda_i^{-1/2}\langle \clr(X),\xi_i \rangle\xi_i+\sum_{i=k+1}^\infty \frac{\lambda_i^{1/2}}{\lambda_i+\alpha}\langle \clr(X),\xi_i \rangle\xi_i
$$
is a solution to the corresponding optimization problem in the clr space. From here, it is again straightforward to verify the equivalence of the corresponding optimization problems. 
\end{proof}

\begin{proof}[Proof of Proposition \ref{prop:Mah_distribution}]
For simplicity, we give the proof for $k\geq 1$. The other case is shown using the same argumentation. As in the proof of Proposition \ref{prop:L2_B2_equivalence_projector_opertor} and using the definition of Bayes-space inner product, we obtain that 
$$
M_{\alpha,k,B^2}^2(X)=\sum_{i=1}^k\frac{1}{\lambda_i}z_i^2+\sum_{i=k+1}^\infty\frac{\lambda_i}{(\lambda_i+\alpha)^2}z_i^2,
$$
where $z_i=\langle\clr(X),\xi_i\rangle$, $i\geq 1$. Gaussianity of $\clr(X)$ gives that the scores $z_i\sim \mathcal{N}(0,\lambda_i)$, $i\geq 1$ are mutually independent. Substituting $\eta_i=\lambda_i^{-1/2}z_i$, $i\geq1$ yields the desired claim. \\
\end{proof}

\begin{proof}[Proof of Proposition \ref{prop:connection_to_other_Mah_dist}]
    Claim (i) follows directly from the \cite[Proposition 2]{berrendero2020}. Using the same reasoning
    $$
M_{\alpha,k,B^2}^2(X)=\sum_{i=1}^k\frac{\langle\clr(X),\xi_i\rangle^2}{\lambda_i}+\sum_{i=k+1}^{\infty}\frac{\lambda_i}{(\lambda_i+\alpha)^2}\langle\clr(X),\xi_i\rangle^2.
    $$
   As $0\leq \langle\clr(X),\xi_i\rangle^2\leq \|\clr(X)\|^2$, we have that the second part in the upper sum is bounded by
   $$
0\leq \sum_{i=k+1}^{\infty}\frac{\lambda_i}{(\lambda_i+\alpha)^2}\langle\clr(X),\xi_i\rangle^2\leq  \|\clr(X)\|^2\sum_{i=k+1}^{\infty}\frac{\lambda_i}{(\lambda_i+\alpha)^2},
$$
where the upper sequence converges uniformly for all $\alpha>0$. Denoting $f(i,\alpha)=\frac{\lambda_i}{(\lambda_i+\alpha)^2}$, \citet[Theorem 1]{Hildebrandt1912} implies that the limit and sum commute. Additionally, as for every $i$ $\displaystyle\lim_{\alpha\to\infty}f(i,\alpha)=0$ gives 
$$
0\leq \lim_{\alpha\to \infty}\sum_{i=k+1}^{\infty}\frac{\lambda_i}{(\lambda_i+\alpha)^2}\langle\clr(X),\xi_i\rangle^2\leq 0,
$$
thus completing the proof of (ii).
\end{proof}

\begin{proof}[Proof of Corollary \ref{cor:cor_1}]
We start this proof by showing $ \| X_{\mathrm{st},\hat{C}_{B^2,X}}^{\alpha,k} \|_{B^2} \rightarrow_\mathrm{a.s.} \|  X_{\mathrm{st},C_{B^2,X}}^{\alpha,k} \|_{B^2}$. First, note that Proposition \ref{prop:L2_B2_equivalence_projector_opertor} further implies an explicit representation for the solution of \eqref{eq:standardization_B^2_final} based on the underlying operators. From this, it follows that 
\begin{align*} 
    \| X_{\mathrm{st},\hat{C}_{B^2,X}}^{\alpha,k} \|_{B^2}  -   \|  X_{\mathrm{st};C_{B^2,X}}^{\alpha,k} \|_{B^2}  
    \leq&~  \| X_{\mathrm{st},\hat{C}_{B^2,X}}^{\alpha,k}   \ominus X_{\mathrm{st};C_{B^2,X}}^{\alpha,k}  \|_{B^2} \\
    =&~ \|  \hat{C}_{B^2,X}^{1/2}(\hat{C}_{B^2,X} \oplus \alpha \hat{W}(k) )^{-1}(X \ominus \hat{\mu}_{B^2,X} ) \\
    &~ \ominus C^{1/2}_{B^2,X}(C_{B^2,X} \oplus \alpha W(k))^{-1}X  \|_{B^2}\\ 
    \leq &~  \|  \left( \hat{C}_{B^2,X}^{1/2}(\hat{C}_{B^2,X} \oplus \alpha \hat{W}(k) )^{-1} \ominus C^{1/2}_{B^2,X}(C_{B^2,X} \oplus \alpha W(k))^{-1}\right) X \|_{B^2} \\
    &~ + \| \hat{C}_{B^2,X}^{1/2}(\hat{C}_{B^2,X} \oplus \alpha \hat{W}(k) )^{-1} \hat{\mu}_{B^2,X}\|_{B^2}.
\end{align*}
The next part of the proof establishes consistency of both terms in the inequality above, i.e.
\begin{itemize}
        \item[(i)] $\|  \left( \hat{C}_{B^2,H}^{1/2}(\hat{C}_{B^2,H} \oplus \alpha \hat{W}(k) )^{-1} \ominus C^{1/2}_{B^2,X}(C_{B^2,X} \oplus \alpha W(k))^{-1}\right) X \|_{B^2} \rightarrow_\mathrm{a.s.} 0$,
        \item[(ii)] $\| \hat{C}_{B^2,H}^{1/2}(\hat{C}_{B^2,H} \oplus \alpha \hat{W}(k) )^{-1} \hat{\mu}_{B^2,H}\|_{B^2} \rightarrow_\mathrm{a.s.} 0$.
    \end{itemize}
By focusing on term (i) and introducing the operator norm in the Bayes space $\| \cdot \|_{\mathrm{op},B^2}$ we can deduce
\begin{align*}
&~ \|  \left( \hat{C}_{B^2,}^{1/2}(\hat{C}_{B^2,X} \oplus \alpha \hat{W}(k) )^{-1} \ominus C^{1/2}_{B^2,X}(C_{B^2,X} \oplus \alpha W(k))^{-1}\right) X \|_{B^2} \\
\leq&~ \|  \left( \hat{C}_{B^2,X}^{1/2}(\hat{C}_{B^2,X} \oplus \alpha \hat{W}(k) )^{-1} \ominus C^{1/2}_{B^2,X}(C_{B^2,X} \oplus \alpha W(k))^{-1}\right)\|_{\mathrm{op},B^2} \| X \|_{B^2}.
\end{align*}
Here, for the operator norm above observe that
\begin{align*}
    &~ \|  \left( \hat{C}_{B^2,X}^{1/2}(\hat{C}_{B^2,X} \oplus \alpha \hat{W}(k) )^{-1} \ominus C^{1/2}_{B^2,X}(C_{B^2,X} \oplus \alpha W(k))^{-1}\right)\|_{\mathrm{op},B^2} \\
    = &~ \|  \left( \hat{C}_{B^2,X}^{1/2}(\hat{C}_{B^2,X} \oplus \alpha \hat{W}(k) )^{-1} \ominus \hat{C}^{1/2}_{B^2,X}(C_{B^2,X} \oplus \alpha W(k))^{-1} \right) \\
     &~  \oplus \left( \hat{C}^{1/2}_{B^2,X}(C_{B^2,X}\oplus \alpha W(k))^{-1} \ominus C^{1/2}_{B^2,X}(C_{B^2,X} \oplus \alpha W(k))^{-1}\right)\|_{\mathrm{op},B^2} \\
     \leq &~ \| \hat{C}^{1/2}_{B^2,X} \|_{\mathrm{op},B^2} \|(\hat{C}_{B^2,X} \oplus \alpha \hat{W}(k) )^{-1}\ominus(C_{B^2,X} \oplus \alpha W(k))^{-1} \|_{\mathrm{op},B^2} \\
     &~ + \|\hat{C}^{1/2}_{B^2,X} \ominus  C^{1/2}_{B^2,X}\|_{\mathrm{op},B^2}  \|(C_{B^2,X}\oplus \alpha W(k))^{-1} \|_{\mathrm{op},B^2}  
\end{align*}
First, since $\hat{C}_{B^2,X} \rightarrow_\mathrm{a.s.} C_{B^2,X}$, there exists a bound on the operator norm of $\hat{C}_{B^2,X}$. Then, we have that 
\begin{align*}
   \| (\hat{C}_{B^2,X} \oplus \alpha \hat{W}(k) ) \ominus (C_{B^2,X} \oplus \alpha W(k))\|_{\mathrm{op},B^2} =&~ \| \left(\hat{C}_{B^2,X} \ominus C_{B^2,X} \right) \oplus \alpha \left( \hat{W}(k) \ominus W(k) \right)\|_{\mathrm{op},B^2} \\
   \leq&~ \| \hat{C}_{B^2,X} \ominus C_{B^2,X} \|_{\mathrm{op},B^2} + \alpha  \|  \hat{W}(k) \ominus W(k)\|_{\mathrm{op},B^2}.
\end{align*}
As the projector onto the finite-dimensional subspace spanned by the first $k$-principal functions is a strongly consistent estimator, provided that $\lambda_k>\lambda_{k+1}$ (see e.g. \cite{hall2006}), $\hat{W}(k)$ is also strongly consistent for $W(k)$. Therefore $\hat{C}_{B^2,X} \oplus \alpha \hat{W}(k) \rightarrow_\mathrm{a.s.} C_{B^2,X} \oplus \alpha W(k)$. Furthermore, we have $\mathrm{sup}_n \| (\hat{C}_{B^2,X} \oplus \alpha \hat{W}(k))^{-1}\|_{\mathrm{op},B^2} \leq \alpha^{-1}$. This fulfills the assumptions of Lemma 8 in \cite{berrendero2020} and we get $ \|(\hat{C}_{B^2,X} \oplus \alpha \hat{W}(k) )^{-1}\ominus(C_{B^2,X} \oplus \alpha W(k))^{-1} \|_{\mathrm{op},B^2}  \rightarrow_\mathrm{a.s.} 0$. Following these derivations, we have
\begin{align}\label{proof:cor3.1_i1}
    \| \hat{C}^{1/2}_{B^2,X} \|_{\mathrm{op},B^2} \|(\hat{C}_{B^2,X} \oplus \alpha \hat{W}(k) )^{-1}\ominus(C_{B^2,X} \oplus \alpha W(k))^{-1} \|_{\mathrm{op},B^2} \rightarrow_\mathrm{a.s.} 0.
\end{align}
Second, because of the consistency of $\hat{C}^{1/2}_{B^2,X}$ and the continuity of the square root we also have $\|\hat{C}^{1/2}_{B^2,X} \ominus  C^{1/2}_{B^2,X}\|_{\mathrm{op},B^2} \rightarrow_\mathrm{a.s.} 0$. Since for every $Y \in \mathcal{B}^2(I): \|(C_{B^2,X}\oplus \alpha W(k))^{-1} \| \leq \alpha^{-1}$, 
\begin{align}\label{proof:cor3.1_i2}
     \|\hat{C}^{1/2}_{B^2,X} \ominus  C^{1/2}_{B^2,X}\|_{\mathrm{op},B^2}  \|(C_{B^2,X}\oplus \alpha W(k))^{-1} \|_{\mathrm{op},B^2} \rightarrow_\mathrm{a.s.} 0.
\end{align}
Finally, \eqref{proof:cor3.1_i1} and \eqref{proof:cor3.1_i2} proof claim (i). For claim $(ii)$ we can establish
\begin{align*}
    \| \hat{C}_{B^2,X}^{1/2}(\hat{C}_{B^2,X} \oplus \alpha \hat{W}(k) )^{-1} \hat{\mu}_{B^2,X}\|_{B^2} \leq \| \hat{C}_{B^2,X}^{1/2}\|_{\mathrm{op},B^2} \|(\hat{C}_{B^2,X} \oplus \alpha \hat{W}(k) )^{-1}\|_{\mathrm{op},B^2} \| \hat{\mu}_{B^2,X}\|_{B^2}.
\end{align*}
In a similar fashion as before, we can determine that both operator norms above are bounded. Because of the assumption of this corollary we further have $\|\hat{\mu}_{B^2,X} \|_{B^2} \rightarrow_\mathrm{a.s.} 0$ which concludes to
\begin{align*}
    \| \hat{C}_{B^2,X}^{1/2}(\hat{C}_{B^2,X} \oplus \alpha \hat{W}(k) )^{-1} \hat{\mu}_{B^2,X}\|_{B^2} \rightarrow_\mathrm{a.s.} 0.
\end{align*}
The proofs of (i) and (ii) now show that the initial claim $ \| X_{\mathrm{st},\hat{C}_{B^2,X}}^{\alpha,k} \|_{B^2} \rightarrow_\mathrm{a.s.} \|  X_{\mathrm{st};C_{B^2,X}}^{\alpha,k} \|_{B^2}$ holds. The remaining proof of this corollary now directly follows from Proposition \eqref{prop:Mah_distribution}. 
\end{proof}

\begin{proof}[Proof of  
Proposition~\ref{prop:Mah_distance_in_a_basis}]
Denote $Y_i=\clr(X_i)=\sum_{i=1}^MC_{i,j}\tilde{\phi}_i(t)$, where $\tilde{\phi}_i=\clr(\phi_i)\in L_0^2(I)$, $i=1,\dots,n$. Let further $\hat C_{H,\clr,X}$ be the sample covariance of the transformed sample, calculated at subsample $H$, and let $(\hat{\lambda}_i,\hat{\xi}_i)$, $i=1,\dots,M$, be the corresponding eigenpairs, where for simplicity of the notation, we will drop the subscript $H$.  

Since by assumption $Y_i\in\mathrm{span}(\tilde{\phi}_1,\dots,\tilde{\phi}_M)$, $i=1,\dots,n$, then $\bar{Y}_{H}\in\mathrm{span}(\tilde{\phi}_1,\dots,\tilde{\phi}_M)$. Similarly, $\hat C_{H,\clr,X}:L_2(I)\to  \mathrm{span}(\phi_1,\dots,\phi_M)$, further giving that $\hat{\xi}_i=\textbf{u}_i' \boldsymbol{\tilde\Phi}$, for some vector $\textbf{u}_i$, $i=1,\dots,M$; remember that $\hat{\lambda}_i\hat{\xi}_i=\hat C_{H,\clr,X}\hat{\xi}_i$, where we denote $ \boldsymbol{\tilde\Phi}=(\tilde{\phi}_1,\dots,\tilde{\phi}_M)'$. Furthermore, since $\{{\phi}_1,\dots,{\phi}_M\}$ is orthonormal in $\mathcal{B}^2(I)$, then $\{\tilde{\phi}_1,\dots,\tilde{\phi}_M\}$ is orthonormal in $L^2(I)$, further giving that  $\textbf{u}_1,\dots,\textbf{u}_M$ are orthogonal vectors of unit norm. For $j=1,\dots,M$, the the following statements are equivalent: 
\begin{align*}
\hat C_{H,\clr,X}\hat{\xi}_j=\hat{\lambda}_j\hat{\xi}_j &\iff \frac{1}{h}\sum_{i\in H}\langle Y_i-\bar{Y}_{H},\hat{\xi}_j\rangle ( Y_i-\bar{Y}_{H})=\hat{\lambda}_j\hat{\xi}_j\\
&\iff \frac{1}{h}\sum_{i\in H}\langle (\textbf{e}_i-\frac{1}{h}\mathbf{1}_{H_0})'\textbf{C}\tilde{\boldsymbol{\Phi}}, \tilde{\boldsymbol{\Phi}}'\textbf{u}_j\rangle(\textbf{e}_i-\frac{1}{h}\mathbf{1}_{H_0})'\textbf{C}\tilde{\boldsymbol{\Phi}}=\hat{\lambda}_j\textbf{u}_j'\tilde{\boldsymbol{\Phi}}\\
&\iff \tilde{\boldsymbol{\Phi}}'\textbf{C}'\frac{1}{h}\sum_{i\in H}(\textbf{e}_i-\frac{1}{h}\mathbf{1}_{H})(\textbf{e}_i-\frac{1}{h}\mathbf{1}_{H})'\textbf{C}\textbf{u}_j=\hat{\lambda}_j\tilde{\boldsymbol{\Phi}}'\textbf{u}_j\\
&\iff \tilde{\boldsymbol{\Phi}}'\textbf{C}'\frac{1}{h}(\textbf{I}_{n,H}-\frac{1}{h}\textbf{J}_{n,H})'\textbf{C}\textbf{u}_j=\hat{\lambda}_j\tilde{\boldsymbol{\Phi}}'\textbf{u}_j\\
&\iff \textbf{C}'\frac{1}{h}(\textbf{I}_{n,H}-\frac{1}{h}\textbf{J}_{n,H})'\textbf{C}\textbf{u}_j=\hat{\lambda}_j\textbf{u}_j,
\end{align*}
where the final equivalence holds as $\tilde{\boldsymbol{\Phi}}(t)'\textbf{C}'\frac{1}{h}(\textbf{I}_{n,H}-\frac{1}{h}\textbf{J}_{n,H})'\textbf{C}\textbf{u}_j=\hat{\lambda}_j\tilde{\boldsymbol{\Phi}}(t)'\textbf{u}_j$, for every $t\in I$. This implies that for $j=1,\dots,M$, $(\hat{\lambda}_j,\textbf{u}_j)$ is an eigenpair of the covariance $\textbf{C}_{H}$ of $\{C_{i,j};i\in H\}$, additionally giving a computationally elegant way of calculating the robust SFPCs. 

Observe first that  $(\textbf{I}_{n,H}-\frac{1}{h}\textbf{J}_{n,H})$ is a symmetric idempotent matrix, further making $\textbf{C}_{H}$ a symmetric, positive semi-definite matrix. The following equalities now hold:
\begin{align*}
    M_{\alpha,k,B^2}(X_i,\hat{\mu}_{B^2,H};\hat{C}_{B^2,H,c_H})
    &=\sum_{j=1}^k\frac{1}{\hat{\lambda}_j}\langle\hat{\xi}_j,Y_i-\bar{Y}_{H}\rangle^2+\sum_{j=k+1}^M\frac{\hat{\lambda}_j}{(\hat{\lambda}_j+\alpha)^2}\langle\hat{\xi}_j,Y_i-\bar{Y}_{H}\rangle^2\\
    &=\sum_{j=1}^k\left(\frac{1}{\hat{\lambda}_j}\mathbb{I}_{j\leq k}+\mathbb{I}_{j> k}\frac{\hat{\lambda}_j}{(\hat{\lambda}_j+\alpha)^2}\right)\langle\textbf{u}_j'\tilde{\boldsymbol{\Phi}},\tilde{\boldsymbol{\Phi}}'\textbf{C}'(\textbf{e}_i-\frac{1}{h}\mathbf{1}_{H})\rangle^2\\
    &=\sum_{j=1}^M\left(\frac{1}{\hat{\lambda}_j}\mathbb{I}_{j\leq k}+\mathbb{I}_{j> k}\frac{\hat{\lambda}_j}{(\hat{\lambda}_j+\alpha)^2}\right)\left((\textbf{e}_i-\frac{1}{h}\mathbf{1}_{H})'\textbf{C}\textbf{u}_j \right)^2\\
    &=(\textbf{e}_i-\frac{1}{h}\mathbf{1}_{H})'\textbf{C}\sum_{j=1}^M\left(\frac{1}{\hat{\lambda}_j}\mathbb{I}_{j\leq k}+\mathbb{I}_{j> k}\frac{\hat{\lambda}_j}{(\hat{\lambda}_j+\alpha)^2}\right)\textbf{u}_j\textbf{u}_j'\textbf{C}'(\textbf{e}_i-\frac{1}{h}\mathbf{1}_{H})\\
    &=(\textbf{e}_i-\frac{1}{h}\mathbf{1}_{H})'\textbf{C}\textbf{C}_{H}(\textbf{C}_{H}+\alpha\textbf{V}(k))^{-2}\textbf{C}'(\textbf{e}_i-\frac{1}{h}\mathbf{1}_{H}).
\end{align*}
\end{proof}

\begin{proof}[Proof of Lemma \ref{sup_lemma:consistency}]
For the simplicity of the notation, we derive the proof in the clr space, where we abuse notation and write $X_i$ to be the clr transformation of the $i$th observation, $i=1,\dots,N$.  We begin deriving a relationship between overall sample covariance function $\hat{\gamma}(s,t) = \frac{1}{N}\sum_{i=1}^{N} \left(X_i(s) - \hat{\mu}(s)\right)\left(X_i(t) - \hat{\mu}(t)\right)$ and the trimmed covariance function $\hat{\gamma}_{\mathrm{clr},H,1}(s,t) = \frac{1}{h} \sum_{i=1}^h (X_i(s) - \hat{\mu}_{\mathrm{clr},H}(s))(X_i(t) - \hat{\mu}_{\mathrm{clr},H}(t))$, where wlog for the simplicity of the notation, we assume that the $H$-subset of observations used to robustly estimate covariance consists of the initial $h$ observations, $X_1,\dots,X_h$. We further define the corresponding  sample means $\hat{\mu}(s) = \frac{1}{N} \sum_{i=1}^{N} X_i(s),~ \hat{\mu}_{\mathrm{clr},H}(s) = \frac{1}{h} \sum_{i=1}^{h} X_i(s)$. Let further $\tilde{\gamma}_1(s,t) = \frac{1}{N-h} \sum_{i=h+1}^{N} \left(X_i(s) - \tilde{\mu}(s)\right)\left(X_i(t) - \tilde{\mu}(t)\right)$ with $\tilde{\mu}(s) = \frac{1}{N-h}\sum_{i=h+1}^{N} X_i(s)$. Then,
    \begin{align*}
        \hat{\gamma}(s,t) =&~ \frac{1}{N} \sum_{i=1}^{N} \left(X_i(s) - \hat{\mu}(s)\right)\left(X_i(t) - \hat{\mu}(t)\right) \\
        =&~ \frac{1}{N} \left(  \sum_{i=1}^{h} \left(X_i(s) - \hat{\mu}(s)\right)\left(X_i(t) - \hat{\mu}(t)\right) +  \sum_{i=h+1}^{N} \left(X_i(s) - \hat{\mu}(s)\right)\left(X_i(t) - \hat{\mu}(t)\right)\right) \\
        =&~ \frac{1}{N} \Bigg(  \sum_{i=1}^{h} \left(X_i(s) - \hat{\mu}_{\mathrm{clr},H}(s) + \hat{\mu}_{\mathrm{clr},H}(s) - \hat{\mu}(s)\right)\left(X_i(t) - \hat{\mu}_{\mathrm{clr},H}(t) + \hat{\mu}_{\mathrm{clr},H}(t)- \hat{\mu}(t)\right) \\ &~~ +  \sum_{i=h+1}^{N} \left(X_i(s) - \tilde{\mu}(s) + \tilde{\mu}(s) - \hat{\mu}(s)\right)\left(X_i(t) - \tilde{\mu}(t) + \tilde{\mu}(t)- \hat{\mu}(t)\right)\Bigg) \\
        =&~ \frac{1}{N} \Bigg(  \sum_{i=1}^{h} \Big( (X_i(s) - \hat{\mu}_{\mathrm{clr},H}(s))(X_i(t) - \hat{\mu}_{\mathrm{clr},H}(t)) + (X_i(s) - \hat{\mu}_{\mathrm{clr},H}(s))(\hat{\mu}_{\mathrm{clr},H}(t) - \hat{\mu}(t)) \\
        &~~~~~~~ + (\hat{\mu}_{\mathrm{clr},H}(s)- \hat{\mu}(s))(X_i(t) - \hat{\mu}_{\mathrm{clr},H}(t)) + (\hat{\mu}_{\mathrm{clr},H}(s)- \hat{\mu}(s))(\hat{\mu}_{\mathrm{clr},H}(t)- \hat{\mu}(t))\Big ) \\ 
        &~~ +  \sum_{i=h+1}^{N}\Big( (X_i(s) - \tilde{\mu}(s))(X_i(t) - \tilde{\mu}(t)) + (X_i(s) - \tilde{\mu}(s))(\tilde{\mu}(t)- \hat{\mu}(t)) \\
        &~~~~~~~ + (\tilde{\mu}(s)- \hat{\mu}(s))(X_i(t) - \tilde{\mu}(t)) + (\tilde{\mu}(s) - \hat{\mu}(s))(\tilde{\mu}(t)- \hat{\mu}(t))\Big)\Bigg ) \\
        =&~ \frac{1}{N}\Bigg( h \hat{\gamma}_{\mathrm{clr},H,1}(s,t) + (h\hat{\mu}_{\mathrm{clr},H}(s) - h\hat{\mu}_{\mathrm{clr},H}(s))(\hat{\mu}_{\mathrm{clr},H}(t) - \hat{\mu}(t)) \\
        &~~~~~~~  + (\hat{\mu}_{\mathrm{clr},H}(s) - \hat{\mu}(s))(h\hat{\mu}_{\mathrm{clr},H}(t) - h\hat{\mu}_{\mathrm{clr},H}(t)) + h(\hat{\mu}_{\mathrm{clr},H}(s)- \hat{\mu}(s))(\hat{\mu}_{\mathrm{clr},H}(t)- \hat{\mu}(t)) \\
        &~~ + (N-h) \tilde{\gamma}(s,t) + ((N-h)\tilde{\mu}(s) - (N-h)\tilde{\mu}(s))(\tilde{\mu}(t)- \hat{\mu}(t)) \\
        &~~~~~~~ + (\tilde{\mu}(s)- \hat{\mu}(s))((N-h)\tilde{\mu}(t) - (N-h)\tilde{\mu}(t)) + (N-h)(\tilde{\mu}(s)- \hat{\mu}(s))(\tilde{\mu}(t)- \hat{\mu}(t))\Bigg) \\
        =&~ \frac{1}{N}\Bigg( h \hat{\gamma}_{\mathrm{clr},H,1}(s,t) + \underbrace{h(\hat{\mu}_{\mathrm{clr},H}(s)- \hat{\mu}(s))(\hat{\mu}_{\mathrm{clr},H}(t)- \hat{\mu}(t))}_\text{=: $I$} \\
        &~~ + (N-h) \tilde{\gamma}(s,t) + \underbrace{(N-h)(\tilde{\mu}(s)- \hat{\mu}(s))(\tilde{\mu}(t)- \hat{\mu}(t))}_\text{=: $II$}\Bigg). \\
    \end{align*}
    Using the relationship between the sub-sample means, we have 
    \begin{equation}\label{eq:means1}
    \hat{\mu}(s) = \frac{1}{N} \sum_{i=1}^{N} X_i(s) = \frac{1}{N}\left(\sum_{i=1}^h X_i(s) + \sum_{i=h+1}^{N} X_i(s)\right) = \frac{1}{N}(h \hat{\mu}_{\mathrm{clr},H}(t) + (N-h) \tilde{\mu}(s)).    
    \end{equation}
    Plugging \eqref{eq:means1} into $I$  results in
    \begin{align*}
        h(\hat{\mu}_{\mathrm{clr},H}(s)- \hat{\mu}(s))(\hat{\mu}_{\mathrm{clr},H}(t)- \hat{\mu}(t)) =&~ h(\hat{\mu}_{\mathrm{clr},H}(s)- \frac{1}{N}(h \hat{\mu}_{\mathrm{clr},H}(s) - (N-h) \tilde{\mu}(s))) \\
        &~ \cdot (\hat{\mu}_{\mathrm{clr},H}(t)- \frac{1}{N}(h \hat{\mu}_{\mathrm{clr},H}(t) - (N-h) \tilde{\mu}(t))) \\
        =&~ h\left(\frac{N-h}{N}\hat{\mu}_{\mathrm{clr},H}(s) - \frac{N-h}{N_1}\tilde{\mu}(s)\right)\\
        &~ \cdot \left(\frac{N-h}{N}\hat{\mu}_{\mathrm{clr},H}(t) - \frac{N-h}{N}\tilde{\mu}(t)\right) \\
        =&~ \frac{h(N-h)^2}{N^2}(\hat{\mu}_{\mathrm{clr},H}(s) - \tilde{\mu}(s))(\hat{\mu}_{\mathrm{clr},H}(t) - \tilde{\mu}(t)). 
    \end{align*}
    Similarly, plugging \eqref{eq:means1} into $II$ we get
    \begin{align*}
        (N-h)(\tilde{\mu}(s)- \hat{\mu}(s))(\tilde{\mu}(t)- \hat{\mu}(t)) =&~ 
        (N-h)(\tilde{\mu}(s)- \frac{1}{N}(h \hat{\mu}_{\mathrm{clr},H}(s) - (N-h) \tilde{\mu}(s)))\\
        &~ \cdot(\tilde{\mu}(t)- \frac{1}{N}(h \hat{\mu}_{\mathrm{clr},H}(t) - (N-h) \tilde{\mu}(t))) \\ 
        =&~ (N-h)\left(\frac{h}{N}\tilde{\mu}(s)- \frac{h}{N}\hat{\mu}_{\mathrm{clr},H}(s)\right)\\
        & \cdot \left(\frac{h}{N}\tilde{\mu}(t)- \frac{h}{N}\hat{\mu}_{\mathrm{clr},H}(t)\right) \\
        =&~ \frac{(N-h)h^2}{N^2}(\tilde{\mu}(s) - \hat{\mu}_{\mathrm{clr},H}(s))(\tilde{\mu}(t) - \hat{\mu}_{\mathrm{clr},H}(t)) \\
        =&~ \frac{(N-h)h^2}{N^2}( \hat{\mu}_{\mathrm{clr},H}(s) - \tilde{\mu}(s))( \hat{\mu}_{\mathrm{clr},H}(t) - \tilde{\mu}(t)). 
    \end{align*}
    Adding $I$ and $II$ yields
    \begin{align*}
        I + II =&~ \frac{h(N-h)^2}{N^2}(\hat{\mu}_{\mathrm{clr},H}(s) - \tilde{\mu}(s))(\hat{\mu}_{\mathrm{clr},H}(t) - \tilde{\mu}(t)) + \frac{(N-h)h^2}{N_1^2}( \hat{\mu}_{\mathrm{clr},H}(s) - \tilde{\mu}(s))( \hat{\mu}_{\mathrm{clr},H}(t) - \tilde{\mu}(t)) \\
        =&~ \frac{(N-h)h}{N^2}[(N-h) + h](\hat{\mu}_{\mathrm{clr},H}(s) - \tilde{\mu}(s))(\hat{\mu}_{\mathrm{clr},H}(t) - \tilde{\mu}(t)) \\
        =&~ \frac{(N-h)h}{N}\underbrace{(\hat{\mu}_{\mathrm{clr},H}(s) - \tilde{\mu}(s))}_{=: \Delta(s)}\underbrace{(\hat{\mu}_{\mathrm{clr},H}(t) - \tilde{\mu}(t))}_{=: \Delta(t)}
    \end{align*}
    Inserting this into our initial derivation gives us
    \begin{align*}
        \hat{\gamma}(s,t) 
        =&~ \frac{1}{N}\left( h \hat{\gamma}_{\mathrm{clr},H,1}(s,t) + (N-h) \tilde{\gamma}(s,t) + \frac{N(N-h)h}{N^2}\Delta(s)\Delta(t)\right) \\
        =&~ \frac{h}{N} \hat{\gamma}_{\mathrm{clr},H,1}(s,t) + \frac{N-h}{N}\tilde{\gamma}(s,t) + \frac{N-h}{N}\frac{h}{N}\Delta(s)\Delta(t).
    \end{align*}
    Equivalently, this means 
    $$ \hat{\gamma}(s,t)  - \hat{\gamma}_{\mathrm{clr},H,1}(s,t) = -\frac{N - h}{N} \hat{\gamma}_{\mathrm{clr},H,1}(s,t) + \frac{N-h}{N}\tilde{\gamma}(s,t) + \frac{N-h}{N}\frac{h}{N}\Delta(s)\Delta(t).$$

    Before we continue with establishing the consistency of the sub-sampled covariance operator, we introduce another representation of the Hilbert-Schmidt norm. Recall that a Hilbert-Schmidt operator $T$ acting on $L^2(I)$ is an integral operator with kernel $k$ on $L^2(I \times I)$. Since $L^2$ is separable, a countable orthonormal basis $\{e_i \}_{i \geq 1}$ exists.  By Fubini's theorem, providing $k$ is continuous, we can define a function $k_s(t) := k(s,t)$ and further conclude $ T(e_i)(s) = \int_I k(s,t)e_i(t) \mathrm{d}t = \int_I k_s(t) e_i(t) \mathrm{d}t = \langle k_s,e_i\rangle_2^2$ for $i \geq 1.$ Since $\int_I \sum_{i \geq 1} \langle k_s, e_i \rangle_2^2 \mathrm{d}s = \| k \|_{L^2(I \times I)}^2 < \infty$, by dominated convergence we get $\sum_{i\geq1} \| T(e_i) \|^2_2 = \sum_{i\geq1} \int_I \langle k_s,e_i\rangle_2^2 \mathrm{d}s =  \int_I \sum_{i\geq1} \langle k_s,e_i\rangle_2^2 \mathrm{d}s$. Here, we have $\sum_{i\geq1} \langle k_s,e_i\rangle_2^2 = \|k_s\|_2^2 = \int_I k(s,t)^2 \mathrm{d}t$.
    Finally, we can write $\| T \|^2 = \sum_{i \geq 1} \| T(e_i)\|^2_2 = \int_I \int_I k^2(s,t) \mathrm{d}s \mathrm{d}t$. 
    Continuing from here, we get
    \begin{align*}
        \mathbb{E} \left[  \|\hat{C} - \hat{C}_{\mathrm{clr},H,1} \|^2_{\mathrm{HS}}\right] =&~ \mathbb{E} \left[  \int_I \int_I \left(\hat{\gamma}(s,t) - \hat{\gamma}_{\mathrm{clr},H,1}(s,t) \right)^2 \mathrm{d}s \mathrm{d}t \right] \\
        =&~ \mathbb{E} \Bigg[  \int_I \int_I \Bigg(-\frac{N - h}{N} \hat{\gamma}_{\mathrm{clr},H,1}(s,t) + \frac{N-h}{N}\tilde{\gamma}(s,t) \\
        &~~ + \frac{N-h}{N}\frac{h}{N}\Delta(s)\Delta(t) \Bigg)^2 \mathrm{d}s \mathrm{d}t \Bigg] \\
        &\leq~ 4 \mathbb{E} \int_I \int_I \left( -\frac{N-h}{N}\right)^2 \hat{\gamma}_{\mathrm{clr},H,1}(s,t)^2 + \left(\frac{N-h}{N}\right)^2\tilde{\gamma}(s,t)^2 \\
        &~~ + \left(\frac{N-h}{N}\right)^2\left(\frac{h}{N}\right)^2\Delta(s)^2\Delta(t)^2 \mathrm{d}s \mathrm{d}t\\
        =&~ 4 \left(\frac{N - h}{N}\right)^2\mathbb{E}\Bigg[ \int_I \int_I \hat{\gamma}_{\mathrm{clr},H,1}(s,t)^2 \mathrm{d}s \mathrm{d}t + \int_I \int_I \tilde{\gamma}(s,t)^2 \mathrm{d}s \mathrm{d}t \\
        &~~ + \int_I \int_I \left(\frac{h}{N}\right)^2\Delta(s)^2\Delta(t)^2 \mathrm{d}s \mathrm{d}t \Bigg] \\
        =&~ 4 \left(\frac{N - h}{N}\right)^2\mathbb{E}\Bigg[ \frac{1}{h} \sum_{i=1}^h \|X_i - \hat{\mu}_{\mathrm{clr},H} \|^2_2 + \frac{1}{N - h} \sum_{i=h+1}^{N} \|X_i - \tilde{\mu} \|^2_2 \\
        &~~ + \left(\frac{h}{N}\right)^2 \| \Delta \|_2^4 \Bigg].
    \end{align*}
    As by assumption $\mathbb{E}\|X_i\|^4<\infty$, and $X_i$, $i=1,\dots,n$ are mutually independent, then all of the upper expectations are uniformly bounded. 
    Moreover $\frac{N - h}{N}=\rho N^{\delta-1} \rightarrow 0$ for $N \rightarrow \infty$, as $\delta < 1$. Thus, 
    \begin{align*}
        \frac{\mathbb{E} \left[  \|\hat{C} - \hat{C}_{\mathrm{clr},H,1} \|^2_{\mathrm{clr},\mathrm{HS}}\right]}{(N - h)/ N } 
        \rightarrow 0 \text{ for } N \rightarrow \infty.
    \end{align*}
    Take now any $\varepsilon>0$ and choose $N$ large enough such that $\varepsilon>(N - h)/ N $. 
    Markov inequality now gives 
    \begin{align*}
        \mathbb{P}( \|\hat{C} - \hat{C}_{\mathrm{clr},H,1}\|_{\mathrm{HS}}^2 > \varepsilon) \leq \frac{\mathbb{E}\left[ \|\hat{C} - \hat{C}_{\mathrm{clr},H,1}\|_{\mathrm{HS}}^2\right]}{\varepsilon} \leq \frac{\mathbb{E} \left[  \|\hat{C} - \hat{C}_{\mathrm{clr},H,1} \|^2_{\mathrm{HS}}\right]}{(N - h)/ N },
    \end{align*}
    showing that $ \|\hat{C} - \hat{C}_{\mathrm{clr},H,1}\|_{\mathrm{HS}}^2 \overset{P}{\rightarrow} 0$. Observe that the statement of the Lemma can be broadened in a sense where we assume that the outliers originate from a certain mixture distribution with finite fourth moments, or for the setting where there is a finite (fixed) number of uncontrolled outliers. 
    Consistency of the sample covariance under the non-contaminated model and the triangle inequality of the HS norm give $\|{C}_{\clr,X} - \hat{C}_{\mathrm{clr},H,1}\|_{\mathrm{HS}} \overset{P}{\rightarrow} 0$. This finally leads to
    \begin{align*}
        \| C_{B^2,X} \ominus \hat{C}_{B^2,H,1}\|_{B^2,\mathrm{HS}} = \|\hat{C} - \hat{C}_{\mathrm{clr},H,1}\|_{\mathrm{HS}} \overset{P}{\rightarrow} 0,\quad N\to\infty.
    \end{align*}
    Claim $(ii)$ follows from \citet[Lemma 1]{Hormann2011ConsistencyOT}. 
    \end{proof}

\bibliographystyle{chicago}
\bibliography{references}  %%% Uncomment this line and comment out the ``thebibliography'' section below to use the external .bib file (using bibtex) .





% \end{thebibliography}